\documentstyle[preprint,prl,aps]{revtex}
\begin{document}

\input epsf

\preprint{\vbox{\hbox {July 1998} \hbox{IFP-759-UNC} } }

\title{\bf Quarks and Leptons Beyond the Third Generation.}
\author{\bf Paul H. Frampton}
\address{Department of Physics and Astronomy,}
\address{University of North Carolina, Chapel Hill, NC  27599-3255}
\author{\bf P.Q. Hung}
\address{Physics Department,}
\address{University of Virginia, Charlottesville, VA 22901}
\author{\bf Marc Sher}
\address{Physics Department}
\address{College of William and Mary, Williamsburg VA  23187}
\maketitle
\begin{abstract}
The possibility of additional quarks and leptons beyond the three generations
already established is discussed.  The make-up of this Report is (I)
Introduction:  the motivations for believing that the present litany of
elementary fermions is not complete; (II) Quantum Numbers:  possible
assignments for additional fermions; (III) Masses and Mixing Angles:  mass
limits from precision electroweak data, vacuum stability and perturbative gauge
unification; empirical constraints on mixing angles; (IV) Lifetimes and Decay
Modes:  their dependence on the mass spectrum and mixing angles of the
additional quarks and leptons; the possibility of exceptionally long lifetimes;
(V) Dynamical Symmetry Breaking:  the significance of the top quark and other
heavy fermions for alternatives to the elementary Higgs Boson; (VI) CP
Violation: extensions to more generations and how strong CP may be solved by
additional quarks; (VII) Experimental Searches:  present status and future
prospects; (VIII) Conclusions.
\end{abstract}
\newpage

\section{Introduction.}

The elementary spin-half fermions as we now know them are the quarks
and leptons. The principal constituents of normal atoms and normal matter
are the electron, as well as the up and down quarks which comprise the
valence quarks of the protons and neutrons in the atomic nucleus. In
addition, there is the electron neutrino which was first postulated
by Pauli\cite{1a} in 1931 to explain conservation of energy and
angular momentum in nuclear $\beta$-decay, and which was eventually 
discovered in 1956 by Reines and Cowan\cite{1b}.

These quarks and leptons - the up and down quarks, the electron and 
{\it its} neutrino - comprise what is now called the first generation.
The first intimation that Nature is more complex came with the discovery of
the muon in 1937\cite{1c,1d}. The muon appears identical to the electron
except for its mass which is $\sim 200$ times heavier. The muon appeared
so surprising that there was a famous comment by I.I. Rabi\cite{1e}:
"Who ordered that?" 

The fact that the muon neutrino differs from the electron
neutrino was established in 1962\cite{1f}. The strange quark
had already been discovered implicitly through the discovery
of strange particles beginning in 1944\cite{1g,1gg,1ggg}. 
Completion of the second family with the charmed quark,
predicted in 1970\cite{1h}, was accomplished in 1974\cite{1i,1ii,1j,1jj}.
At first only {\it hidden} charm was accessible but
two years later explicit charm was detected\cite{1k}.

By this time, a renormalizable gauge theory was available
\cite{1m,1n,1o,1p,1q} based on the first two generations
and incorporating the Cabibbo mixing\cite{1r} between
the two generations.

The situation became even more challenging to theorists when
experimental discovery of the third generation of quarks
and leptons began with the tau lepton, discovered in 1975\cite{1s}
in $e^+e^-$ scattering at SLAC. Next was the bottom quark in 1977
\cite{1t,1u}. The top quark, at $\sim 175\ $GeV much heavier than originally
expected, was finally discovered in 1995\cite{1v,1w}. Together
with the $\tau$ neutrino which presumably participates
(its distinct identity - while not questioned -
is not fully demonstrated) in tau decay, this completed the
third generation.

Since the present review is dedicated to the premise of further
quarks and leptons beyond the third generation, it is worthwhile
to recall to what extent and how the third generation
was anticipated from the existence of the first
two generations, why it is regarded as the end of the
litany of quarks and leptons 
and what loopholes there are in the latter arguments. 

One early theoretical anticipation of a third generation
was the paper of Kobayashi and Maskawa\cite{1x} who pointed out,
at a time (1973) when only three flavors u, d, s of quark
were established, that the existence of six flavors
in three generations would allow the standard model naturally
to accommodate CP violation. 

Study of the formation of the light elements
(Hydrogen, Deuterium, Helium, and Lithium) in
the early universe was started earlier
in the 1960's \cite{1y,1z}.
In the 1970's tighter constraints were
found based on the steadily-improving estimates 
of the primordial abundances of these
light isotopes. Since the expansion rate of the universe in this
era of  Big-Bang Nucleosynthesis, and hence the
abundances, depends sensitively on the number of 
light neutrinos it was then possible
to limit the acceptable number. The group
of Schramm {\it et al.}\cite{1aa,1ab,1ac}
found in this way that the number of generations
should not be greater than four\cite{1aa}, or in some
analyses not greater than three (see {\it e.g.} 
footnote 4 on page 242 of\cite{1ab}); it is
surely remarkable that such a strong
constraint was found from early universe
considerations already in 1979, a decade
before the situation was clarified using
colliders.

A current plot of the primordial $^4He$ abundance
(whose exact value is still controversial in 1998)
versus neutrino number $n(\nu)$ 
for $n(\nu) = 3.0, 3.2, 3.4$ is
given in Figure $1$. The main point is that the neutrino number
from cosmology is by now tied very closely to the high-energy
experimental value in what is the strongest known link between
particle theory and cosmology.

\begin{figure} 

\centerline{ \epsfysize 5in \epsfbox{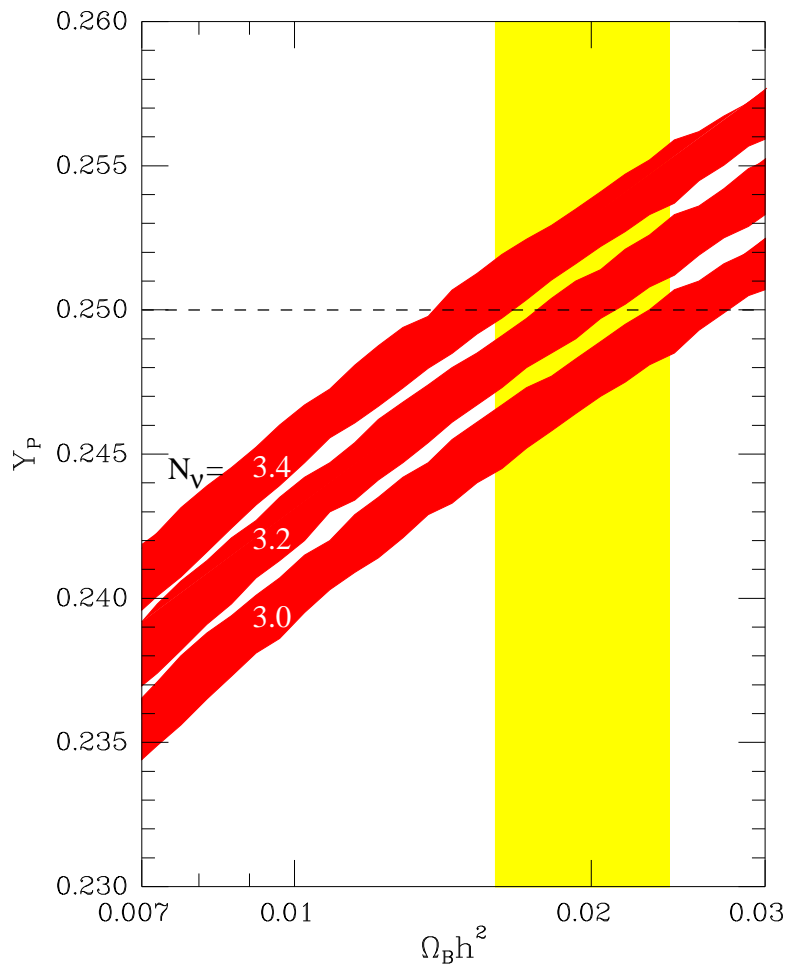}  }
\vskip .25in
\caption{Helium-4 production for $N_\nu=3.0,3.2,3.4$.  The vertical band indicates the baryon
density consistent with $(D/H)_P = (2.7\pm 0.6) \times 10^{-5}$ and the horizontal line indicates
a primeval Helium-4 abundance of $25\%$.  The widths of the curves indicate the two-sigma
theoretical uncertainty. Figure from Ref. \protect\cite{schrammturner}.}

\end{figure}

In 1989, there came an experimental {\it epiphany}
concerning the number of generations, or more
precisely, the number of light neutrinos.
This arose from the measurement of the Z width
at SLAC\cite{1aC,1aC2} and especially at CERN
\cite{1aCC,1aCCC,1aCCCC,1aCCCCC}.
The answer from this source is
indisputably equal to {\bf three}. The argument is
simple: 
One can measure the total width of the $Z$ to high accuracy, and then
subtract the visible width to get the invisible width.  Identifying the
invisible width with neutrino
decays leads to\cite{1aCCCCCC}:
\begin{equation}
n(\nu) = 3.00 \pm 0.025  \label{Nnu}
\end{equation}
This provides compelling proof that there are only
three conventional neutrinos with mass below
$M_Z/2 \simeq 45$ GeV. And, by extrapolation, it leads to
the idea that there are only three quark-lepton families.
Since this ties in nicely with the KM mechanism
of CP violation and with the Big-Bang-Nucleosynthesis,
the overall picture looks very attractive.

However, this finality was not universally accepted\cite{1ad,1ae}.
There are other reasons for entertaining the
possibility of further quarks and leptons:

\begin{itemize}

\item In many grand-unified theories (GUTs), there naturally occur
additional fermions. Although the minimal $SU(5)$ GUT
can contain only the basic three families, extension
to any higher GUT such as $SO(10)$ or $E(6)$ inevitably adds
new fermions. In $SO(10)$ this may be only right-handed neutrinos
(leptons) but in $E(6)$ there are also non-chiral color triplets(quarks) and 
color singlet(leptons).
Although there is no direct evidence for GUTs
they are attractive theoretically and suggestive of how the
standard nodel is extended.

\item Models of CP violation which solve the strong CP problem without axions
generically require additional quarks.

\item For some models of dynamical symmetry breaking\cite{1af,1ag,1ah,1ai}
the top quark mass, although higher than originally expected,
is still not quite high enough to play
its role in electroweak symmetry breaking. This
might be interpreted as evidence for even heavier quarks.

\item It has been shown that a non-supersymmetric model with four generations
can have successful unification of the gauge couplings at the unification
scale.  

\item  In recently-popular models of gauge-mediated supersymmetry
breaking, additional vectorlike quarks and leptons arise automatically.  In
addition, in models in which higher dimensions arise at the TeV
scale\cite{dienes}, it has been shown\cite{carone} that if some standard model
fields live in the higher dimensional space, low-scale gauge unification can be
obtained--the Kaluza-Klein excitations of these fields, which could be
rather light, must be vectorlike.

\end{itemize}

None of these reasons is fully compelling
but each is suggestive that one should
keep alive the study of this issue. Our hope
is that this review will play a role in
encouraging further thought about this open question.

The present review contains the following subsections:
Section II is on the possible quantum numbers of
additional quarks and leptons; Section III
discusses their masses and mixing angles;
Section IV deals with lifetimes and decay modes;
Dynamical symmetry breaking is in Section V; CP violation
is treated in Section VI,
and finally in Section VII there is a treatment of the experimental
situation and in Section VIII are the conclusions.

\newpage

\section{Quantum Numbers.}

When we add fermions to the standard model, there are choices in the
possible quantum numbers.

Under the color $SU(3)_C$ group we refer to {\it color triplets}
as {\it quarks}, color antitriplets as antiquarks. Color singlets
which do not experience the strong interaction are generically referred
to as {\it leptons}. Higher representations of color such
as {\bf $6$, $\bar{6}$, $8$,...} may be called quixes, antiquixes, queights,
and so on. Such exotic color states are necessary in some models 
to cancel chiral anomalies. For example, in chiral color\cite{2aa,2aaa}
one version (called Mark II in\cite{2aa}) involves three conventional
fermion generations, an extra $Q = 2/3$ quark, and an $SU(3)$ sextet
fermion or quix. The extended gauge group of chiral color is
$SU(3)_L \times SU(3)_R \times SU(2)_L \times U(1)_Y$
and we may list the fermions by their $(SU(3)_L,SU(3)_R,Q)$
quantum numbers. There are three colored weak doublets
\begin{equation}
3[(3,1,\frac{2}{3}) + (3,1,-\frac{1}{3})],
\end{equation}
eight colored weak doublets
\begin{equation}
4(1,\bar{3},-\frac{2}{3}) + 3(1,\bar{3},+\frac{1}{3}) + (3, 1, \frac{2}{3}),
\end{equation}
a weak singlet quix
\begin{equation}
(\bar{6},1,-\frac{1}{3}) + (1,6,\frac{1}{3}),
\end{equation}
and three charged leptons and their neutrinos. The
quix plays an essential role in anomaly cancellation.
But in this review we restrict our attention only to quarks and leptons
because while more exotic color states are a logical possibility
it is one which is difficult to categorize systematically.

Quarks and leptons may be either chiral or non-chiral. The latter are sometimes alternatively
called vector-like. Let us therefore define the meaning
of these adjectives.

Chiral fermions are, for present purposes, spin-$\frac{1}{2}$
paricles where the left and right components transform differently
under the electroweak gauge group $SU(2)_L \times U(1)$. All
the fermions of the standard model are chiral. This
means that they are strictly
massless before the electroweak symmetry is broken.

The simplest generalization of the standard model is
surely
to add a fourth sequential family. Of course, a fourth light neutrino
is an immediate phenomenological problem with the invisible $Z$ width,
but the addition of a right-handed neutrino can resolve this.

More generally we may add a {\it chiral doublet } quark or lepton
where the left-handed components transform as a doublet of $SU(2)_L$
and the right-handed components as singlets. A chiral doublet of quarks
is:
\begin{equation}
\left( \begin{array}{c} U \\ D \end{array} \right)_L;
\hspace{0.2in}U_R;\hspace{0.2in}D_R  \label{chdoub}
\end{equation}  
while a chiral doublet of leptons is
\begin{equation}
\left( \begin{array}{c} N \\ E \end{array} \right)_L;
\hspace{0.2in}N_R;\hspace{0.2in}E_R
\end{equation}  

Equally possible are chiral singlets such as
\begin{equation}
U_L
\end{equation}
or
\begin{equation}
D_L
\end{equation}
for quarks, or for leptons
\begin{equation}
N_L
\end{equation}
or
\begin{equation}
E_L
\end{equation}

Of course, with chiral doublets or singlets, there is a constant danger
of chiral anomalies. In the standard model, there is a spectacular cancellation
in each generation between the anomalies of the chiral doublets of 
quarks and leptons. In other models one
sometimes adds mirror chiral doublets to cancel anomalies. For
example, a {\it mirror chiral doublet of quarks} is
\begin{equation}
U_L; \hspace{0.2in} D_L;\hspace{0.2in} \left(\begin{array}{c} U \\ D 
\end{array} \right)_R
\end{equation}

There can also be non-chiral (also known as {\it vector-like}) fermions, where
the right and left components transform similarly under
the electroweak $SU(2) \times U(1)$ group. For example a vector-like
quark doublet is
\begin{equation}
\left( \begin{array}{c} U \\ D \end{array} \right)_L;\hspace{0.2in} 
\left( \begin{array}{c} U \\ D \end{array} \right)_R
\end{equation}

A vector-like doublet is not to be confused with the doublets occurring in the
left-right model\cite{2a,2b,2c} where the gauge group is extended to
$SU(2)_L \times SU(2)_R \times U(1)_{B-L}$.
The U and D quarks transform there chirally as in (\ref{chdoub}) above
under $SU(2)_L$, but the right-handed singlets 
$U_R$ and $D_R$ of $SU(2)_L$ are then 
assumed to transform
as a doublet under the additional gauge group $SU(2)_R$.

General patterns for adding anomaly-free charge-vectorial chiral sets
of fermions which acquire mass by coupling to the Higgs doublet of the standard model
have been studied in \cite{2cc} and further developed in \cite{2ccc,2cccc}.
An extensive analysis of the possible quantum numbers for additional fermions
can be found in Ref. \cite{flvj}.  They looked at the general structure of
exotic generations given the gauge and Higgs structure of the standard model.
A similar analysis for left-right symmetric models was carried out in Ref.
\cite{choi}.

In {\it grand unified theories} (GUTs) all types of additional fermions are possible.
For example, in $SU(5)$ there may be no additional fermions. But in $SO(10)$
there must be at least an additional chiral right-handed neutrino in each family.
In $E(6)$ each {\bf 27} of fermions contains not only two extra 
neutrino-like states but also a ${\bf 5 + \bar{5}}$ of $SU(5)$ which
contains a non-chiral singet of quarks and a non-chiral doublet of leptons. 

In superstring models, and M-theory models, the additional fermions
are even less constrained. For example $E(6)$ and its content has
been a familiar superstring possibility since the beginning\cite{1d},
and any gauge group contained in at least $O(44)$ is possible.
Families and anti (mirror) families often occur in superstrings.

Some extensions of the standard model require extra chiral fermions to
cancel anomalies {\it e.g.}

\begin{itemize}

\item As already mentioned, chiral color\cite{2aa}, anomaly cancellation dictates addition
of quixes.

\item In some GUTs, {\it e.g.} $SU(15)$ anomaly cancellation requires\cite{2dd}
inclusion of mirror fermions.

\end{itemize}
More interesting are the types of additional quarks and leptons
appearing in extensions of the standard model motivated by
attempting to explain shortcomings of the model itself.

Examples are:
\begin{itemize}

\item Attempts to solve the strong CP problem without an axion\cite{6aa,6ab}
can introduce a non-chiral
doublet of quarks, or non-chiral singlets (but not both),

\item In trying to explain the three families through anomaly
cancellation, the 331-model\cite{2e,2f} extends the individual families of
the standard model by adding non-chiral singlets of quarks. The
charges of the additional quarks differ between the families. 
In a sense, this is not ``Quarks and Leptons beyond the Third Generation"
since the quarks are being added to the discovered generations.
Nevertheless, our framework is sufficiently general to accommodate
this possibility - as additional {\it non-chiral singlets} of quarks.

\item In the non-supersymmetric standard model, gauge unification of the
couplings fail--the three couplings do not meet at a point.  It has been
noted\cite{pqh} that extending the model to allow a fourth generation introduces
enough flexibility that a successful unification of couplings can occur.

\item A potential problem of the minimal supersymmetric standard model is that
the scalar quark masses must be very nearly degenerate to avoid large
tree-level flavor-changing neutral currents.  This degeneracy is very
unnatural, using the conventional mechanism of gravitationally mediated
supersymmetry breaking (where the supersymmetry breaking part of the
Lagrangian is transmitted to the known sector via gravitational
interactions).  However, in gauge-mediated supersymmetry breaking, the
supersymmetry breaking is transmitted to the known sector via the gauge
interactions of ``messenger" fields.  Since the gauge interactions are
flavor-blind, the degeneracy of squark masses is natural.  The messenger
fields are, in the simplest case, composed of a $5+\bar{5}$ of $SU(5)$ (or of
several such fields), which constitute a vectorlike $Q=-1/3$ quark and a
vectorlike lepton.

\item  There has been recent excitement about the possibility that 
additional spacetime dimensions could be compactified at (or even well
below) the TeV scale\cite{dienes}.  In such models, the Kaluza-Klein excitations
must be vectorlike (to avoid having, for example, too many light neutrinos).

\end{itemize}

In this Report, we will primarily concentrate on chiral quarks and leptons, or vectorlike quarks 
and leptons, since they appear in the majority of models.  However, when appropriate, we will
note how our various constraints and bounds will apply to mirror quarks.

\newpage

\section{Masses and Mixing Angles.}  

One of the most unsatisfactory features of the standard model is the apparent
arbitrariness of the masses and mixing angles of the known fermions.  Although
masses and mixing angles can be accommodated in the standard model (with addition of
right handed neutrino fields, if necessary), there is no understanding of 
their values.  An entire industry of model-building has developed in an attempt to
provide some theoretical guidance, ranging from flavor symmetries to relationships
from grand unification, but no model seems particularly compelling.

In the case of additional fermions, the masses and mixing angles also are arbitrary.
Nonetheless, some general features can be found.   Phenomenological bounds can be
obtained from high precision electroweak studies, theoretical bounds can be
obtained from requiring the stability of the standard model vacuum and from 
 requiring that perturbation theory be valid (up to some
scale).  In this Section, these bounds are explored in some
detail.  We will start with a discussion of the phenomenological
constraints from precision electroweak studies.  Then we will consider bounds from
vacuum stability and perturbation theory. In the vast majority of analyses of these
bounds, the authors focused on bounds to the top quark mass, since its mass was
unknown until relatively recently, thus we will first look at constraints on the top
quark mass, and then generalize the results to find bounds on masses of additional
quarks and leptons.  Finally, we will discuss plausible models for the mixing angles.

\subsection{Precision Electroweak Constraints}

\subsubsection{Chiral Fermions}

In the past decade, high precision electroweak measurements have led to remarkable
constraints on potential physics beyond the standard model.   The most important of
these is the $\rho$ parameter.  As originally pointed out by
Veltman\cite{veltman,veltman1}, the tree level mass relation
\begin{equation}
\rho\equiv {M^2_W\over M_Z^2\cos^2\theta_W}=1
\end{equation}
is very sensitive to non-standard model physics (ruling out, for example,
significant vacuum expectation values for Higgs triplets).  Since the relation is
good to better than $1\%$, it is assumed that deviations from the relation are due
to electroweak radiative corrections, which are sensitive to new particles in loops.

An extensive, detailed analysis of electroweak radiative corrections can be
found (with a long list of references) in the work of Peskin and
Takeuchi\cite{peskin,peskin1}.  They define the S and T parameters as
\begin{eqnarray}
\alpha S&\equiv 4e^2[\Pi^\prime_{33}(0)-\Pi^\prime_{3Q}(0)]\cr
\alpha T&\equiv {e^2\over
\sin^2\theta_WM^2_W}[\Pi_{11}(0)-\Pi_{33}(0)]
\end{eqnarray}
where $\alpha$ is the fine-structure constant, $\Pi_{XY}(q^2)$ is the
vacuum-polarization amplitude with $(XY)=(11),(22),(33),(3Q),(QQ)$, and
$\Pi^\prime ={\Pi(q^2)-\Pi(0)\over q^2}$.   Roughly, $T$ is a measure of the
deviation of the $\rho$ parameter from unity, coming from isospin violating
contributions.   $S$ is an isospin symmetric quantity which measures the momentum
dependence of the vacuum polarization; it is roughly the ``size" of the new physics
sector\footnote{An alternative representation, using parameters $\epsilon_1$,
$\epsilon_2$ and $\epsilon_3$ can be found in Ref. \cite{altbarb}.}

As an example, Peskin and Takeuchi consider the case of a chiral lepton doublet,
$E$ and $N$.  In the limit $M_N, M_E >>M_Z$, they show that if the mass splitting
is small, then the contribution for $S$ is just ${1\over 6\pi}$, and the
contribution for $T$ is ${|\Delta M^2|\over 12\pi\sin^2\theta_W M^2_W}$.
For a chiral quark doublet, these contributions are  tripled.  As stated above,
one can see that $T$ is a measure of the isospin splitting, while $S$ is a measure
of the size of the new sector.  Thus, for a complete degenerate fourth generation, the
contribution to $S$ is
\begin{equation}\label{stheory}
S={2\over 3\pi}\sim 0.21
\end{equation}
while the contribution to $T$ is
\begin{equation}
T={|\Delta M_L^2|\over 12\pi\sin^2\theta_W M^2_W}+{|\Delta M_Q^2|\over
4\pi\sin^2\theta_W M^2_W}
\end{equation}
It is important to note that these results were obtained under the assumption that
the extra fermion masses were much greater than that of the $Z$, and that the
mass splitting was small.  Peskin and Takeuchi give more exact expressions.

What are the experimental bounds?  The values can be determined from the data on
Z-pole asymmetries, $M_Z$, $M_W$, $R_l$, $m_{top}$, $R_b$ and Z-decay widths.  The
contribution from new physics, $S_{new}=S-S_{SM}$ and $T_{new}=T-T_{SM}$, can be
determined\cite{1aCCCCCC,hagiwara,erler}.  The value of this contribution depends on the Higgs
and top quark masses (which affect $S_{SM}$, for example).  For a top mass of $175$
GeV and a Higgs mass between $M_Z$ and $1$ TeV, Erler and Langacker\cite{erler} find 
\begin{eqnarray}\label{sexp}
S_{new}&= -0.20^{+0.40}_{-0.33}\ (3\sigma)
\end{eqnarray}
One can immediately see a apparent conflict
between Eqs.
\ref{stheory} and \ref{sexp}.   Independent of the mass, a fourth generation
chiral multiplet is roughly $3$ standard deviations off.  This led Erler and Langacker to claim that a fourth sequential family is excluded at the $99.8\%$ confidence level.

However, there are several reasons why we believe it is premature to exclude a fourth sequential family.   First, of course, is the fact that many $3\sigma$ effects in recent years have disappeared.   Second, the results for $S$ are based on virtual heavy fermion loops.   {\it Any additional new physics will likely make a similar contribution.}  For example, Erler and Langacker show that the allowed range for $S$ in minimal SUSY is $-0.17^{+0.17}_{-0.12}$, where the error bars are $1\sigma$, and this is only in conflict by 2.2 standard deviations.
Third, it has been noted\cite{terning} that Majorana neutrino masses (which may be needed to give
neutrino masses in the right range) lower $S$, as do models which involve mixing of two scalar doublets\cite{lavourali}, and this could reduce the
discrepancy further.   Finally, the result above assumed a degenerate family.  If one uses the exact expressions, takes $M_H=180$ GeV, $M_D=M_U=150$ GeV, $M_N=100$ GeV and $M_E=200$ GeV, for example, one finds the contribution to $S$ to be approximately $0.11$, rather than $0.21$.  This also would lower the discrepancy to slightly more than $2\sigma$.

For $S\sim 0.2$, the $2\sigma$ upper bound on $T$ is approximately $0.2$, leading to bounds on
the mass splitting from Eq. \ref{stheory}.   For quarks, the splitting must be less\cite{erler}
than $54(M_W/M_D)$ GeV and for leptons must be less than $162(M_W/M_E)$ GeV, at the $2\sigma$ level.   Note
that for quark masses above $200$ GeV, the splitting must be less than 10 percent.

Of course, the extra fermions must not contribute significantly to the width of the
$Z$, and their masses are thus bounded from below by $M_Z/2$.  Other experimental
bounds will be discussed in Section VII.

\subsubsection{Non-chiral Doublets}

Vector-like fermions do not contribute in leading order to $S$ and $T$, and thus
the values of these parameters do not constrain their masses.  However, since
vector-like doublets do not couple to the Higgs boson, the mass terms
involving the $E$ and the $N$ cannot violate isospin invariance, and thus
the masses must be degenerate at tree-level, as must the masses of the $U$ and
$D$.  Even if one adds a singlet Higgs field, the degeneracy will remain.  Only
a Higgs triplet can split this degeneracy, but Higgs triplet vacuum expectation
values are severely constrained by the $\rho$ parameter.  We conclude (in the
absence of sizeable mixing with lighter generations) that {\it the masses of
states in a vector-like doublet are degenerate at tree level}.  The masses will
be split by a few hundred MeV due to electroweak radiative corrections--this
calculation will be done in the next Section. 

\subsubsection{Other Fermions}

Non-chiral singlets will have arbitrary mass terms and arbitrary couplings to any
Higgs singlets.  No constraints can be placed on their masses.  
\newpage

\subsection{Vacuum Stability Bounds}

Upper bounds on fermion masses can be obtained from the requirement that
fermionic corrections to the effective potential not destabilize the standard
model vacuum.  We will first discuss the effective potential, and its
renormalization group improvement.  Then, the bounds from the requirement of
vacuum stability will be discussed, first for the top quark mass, and then for
additional quarks and leptons.

\subsubsection{The Effective Potential}

An extensive review of the effective potential and bounds from vacuum
stability appeared in 1989\cite{sher}.  Since then, the potential has been
improved, including a proper renormalization-group improvement of scalar loops, and
the bounds have been refined to much higher precision.  In addition, the discovery
of the top quark has narrowed the region of parameter space that must be
considered.  In this section, we discuss the effective potential and its
renormalization-group improvement.

It is easy to see how bounds on fermion masses can arise.  The one-loop effective
potential, as originally written down by Coleman and Weinberg\cite{cw} can be
written, in the direction of the physical $\phi$ field, as
\begin{equation}
V(\phi)=V_0+V_1
\end{equation}
where
\begin{equation}\label{first}
V_0=-{1\over 2}\mu^2\phi_c^2+{1\over 4}\lambda \phi_c^4
\end{equation}
and \begin{equation}
V_1={1\over 64\pi^2}\sum_i (-1)^F\eta_i {\cal M}^4(\phi_c)\ln{{\cal
M}^2(\phi_c)\over M^2}
\end{equation}
where the sum is over all particles in the model, F is the fermion number, $\eta_i$
is the number of degrees of freedom of the field i, and ${\cal M}^2(\phi_c)$ is the
mass that the field has in the vacuum in which the scalar field has a value
$\phi_c$.  In the expression for $V_1$, we have ignored terms which can be absorbed
into $V_0$--these will be fixed by the renormalization procedure.  In the standard
model, we have for the W-boson, ${\cal M}^2(\phi_c)={1\over 4}g^2\phi_c^2$, for the
Z-boson, ${\cal M}^2(\phi_c)={1\over 4}(g^2+g^{\prime 2})\phi_c^2$, for the Higgs
boson, ${\cal M}^2(\phi_c)=-\mu^2+3\lambda\phi_c^2$, for the Goldstone bosons,
${\cal M}^2(\phi_c)=-\mu^2+\lambda\phi_c^2$ and for the top quark ${\cal
M}^2(\phi_c)={1\over 2} h^2\phi_c^2$.  For a very large values of $\phi$, quadratic
terms are negligible and the potential becomes
\begin{equation}\label{second}
V={1\over 4}\lambda\phi^4+B\phi^2\ln(\phi^2/M^2)
\end{equation}
where
\begin{equation}\label{bvalue}
B={3\over 64\pi^2}[4\lambda^2+{1\over 16}(3g^4+2g^2g^{\prime 2}+g^{\prime 4})-h^4]
\end{equation}
One can see that if the top quark is very heavy, then $h$ is large and thus $B$ is
negative.  In this case, the potential is unbounded from below at large values of
$\phi$.  This is the origin of the instability of the vacuum caused by a heavy
quark.

Although this form of the effective potential is well known, it is NOT useful in
determining vacuum stability bounds.  The reason is as follows.  Suppose one
denotes the largest of the couplings in a theory by $\alpha$,  in the standard
model,for example, $\alpha=[\max(\lambda,g^2,h^2)]/(4\pi)$.  The loop expansion is an
expansion in powers of $\alpha$, but is also an expansion in powers of logarithms of
$\phi_c^2/M^2$, since each momentum integration can contain a single logarithmic
divergence, which turns into a $\ln(\phi^2_c/M^2)$ upon renormalization.  Thus the
$n$-loop potential will have terms of order
\begin{equation}
\alpha^{n+1}[\ln(\phi^2/M^2)]^n
\end{equation}
In order for the loop expansion to be reliable, the expansion parameter must be
smaller than one.  $M$ can be chosen to make the logarithm small for any specific
value of the field, but if one is interested in the potential over a range from
$\phi_1$ to $\phi_2$, then it is necessary for $\alpha \ln(\phi_1/\phi_2)$ to be
smaller than one.  In examining vacuum stability, one must look at the potential at
very large scales, as well as the electroweak scale, and the logarithm is generally
quite large.  Thus, any results obtained from the loop expansion are unreliable;
in fact, the bound on the top quark mass can be off by more than a factor of
two.

A better expansion, which does not have large logarithms, comes from solving the
renormalization group equation (RGE) for the effective potential. This equation is
nothing other than the statement that the potential cannot be affected by a change
in the arbitrary parameter, $M$, i.e.  $dV/dM=0$.  Using the chain rule,
this is
\begin{equation}
[M{\partial\over\partial M}+\beta(g_i){\partial\over\partial
g_i}-\gamma\phi{\partial\over\partial\phi}]V=0
\end{equation}
where $\beta=Mdg_i/dM$ and there is a beta function for every coupling and mass
term in the theory.  The $\gamma$ function is the anomalous dimension.

It is important to note that the renormalization group equation is exact and no
approximations have been made.  If one knew the beta functions and anomalous
dimensions exactly, one could solve the RGE exactly and determine the full
potential at all scales.  Although we do not know the exact beta functions and
anomalous dimensions, we do have expressions for them as expansions in couplings. 
Thus, by {\it only} assuming that the couplings are small, the beta functions and
$\gamma$ can be determined to any level of accuracy and $V(\phi)$ can be found. 
The resulting potential will be accurate if $g_i<<1$ and will not require
$g_i\ln(\phi/M)<<1$.

For example, in massless $\lambda\phi^4$ theory, the RGE can be solved exactly to
give
\begin{equation}
V={1\over 4}\lambda'(t,\lambda)G^4(t,\lambda)\phi^4
\end{equation}
where $t=\ln(\phi/M)$ and $\lambda'(t,\lambda)$ is defined to be the solution of the
equation
\begin{equation}
{d\lambda'\over dt}={\beta(\lambda')\over (1+\gamma(\lambda'))}
\end{equation}
with the boundary condition being determined by the renormalization condition. 
$G(t,\lambda)$ is defined as $\exp(-4\int_0^t dt'
(\gamma(\lambda')/(1+\gamma(\lambda')))$.
Note that this potential gives the same result as before in the limit that
$\gamma=0$ and $\beta=$ constant.  Then $G=1$ and $\lambda'=\beta t$ + constant.
With $t=\ln(\phi/M)$ this gives the $\phi^4\ln(\phi/M)$ terms as above.

What about the massive case?  The RGE is given by
\begin{equation}
[M{\partial\over\partial
M}+\beta_\lambda{\partial\over\partial\lambda}+\beta(g_i){\partial\over\partial
g_i}
+\beta_{\mu^2}\mu^2{\partial\over\partial\mu^2}
-\gamma\phi{\partial\over\partial\phi}]V=0
\end{equation}
One is tempted to reduce this equation to a set of ordinary differential equations
as before, giving
\begin{equation}\label{firstrge}
V(\phi)={1\over 2}\mu^2(t)g^2(t)\phi^2+{1\over 4}\lambda(t)G^4(t)\phi^4,
\end{equation}
where the coefficients are running couplings obeying first order differential
equations as in the massless case.

However, this is not correct.  By considering small excursions in field space, one
does not, as in the massless case, reproduce the unimproved one-loop potential. 
This is not surprising.  In the massless theory, the only scale is set by $\phi$,
and thus all logarithms must be of the form $t=\ln(\phi^2/M^2)$.  In the massive
theory, there is another scale, and there will be logarithms of the form
$\ln((-\mu^2+3\lambda\phi^2)/M^2)$.  Thus one cannot easily sum all of the
leading logarithms.  In addition, the scale dependence of the constant term in
the potential (the cosmological constant) can be relevant.

In earlier work (and in the review of Ref. \cite{sher}), it was argued that the
bounds only depend on the structure of the potential at large $\phi$, and thus the
mass term and constant term are irrelevant.  However, in going from $\lambda$ to the
Higgs mass, the structure of the potential near its minimum is important, and thus
using the naive expression above is not as accurate (although it is fairly close). 
This will be discussed more in the next section.

More recently, Bando, et al.\cite{bando} and Ford, et al.\cite{ford}, following
some earlier work by Kastening\cite{kastening}, found a method of including the
additional logarithms found in the massive theory.  In general, they showed that if
one considers the $L$-loop potential, and runs the parameters of that potential
using $L+1$ beta and gamma functions, then all logarithms will be summed up to the
Lth-to-leading order.  The standard model potential, including all leading and
next-to-leading logarithms, is then (in the 't Hooft gauge)
\begin{eqnarray}\label{potfinal}V(\phi)&=-{1\over 2}\mu^2\phi^2+{1\over
4}\lambda\phi^4+ {1\over 16\pi^2}\big[{3\over 2}W^2\left(\ln{W\over M^2}-{5\over
6}\right)+ {3\over 4}Z^2\left(\ln{Z\over M^2}-{5\over 6}\right)\cr
&+{1\over 4}H^2\left(\ln{H\over M^2}-{3\over 2}\right)
+{3\over 4}G^2\left(\ln{G\over M^2}-{3\over 2}\right)
-3T^2\left(\ln{T\over M^2}-{3\over 2}\right)\big]
\end{eqnarray}
with $W\equiv g^2\phi^2/4$, $Z\equiv (g^2+g^{\prime 2})\phi^2/4$, $H\equiv 
-\mu^2+3\lambda\phi^2$, $G\equiv -\mu^2+\lambda\phi^2$ and $T=h^2\phi^2/2$.
All of the couplings in this potential run with $t=\ln{\phi/M}$. Use of two-loop
beta and gamma functions will then give a potential in which all leading and
next-to-leading logarithms are summed over.  It was shown by Casas, et al.\cite{cas}
that the resulting minima and masses are relatively independent of the precise
choice of
$M$, {\it as long as} this potential is used (use of earlier potentials was
inaccurate due to a sensitive dependence on the choice of scale).  It is this
potential that will be used to determine bounds on the top quark and Higgs masses in
the next section.

First, one should comment on the gauge-dependence of the potential.  Bounds on the masses of the top quark and Higgs boson are physical quantities, so how can one draw conclusions based on a gauge-dependent potential?  It has long been known\cite{sher} that the existence (or non-existence) of minima of the potential are gauge-independent; an early calculation of the mass of the Higgs boson in the Coleman-Weinberg model\cite{kang} to two-loops in the $R_{\xi}$ gauge showed that the gauge-dependence drops out in the final result.  A detailed analysis of the gauge-dependence of the bounds on the Higgs and top quark masses has been carried out by Loinaz, Willey, et al.\cite{loinaza,loinazb,loinazc}.  They find a gauge-invariant procedure for determining the bounds, and find that the final result is numerically very close to the procedure discussed below.  In a model with stronger gauge couplings, however, the gauge invariant method might give significantly different results.

\subsubsection{Bounds on the top quark and Higgs masses}

 The first paper to notice that fermionic one-loop corrections
could destabilize the effective potential was by Krive and Linde\cite{krive},
working in the context of the linear sigma model.  Later, independent
investigations by Krasnikov\cite{krasnikov}, Hung\cite{hung}, Politzer and
Wolfram\cite{pw} and Anselm\cite{anselm} all looked at the one-loop,
non-renormalization group improved potential of Eqs. \ref{first} and \ref{second}.,
and required that the standard model vacuum be stable for all values of $\phi$.   The
first of these was that of Krasnikov\cite{krasnikov} who noted that the bound would
be of O(100) GeV, rising to O(1000) GeV if scalar loops were included.  The works of
Politzer and Wolfram\cite{pw} and Anselm\cite{anselm} gave much more precise
numerical results, but ignored scalar loop contributions--thus they obtained
upper bounds of $80-90$ GeV on the top quark mass.  Hung\cite{hung} gave
detailed numerical results and did include scalar loops, thus his upper bound
ranged from $80$ GeV to $400$ GeV as the Higgs mass ranged from $0$ to $700$
GeV.  

All of these results are unreliable because the potential used is not valid for
large values of $\phi$.  In these papers, the instability would occur for large
values of $\phi$, and thus $\ln(\phi/\sigma)$ is large enough that only a
renormalization group improved potential is reliable.  The first attempt to use
an improved potential was the work of Cabibbo, Maiani, Parisi and
Petronzio\cite{cmpp}.  They included the scale dependence of the Yukawa and
gauge couplings, and required that the effective scalar coupling be positive
between the weak scale and the unification scale.  Although they didn't use the
language of effective potentials, this procedure turns out to be very close to
that used by considering the full renormalization group improved effective
potential.  Similar results, using the language of effective potentials, was
later obtained by Flores and Sher\cite{flores}.

Use of the renormalization-group improved potential will weaken the bounds. 
The beta function for the top quark Yukawa coupling is negative, and thus the
coupling falls as the scale increases.  Thus, the effects of fermionic
corrections will decrease at larger scales.  Compared with the bounds that one
would obtain by ignoring the renormalization-group improvement, the decrease in
the Yukawa coupling at large scales will weaken the upper bounds.  This effect
is not small; the Yukawa coupling for a quark will fall by roughly a factor of three
between the weak and unification scales.  Note that for additional leptons, the
Yukawa coupling does not fall significantly, thus the bounds obtained by the
non-renormalization-improved potential will not be greatly changed.

The first attempt to bound fermion masses using the full renormalization group
improved effective potential (earlier works, for example, never mentioned
anomalous dimensions) was the 1985 work of Duncan et al.\cite{duncan}.  Their results,
however, used tree level values for the Higgs and top masses, in terms of the scalar
self-coupling and the ``$\overline{MS}$ Yukawa coupling", and found a bound which,
to within a couple of GeV, can be fit by the line
\begin{equation}\label{dunceq}
M_{top}\ <\  80\ {\rm GeV} + 0.54 M_{Higgs}\end{equation}  As we
will see shortly, however, corrections to the top quark mass can be sizeable, as
much as 10 GeV.  A much more detailed analysis, using two loop beta functions and
one-loop corrections to the Higgs and top quark masses (defined as the poles of the
propagator), was carried out in 1989 by Lindner, Zaglauer and Sher\cite{lsz}, and
followed up with more precise inputs in 1993 by Sher\cite{sh93}.  In all of these
papers, the allowed region in the Higgs-top mass plane was given--the allowed region
was always an upper bound on the top mass for a given Higgs mass, or a lower bound
on the Higgs mass for a given top mass.  The allowed region depended on the cutoff
$\Lambda$ at which the instability occurs.  For example, if the instability occurs
for values of $\phi$ above $10^{10}$ GeV, then one concludes that the standard
model vacuum is unstable IFF the standard model is valid up to $10^{10}$ GeV (should
the lifetime of the metastable vacuum be less than the age of the Universe, one
would conclude that the standard model cannot be valid up to $10^{10}$ GeV). 
Thus, all of the bounds depend on the value of $\Lambda$.

In the above papers, the effective potential used was the renormalization group
improved tree-level potential, Eq.\ref{firstrge}.  As discussed in the previous
section, this would be as precise as the precision of the beta functions and
anomalous dimensions (two-loop were used) if the only logarithms were of the form
$\ln({\phi^2\over M^2})$; the resulting potential is exact in terms of the beta and
gamma functions.  However, when scalar loops are included, terms of the form
$\ln({\mu^2+\lambda\phi^2\over M^2})$ arise, and these terms are not summed over. 
In the earlier papers, it was argued that when
$\phi$ is large, the scalar terms are effectively of the form $\ln({\phi^2\over
M^2})$, and thus the difference is irrelevant.  But, in determining the Higgs
boson mass in terms of the potential, the structure of the potential at the
electroweak scale is relevant, and thus the difference in the form of the scalar
loops is relevant.  It turns out that this difference is especially crucial when
the value of $\Lambda$ is relatively small ($1-10$ TeV), and less important when
$\Lambda$ is large ($10^{15-19}$ GeV), thus the results of the above papers are
valid in the large $\Lambda$ case.

To include the proper form of the scalar loops, one must use the form of Ford, et
al.\cite{ford}, discussed in the last section.  This analysis was carried out very
recently by Casas, Espinosa and Quiros\cite{casas} and by Espinosa and
Quiros\cite{espinosa}\footnote{See Altarelli and Isidori\cite{altarelli} for a
similar and independent analysis.}.  A very pedagogical review of the analysis can be
found in  Espinosa's Summer School Lectures\cite{esprev}.   We now
briefly review that analysis and present their results.

Consider the tree level renormalization group improved potential, Eq.
\ref{firstrge}.  At large values of $\phi$, the quadratic term becomes 
negligible, and the question of whether the standard model vacuum is stable is
essentially identical to the question of whether $\lambda(t)$ ever goes negative.
If $\lambda(t)$ goes negative at some scale $\Lambda$, then the instability will
occur at that scale\footnote{For a discussion of the relationship between the
location of the instability and the required onset of new physics, see Ref.
\cite{hungsh}.}.

Casas, et al.\cite{cas,casas} analyzed the question using the full one-loop
renormalization group improved potential, with two-loop beta and gamma functions,
of Eq. \ref{potfinal}.  They showed that the instabilty sets in when
$\tilde{\lambda}$ becomes negative, where $\tilde{\lambda}$ is slightly different
from $\lambda$:
\begin{equation}
\tilde{\lambda}=\lambda-{1\over 16\pi^2}\big[3h^4\left(\ln{h^2\over 2}-1\right)
-{3\over 8}g^4\left(\ln{g^2\over 4}-{1\over 3}\right)-{3\over 16}(g^2+g^{\prime
2})^2\left(\ln{g^2+g^{\prime 2}\over 4}-{1\over 3}\right)\big]
\end{equation}
All that remains is to relate the parameters in the potential to the physical
masses of the Higgs boson and of the top quark.

It is not a trivial matter to extract the Higgs and top quark masses from the
values of $h(t)$ and $\lambda(t)$ used in the potential.  One can write 
\begin{eqnarray}
m_{top}(\mu)&=m_{top}^{pole}(1+\delta_{top}(\mu))={1\over\sqrt{2\sqrt{2}G_F}}h(\mu)\cr
m_H(\mu)&=m_H^{pole}(1+\delta_H(\mu))=\sqrt{{\sqrt{2}\over G_F}\lambda(\mu)}
\end{eqnarray}
where the pole masses are the physical masses of the top and Higgs, and
$\delta_{top}(\mu)$ is the radiative corrections to the
$\overline{MS}$ top quark mass.  Note that the physical Higgs mass is
NOT simply the second derivative of the effective potential, since the potential is
defined at zero external momentum and the pole of the propagator is on-shell;
$\delta_H(\mu)$ accounts for the correction.

The correction $\delta_{top}(\mu)$ receives contributions from QCD, QED and weak
radiative effects, with the QCD corrections being the largest.  The QCD corrections
have been calculated to $O(g_3^2)$ in Ref. \cite{del1} and to $O(g_3^4)$ in Refs.
\cite{del2,del3}, the other corrections were determined in Refs.
\cite{del4,del5,del6}.  The correction $\delta_H(\mu)$ can be found in Refs.
\cite{cas,del7}.  The detailed expressions for these quantities, which correct
several typographical errors in the published works, are summarized in an extensive
review article by Schrempp and Wimmer\cite{sw}.  The largest correction is to the
top quark mass; the leading order term is ${4\over 3}{\alpha_3\over\pi}$, which is
$5\%$, or almost 10 GeV.

\begin{figure} 

\centerline{ \epsfysize 4in \epsfbox{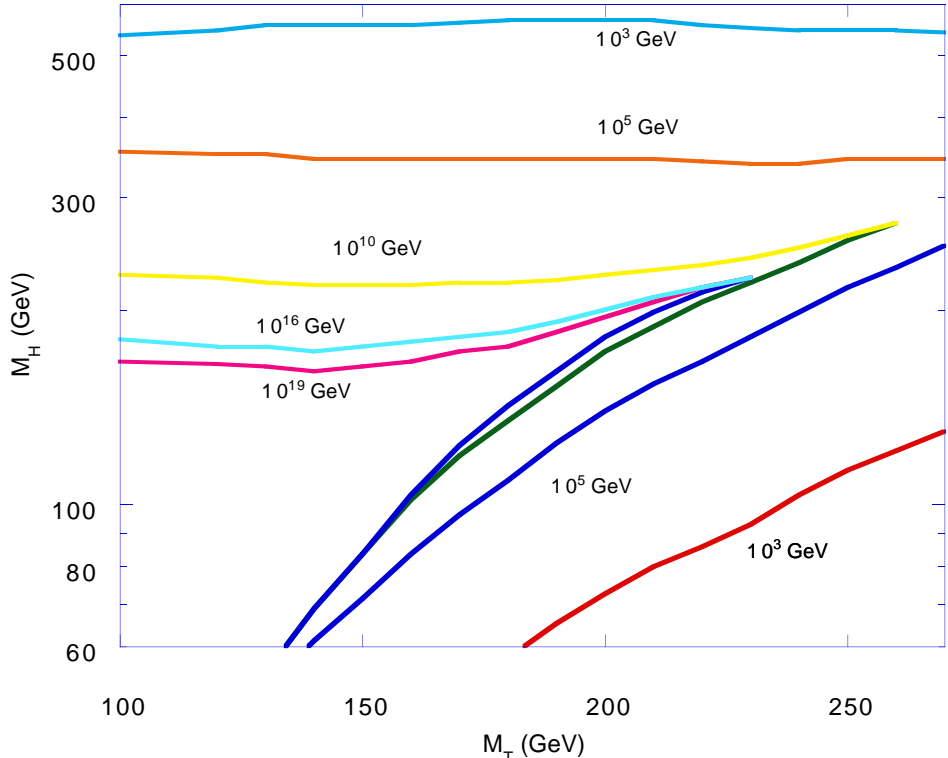}  }
\caption{Perturbativity and stability bounds on the SM Higgs boson.  $\Lambda$ denotes the energy
scale where the particles become strongly interacting.}

\end{figure}

All of these corrections were included in Refs. \cite{casas} and
\cite{espinosa}, and reviewed in Ref. \cite{quiros}.  If one requires stability of the
vacuum up to a scale
$\Lambda$, then there is an excluded region in the Higgs mass-top mass plane.  The
result, for various values of $\Lambda$, is given in Figure 2.  This figure, in
addition, also includes the region excluded by the requirement that the scalar and
Yukawa couplings remain perturbative by the scale $\Lambda$; these bounds will
be discussed in the next section.  The lower part of each curve is the vacuum stability
bound; the upper part is the perturbation theory bound.  The excluded region is
outside the solid lines.  Thus, for a top quark mass of 170 GeV, we see that a
discovery of a Higgs boson with a mass of 90 GeV would imply that the standard model
vacuum is unstable at a scale of
$10^5$ GeV, i.e. if we live in a stable vacuum, the standard model must break down
at a scale below
$10^5$ GeV.  The curves in Figure 2 are approximately straight lines in the
vicinity of $M_{top}\sim 170$ GeV, thus the top mass dependence can be given
analytically\cite{quiros}.  For
$\Lambda=10^{19}$ GeV, we must have
\begin{equation}
M_H(GeV)\quad >\ 133+1.92(M_{top}(GeV)-175)-4.28{\alpha_3(M_Z)-.12\over 0.006}
\end{equation}
and for $\Lambda=1$ TeV,
\begin{equation}
M_H(GeV)\quad >\ 52+0.64(M_{top}(GeV)-175)-0.5{\alpha_3(M_Z)-.12\over 0.006}
\end{equation}
It is estimated\cite{casas,quiros} that the error in the result, primarily due to
the two-loop correction in the top quark pole mass and the effective potential, is
less than $5$ GeV.  In Figure 3, the stability and perturbation theory bounds are
given explicitly as a function of $\Lambda$ for $M_{top}=175$ GeV.

\begin{figure} 

\centerline{ \epsfysize 4in \epsfbox{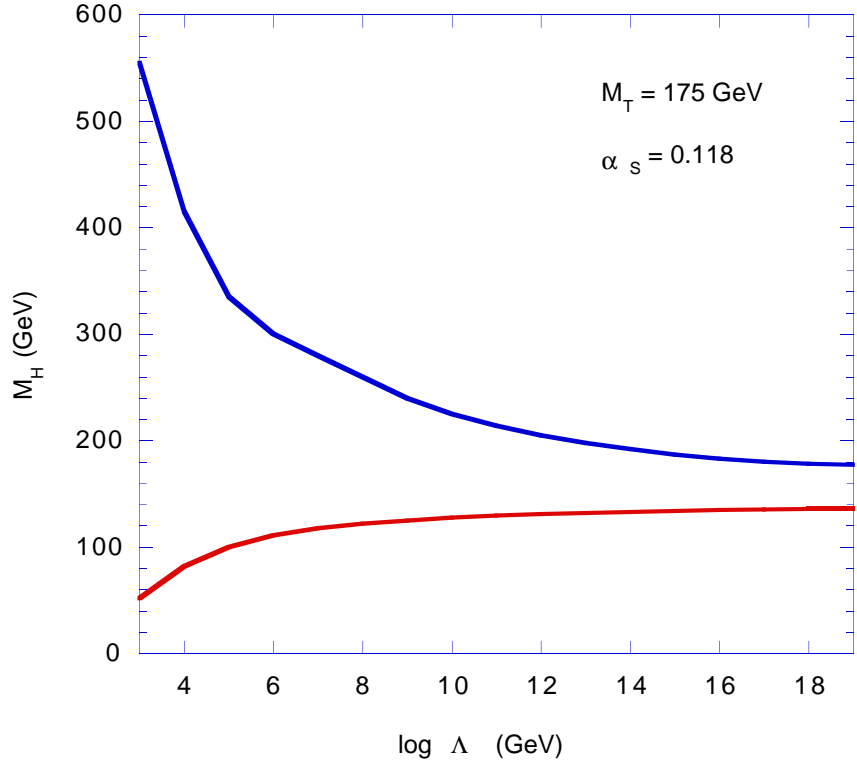}  }
\vskip .25in
\caption{Perturbativity and stability bounds on the SM Higgs boson as a function of $\Lambda$
for $M_{top}=175$ GeV.}

\end{figure}

Of course, it is not formally necessary that we live in a stable
vacuum\footnote{This was first pointed out in Ref. \cite{phf}.}.  Should another
deeper vacuum exist, it is only necessary that the Universe goes into our
metastable vacuum and then stay there for at least 10 billion years.  A detailed
discussion of the finite temperature effective potential and tunnelling
probabilities is beyond the scope of this review; the reader is referred to Refs.
\cite{casas} and \cite{quiros} for the details, as well as a comprehensive list of
references.  In short, the bound in the above paragraph for $\Lambda=10^{19}$ GeV
weakens by $8$ GeV, and for $\Lambda=1$ TeV, weakens by about $25$ GeV.  In all
cases, the bound obtained by requiring that our vacuum have a lifetime in excess of
10 billion years is weaker than the bound obtained by requiring that the Universe
arrive in our metastable vacuum.

We now turn to the question of vacuum stability for models with additional quarks
and leptons.

\subsubsection{Vacuum Stability Bounds and Four Generations}  

In this review, we are concentrating on two possibilities for quarks and leptons
beyond the third generation:  the chiral case and the vector-like case.  In the
latter case, the quarks and leptons cannot couple to Higgs doublets.  They will
thus have no effect on the vacuum stability bounds (except very weakly through
their effects on the two-loop beta functions of the gauge couplings).
We conclude that {\it there are no vacuum stability bounds on the masses of
vector-like quarks and leptons in models with Higgs doublets}.  Should one include a
Higgs singlet, of course, then the vector-like quarks and leptons would couple, and
a bound could be found on their masses which depends on the singlet Higgs mass as
well as the fraction of the mass which comes from the singlet vev (since a bare
mass is possible).  There is one exception to this conclusion.  As noted by
Zheng\cite{zheng}, if one adds a vectorlike doublet {\it and} one or more
vectorlike singlets, then Yukawa couplings can exist.  He studied the
stability bounds in that case (using the tree-level potential and one-loop
beta functions), assuming that the Yukawa couplings were unity, and found that
the bounds are much more stringent than in the three generation case---the
lower bound from vacuum stability and the upper bound from perturbation theory
come together at a scale well below the unification scale.

In the chiral case, more specific bounds can be found.  It is clear from Eq.
\ref{bvalue} that one can naively just replace the $h^4$ term with a summation
over all quark and lepton Yukawa couplings.  This amounts to replacing
$M_{top}$ with
\begin{equation}\label{subs}
\left(M^4_U+M^4_D+{1\over 3}M^4_E+{1\over 3}M^4_N\right)^{1/4}
\end{equation}
where $U$, $D$, $E$ and $N$ are additional $U$-type quarks, $D$-type quarks,
charged leptons and neutrinos, respectively.  The ${1\over 3}$ factor is a color
factor.  This replacement was noted by Babu and Ma\cite{bma} who simply rewrote Eq.
\ref{dunceq} by substituting $M_{top}$ with the above expression (their paper was
written when the top quark mass was believed to be 40 GeV, so the top mass was not
included in the above.)

As discussed above, however, this procedure is not particularly accurate.  A more
detailed analysis, using one-loop beta functions, was carried out by Nielsen et
al.\cite{nielsen}   For simplicity, they assumed that the fourth generation
fermions all had a common mass, $M_4$; since the quarks must have very similar
mass, relaxing this assumption will not significantly affect the results.  If one
assumes that the standard model is valid up to $\Lambda\sim 10^{15}$ GeV, then they
found that there is an upper bound on $M_4$ of only $110$ GeV.  It is easy to see
why this bound is so stringent.  Suppose that $M_4$ were equal to $M_{top}$.  Then
the expression in the above paragraph(ignoring the leptons) would be $2M_{top}$,
and thus the quark contribution would be $3M_{top}$.  One might expect the bounds
to be smaller by a factor of roughly $3$.  In fact, it isn't quite that severe
since the upper line in the allowed region of Figure 2 is not significantly
affected by the presence of additional generations.  Nonetheless, the bound is
quite stringent.

Nielsen et al.\cite{nielsen} argued that CDF bounds\cite{hoffman,CDFbounds} on
stable, color triplets, as well as results from the successful description of
top quark decays (which occur at the vertex), rule out $M_4$ up to $139$ GeV,
and thus the standard model cannot be valid up to $10^{15}$ GeV.  They found
that new physics had to start below approximately $10^{10}$ GeV.  However, this
argument has a flaw.  It is true that the CDF bounds on stable, color triplets
rule out heavy quarks with decay lengths greater than a meter or so, and that
the successful description of top quark decays rule out heavy quarks which decay
at the vertex (at least up to about $150$ GeV), but there is still a window for
decay lengths in which the quarks would have evaded detection.  Depending on
mixing angles, these decay lengths are quite plausible. Nonetheless, theirs is
the most detailed analysis to date of the case in which
$M_4$ is above $140$ GeV (and thus the scale of new physics is well below the
unification scale).

A much more detailed analysis was carried out by Dooling et
al.\cite{kangpark,dooling}   They performed a complete analysis, using the full,
two-loop analysis of Casas, Espinosa and Quiros (discussed in the last section).
They only consider the case in which the standard model is valid up to the
unification scale, and thus only look at the case in which $M_4$ is very light,
typically less than $120$ GeV.   Their work is thus complimentary to that of
Nielsen et al.   They find that the point where the triviality bound and vacuum
stability bound come together (see Figure 3) is (for $\Lambda\sim 10^{19}$ GeV) 
$M_4 < 110$ GeV.  Thus, if the standard model is valid up to the unification scale,
only a narrow region of masses still exists between the LEP lower bound (roughly 1/2
the center-of-mass energy) and the vacuum stability bound of $110$ GeV.

These works did assume that the quark and lepton masses were all degenerate with a
mass $M_4$.  If one relaxes this assumption, then one {\it approximately} can
replace $(8/3)^{1/4}M_4$ with Eq. \ref{subs}.  Clearly it is easier to accommodate
heavy leptons and neutrinos than heavy quarks.  

Hung and Isidori\cite{HuIs} relaxed the assumption of a common $M_4$ and simply
assumed a doublet of degenerate quartks with mass $M_Q$ and a doublet of degenerate
leptons with mass $M_L$.  They found that with $M_L\sim M_W$, $M_Q$ can be
extended to 150 GeV before a Landau pole appears at the Planck mass.  As
$M_L$ is raised, $M_Q$ should correspondingly decrease if one requires
that the Landau pole appear at or above the Planck mass.
\newpage
\subsection{Perturbative Gauge Unification}

The use of the RG equations as a tool to set bounds on and, in particlular,
to ``predict'' particle masses is an ``old'' subject. The discussion
on the bounds has been carried out in previous subsections. This subsection
concentrates instead on the use of the RG equations to ``predict'' various
masses. In particular, we shall pay special attention to the masses of
{\em any extra} family of quarks and leptons. This analysis only
works for chiral fermions. Vector-like fermions, having no Yukawa
coupling to the SM Higgs field, will not have the desired influence
on the evolution of the couplings as we shall see below.

In order to use the RG equations to make ``predictions'' on the
masses, one has to invoke either some experimental necessities
or some theoretical expectations -or rather prejudices- such
as fixed points, gauge unification, etc.  We shall describe
below these concepts along with their consequences. To begin, we
shall list the RG equations at two loops for the minimal SM with three
generations \cite{RG}.

\begin{mathletters}
\begin{eqnarray}
16 \pi^{2} \frac{d\lambda}{dt} =&& 24 \lambda^{2} + 4 \lambda (3 g_{t}^{2}
-2.25 g_{2}^{2}-0.45 g_{1}^{2})\nonumber \\
&&-6 g_{t}^{4}
+(16 \pi^{2})^{-1}\{30 g_t^{6}\nonumber \\ 
&&-[3 g_t^{4}+ 2 g_l^{4} - 80 g_3^{2} g_t^{2}]\lambda\nonumber \\
&&-144 \lambda^{2} g_t^{2} -312
\lambda^{3} -32 g_3^{2} g_t^{4}\}
\end{eqnarray}
\begin{eqnarray}
16 \pi^{2} \frac{d g_t^{2}}{dt} =&& g_t^{2} \{9 g_t^{2}
-16 g_3^{2}-4.5 g_2^{2}-1.7 g_1^{2}+\nonumber \\
&&(8 \pi^{2})^{-1}  [-12 g_t^{4}
+ 6 \lambda^{2} +g_t^{2}\nonumber \\
&&(-12 \lambda + 36 g_3^{2})-108 g_3^{4}] \} 
\end{eqnarray}
\begin{eqnarray}
16 \pi^{2} \frac{d g_1^{2}}{dt} =&&g_1^{4} \{ (41/5)+(16 \pi^{2})^{-1}[
(199/25) g_1^{2} + (27/5) g_2^{2}+\nonumber \\
&&(88/5) g_3^{2}-(17/5) g_t^{2}] \}
\end{eqnarray}
\begin{eqnarray}
16 \pi^{2} \frac{d g_2^{2}}{dt} =&&g_2^{4} \{ -(19/3)+(16 \pi^{2})^{-1}[
(9/5) g_1^{2} + (35/3) g_2^{2}+\nonumber \\
&&24 g_3^{2}-3 g_t^{2}] \}
\end{eqnarray}
\begin{eqnarray}
16 \pi^{2} \frac{d g_3^{2}}{dt} =&&g_3^{4} \{ -14+(16 \pi^{2})^{-1}[
(11/5) g_1^{2} + 9 g_2^{2}\nonumber \\
&&-52 g_3^{2}-4 g_t^2] \}
\end{eqnarray}
\end{mathletters}
To set the notations
straight, our definition of $g_1$ uses the $SU(5)$ convention and is
related to the $U(1)_Y$ coupling $g^{\prime}$ by
$g_1 = \sqrt{3/5} g^{\prime}$.  
In the evolution of
these couplings we will neglect the contributions coming from the lighter
fermions.

In the above RG equations, clearly the important couplings are those of the top
quark Yukawa and of the Higgs quartic couplings, and, to a certain extent,
also the QCD coupling. As we have seen earlier, one important use of such
equations is by following the evolution of $\lambda$ with the initial value
of $g_t$ fixed by its experimental value. Requiring $\lambda$ to be positive (for
vacuum stability reason)
at least up to the Planck scale allows us to set a {\em lower limit}
on the Higgs mass to be $\sim$ 136 GeV.
This use of the RG equations is
rather solid in the sense that it relies only on the vacuum stability
criterion of quantum field theory. Other uses which are discussed below
are more speculative but are quite interesting in that several predictions
can be made and can be tested.

In dealing with RG equations, a natural question that comes to mind
is whether or not there exist stable fixed points. Basically, a
stable fixed point is a point in coupling space to which various
couplings converge regardless of their initial values. This is
an attractive idea that has wide applications in many fields of
physics, such as critical phenomena- to mention just one of
many. In particle physics, there were many speculations concerning
the nature of such fixed points if they truly exist. For example,
Gell-Mann and Low \cite{gellmannlow}, and subsequently, 
Johnson, Wiley and Baker \cite{JWB} have
speculated that quantum electrodynamics might possess an 
ultraviolet stable fixed point which will render QED finite. Other
more ``recent'' speculations dealt with the very interesting subject
of the origin of particle masses- at least of the heavy one(s).

In general, a stable fixed point appears as a {\em zero} of the $\beta$
function which would be meaningful only if one has a full
knowledge of such a function. In the absence of such a knowledge,
one might have to resort to approximations allowed by perturbation
theory. In regions where various couplings can be considered
to be ``small'' enough so that the use of one or two-loop
$\beta$ functions might be justified, one would try to ``run''
the couplings over a large region of energy and see if they
converge to a point for arbitrary initial values. If such
a point is found, say by a numerical study of the RG equations,
one would qualify this as a fixed point. Such an approach
has been pioneered by Pendlenton and Ross \cite{PR}, and, in particular
by Hill \cite{hill1}, where the fixed points are of the infrared nature.
Of relevance to this report is the suggestion by various 
authors that a fourth generation might be needed for the
existence of such a fixed point.

Let us first summarize what has been done for the top quark mass
and subsequently describe works related to the masses of a
fourth generation.  

Pendleton and Ross \cite{PR} were the first to
suggest a relationship between the top quark Yukawa coupling
and the QCD coupling as a result of an infrared (IR) stable
fixed point. To see this, one can combine the one-loop RG
equations for $g_t$ and the QCD coupling $g_3$ (first terms
on the right-hand side of Eq. (38)) to form a RG equation for the
ratio $g_t/g_3$, namely
\begin{equation}
16 \pi^{2} \frac{d(g_t/g_3)}{dt} = \frac{9}{2} g_t^2 - g_3^2 -
\frac{3}{4} (3 g^2 + g^{\prime 2}) - \frac{2}{3} g^{\prime 2}.
\end{equation}
Ignoring the electroweak contributions in Eq. (39), there is
an IR fixed point obtained by setting the right-hand side to 
zero. Pendleton and Ross obtained a relation
\begin{equation}
g_t^2 = g_{t, irs}^2 = \frac{2}{9} g_3^2.
\end{equation}
It turns out that the above relation gives too low of a mass
for the top quark. In fact, the original prediction \cite {PR} using
Eq. (40) and a value of $\alpha_3 = g_3^2/4\pi \sim 1/7$ 
(at a scale of $\sim 2 M_W$) gives a mass of $\sim$ 110 GeV. 
Using the current value of $\alpha_3 \approx 0.12$ (at the Z mass),
the prediction would have been even lower, even after electroweak
corrections are included. It goes without saying that this cannot
be true for we already know that the top quark mass is $\sim$
175 GeV. As pointed out by Hill \cite{hill1} long before the discovery of the
top quark, the Pendleton-Ross fixed point is only a quasi
fixed point in the sense that it can never be reached at the scale of
interest $\sim$ 100 GeV. Hill proposed an intermediate fixed
point that can be found by setting the $\beta$
function in the RG for $g_t$ to zero with a ``slowly
varying'' $g_3$ replaced by a constant taken to be some average
value between two scales: 100 GeV and $10^{15}$ GeV. This
translates into an approximate relation
\begin{equation}
\frac{9}{2} g_t^2 \approx 8 \bar{g_3}^2,
\end{equation}
With $\bar{g_3}^2 \sim 0.7$, Hill made the prediction for the top quark
mass to be $\approx$ 240 GeV. This is now known to be much too large,
although at the time the prediction was made, it appeared to be
a plausible value.

In the above discussion for the top quark mass as a result of an IR
stable fixed point, one feature clearly emerges: a heavy fermion
is needed to drive the evolution toward a fixed point. This point
was made even clearer in  a detailed study of Bagger, Dimopoulos
and Mass\'{o} \cite{BDM}. These authors made two assumptions: the first one
is the existence of a desert between the weak scale $M_W$ and some 
Grand Unified scale $M_X \sim 10^{15}$ GeV and the second one being
that of perturbative unification. The question asked in Ref. \cite{BDM} was
the following: what should the initial values of various
Yukawa couplings at $M_X$ be in order for those couplings to reach the
fixed in a ``physical time'' $t_W = \frac{1}{16 \pi^2} \ln (\frac{M_X}{M_W})
\sim 1/5$? Again, it turns out that ``large'' initial values (at
$M_X$) of Yukawa couplings guarantee that the fixed point is reached
in ``physical time''. What is this fixed point and what does it say
about masses of possible extra generations if they exist? We shall
describe below the salient points of the analysis of Ref. \cite{BDM}.

We begin the discussion of Ref. \cite{BDM} with the following one-loop RG equations for the
quark and lepton Yukawa couplings:
\begin{mathletters}
\begin{equation}
\frac{dT_Q}{dt} = 2(G_Q - T)T_Q - 3 Tr(S_U^2),
\end{equation}
\begin{equation}
\frac{dT_L}{dt} = 2(G_L - T)T_L - 3 Tr(S_E^2),
\end{equation}
\end{mathletters}
where various Yukawa factors are defined as $T_Y = Tr(Y^{\dag}Y)$
with $Y= U, D, E$ or $N$, $T_Q = T_U + T_D$, $T_L = T_E + T_N$,
$T= 3T_Q + T_L$, $S_U = U^{\dag} U - D^{\dag} D$ and
$S_E = E^{\dag} E - N^{\dag} N$. Finally the gauge factors $G$'s are
defined as $G_U = G_D = G_Q = 8 g_3^2 + \frac{9}{4} g_2^2$ and
$G_E = G_N = G_L = \frac{9}{4} g_2^2$ where the contribution from
$g_1$ has been neglected. Notice that $t = \frac{1}{16 \pi^2} 
\ln (\frac{M_X}{M})$. 

\begin{figure} 

\centerline{ \epsfysize 4in \epsfbox{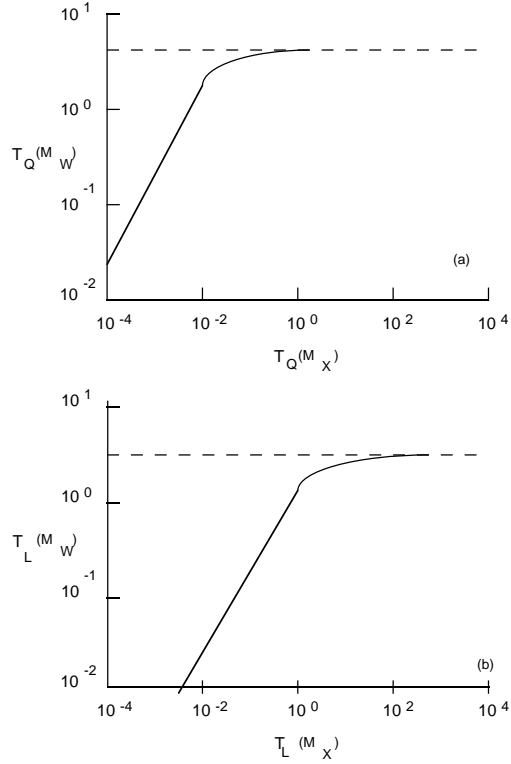}  }
\vskip .25in
\caption{(a) The trace $T_Q$ at $M_W$ as a function of the trace $T_Q$ and $M_X$ for $N_F=8$ and
$T_L=0$.  The dotted line denotes the radial quark fixed point.  For $T_Q(M_X) > 0.1$, the fixed
point is reached in physical time. (b)$T_L(M_W)$ as a function of $T_L(M_X)$ for $N_F=8$ and
$T_Q=0$.}

\end{figure}

To simplify the discussion, Ref. \cite{BDM}
first assumed degenerate quarks and degenerate leptons so
that $S_U=0$ as well as $S_E=0$. If the gauge couplings can be
approximated as constant (or very slowly varying), $G_Q$ and
$G_L$ in Eqs. (42) can be replaced by some averages similar to
the procedure used by Hill \cite{hill1}. Let us denote these
averages by $\bar{G_Q}$ and $\bar{G_L}$.
It is then easy to see that Eqs.(42) have the following two distinct fixed
points: $\bar{G_Q} = T$ (quark radial fixed point) and
$\bar{G_L} = T$ (lepton radial fixed point). Whether or not these
fixed points are reached will depend on the initial values of
$T_Q$ or $T_L$ at the Grand Unified scale $M_X$. If $T(M_X)$
is below some critical value $T(M_X)_{min}$, the fixed point $\bar{G}$ will
not be reached in ``physical time'' $t \sim 1/5$: the value of $T$ at
$M_W$ will be less than the fixed point value $\bar{G} = T$. This fact
has allowed Ref. \cite{BDM} to set an upper limit on heavy fermion masses
with the upper limit being the IR stable fixed point. Setting
$S_Y = T_L = 0$, Ref. \cite{BDM} plotted $T_Q$ at the weak scale as 
a function of its value at $M_X$. This is shown in Fig. 4a.
The result for the lepton case is shown in Fig. 4b.

The figure shows the result for eight families. We are, of
course, concerned only with four families which are still allowed.
within error, by precision electroweak results. For four
families, Ref. \cite{BDM} gave the following upper bounds on $T$:
\begin{mathletters}
\begin{equation}
T_Q \lesssim 2.7, \nonumber
\end{equation}
\begin{equation}
T_L \lesssim 3.4 . \nonumber
\end{equation}
\end{mathletters}
(The above numbers used values of gauge couplings which are now outdated.)
This translates into the following distinct bounds on fermion masses for four families:
\begin{mathletters}
\begin{equation}
\Sigma M_Q^2 \lesssim (290\, GeV)^2,
\end{equation}
\begin{equation}
\Sigma M_L^2 \lesssim (325\, GeV)^2.
\end{equation}
\end{mathletters}
(These bounds would be slightly less if recent values of the gauge couplings are used.)
The above bound for the quarks, for example, would translate roughly into a bound
on the mass of a degenerate fourth generation quark (after subtracting
out the top quark) as $M_{U,D} \lesssim$ 164 GeV. Is this what one
should be aiming for when one tries to look for fourth-generation
quarks? 

\begin{figure} 

\centerline{ \epsfysize 4in \epsfbox{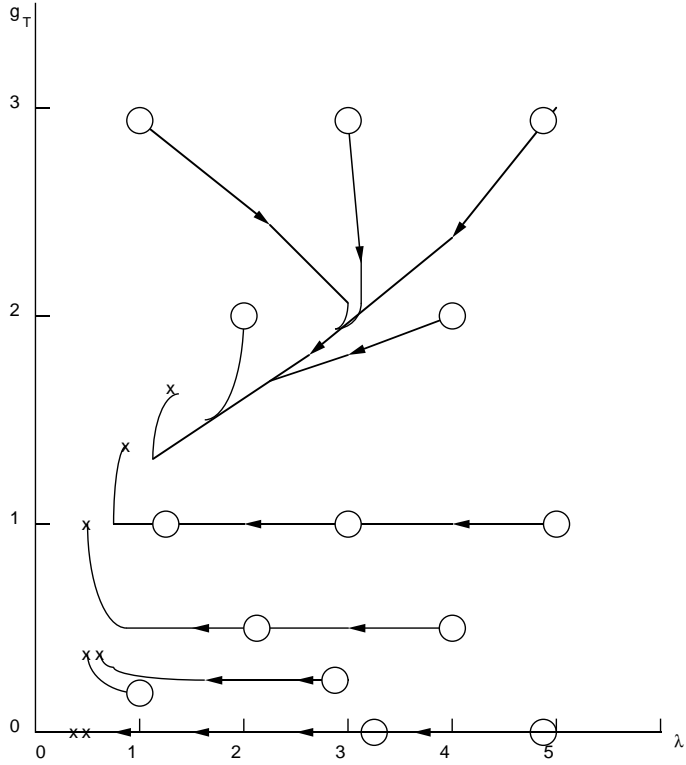}  }
\vskip .25in
\caption{Flow of $\lambda$ and $g_t$ towards fixed points in the standard one-Higgs doublet
model.  Open circles denote initial points.  Crosses denote final fixed points.}

\end{figure}

Hill, Leung and Rao \cite{hill2} made an extensive study of the RG fixed points and their
connections with the mass of the Higgs boson(s) for the one-Higgs doublet
and the two-Higgs doublet cases, and for up to five generations. In this
work, the Higgs quartic coupling(s) is run simultaneously with the various
Yukawa couplings and, as a result, one clearly sees again the interplay between
heavy fermions and the Higgs field. This is shown for example in Figs. 1
and 2 of Ref. \cite{hill2} for the one-Higgs doublet case with three generations, which
are reproduced in Figs. 5 and 6.

\begin{figure} 

\centerline{ \epsfysize 4in \epsfbox{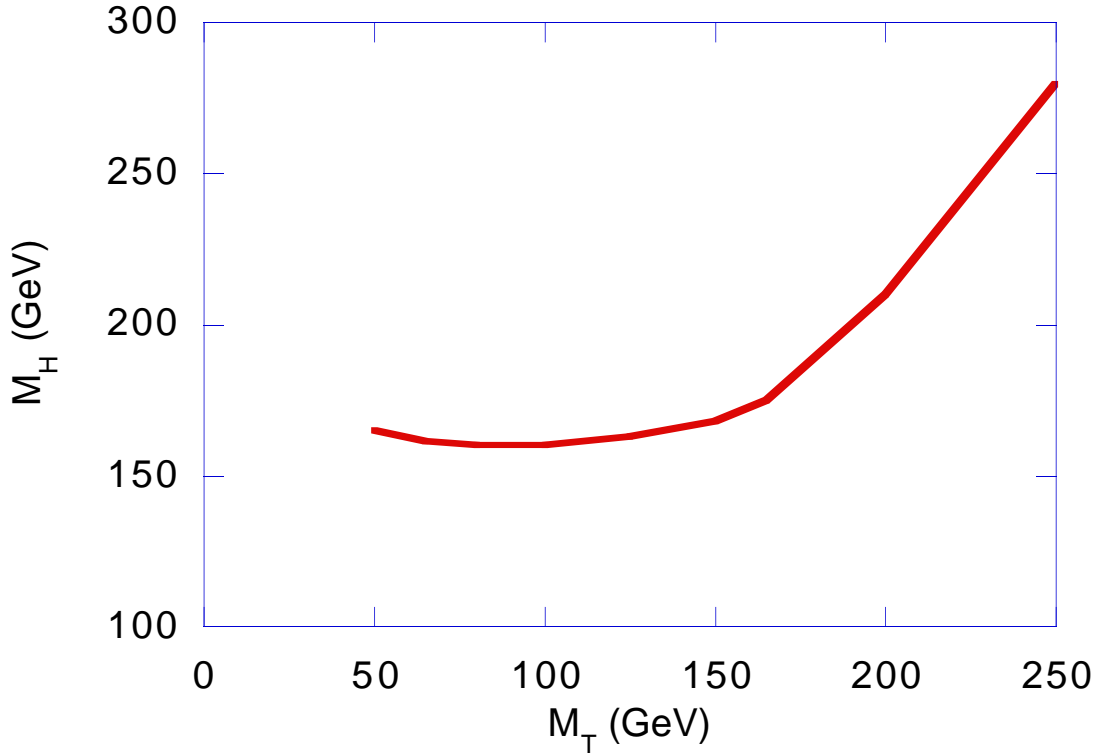}  }
\vskip .25in
\caption{Relation between the Higgs mass and the top quark mass in the standard
 one-Higgs model.}

\end{figure}

Other attempts of using the RG fixed points to construct fermion
mass matrices have been made, for example in Ref. \cite{paschos}.
However the ``predicted'' value for the top quark mass is now
outdated.

A different approach was taken by one of us (P.Q.H.) \cite{pqh} 
concerning the
influence of a possible fourth generation on the evolution of {\em all}
couplings of the SM and not merely the Yukawa couplings. In particular,
the question that was asked was whether or not there can be gauge
unification in the nonsupersymmetric SM and under which conditions this
can be achieved. As we shall see below, it turns out that a fourth
family of chiral fermions will be needed and that their masses are
found to be fairly constrained.

The possibility of coupling-constant unification of the three gauge 
interactions of the Standard Model (SM) is, without any doubt, one of
the most important issues in particle physics. Coupling-constant
unification is a necessary, but not sufficient, condition for
a Grand Unification of the SM \cite{GUT1,GUT2,GUT3}. Such a possibility
is particularly attractive since it would provide a unified explanation
for a number of puzzling features of the SM such as electric charge
quantization for example. 

There are various ways that the three gauge couplings can get unified.
The simplest way is to assume that there is a ``desert'' (i.e. no
new physics) between the electroweak scale and the scale at which
unification occurs. Simply speaking, the three gauge couplings are left
to evolve beyond the electroweak scale under the assumption that
there is {\em no} additional gauge interactions of the type which would
modify the evolution of some of the gauge couplings. A more complicated
way would be to assume that there is one or several intermediate scales
where partial unification among two of the three couplings occurs. This might
well be the case. However we shall restrict ourselves, in this report,
to the simplest scenario of unification with a ``desert'' and search for conditions
under which this can be achieved. This was the approach taken by one of us (P. Q. H.)
\cite{pqh}.

We will proceed in two steps. First, we will present the evolution of the
gauge couplings and show the places where they cross, ignoring any heavy
threshold effects that might be- and should be if there truly is unification- present.
In this discussion, we will show both the minimal SM with three generations
and the one with an extra fourth generation. We shall see that, under
certain restrictions on the masses (fourth generation and Higgs masses), the latter
possibility provides a better ``convergence'' of the three gauge couplings.
By convergence under the quotation marks, we mean that they do not precisely meet
at the same point. The ``true'' convergence will be shown to be accomplished by
the inclusion of heavy threshold effects. In fact it would be senseless to claim
unification without taking into account such effects.

We first summarize the situation with three families. 

The first task is to integrate Eq.(38) numerically and look for the places 
where the couplings meet, disregarding for the moment the possibility that
there might be unification. We then have to set up some kind
of criteria to decide on how close to each other all three couplings have to be in order
for them to have  a chance of actually converging to a single scale, once heavy
particle thresholds, such as those of the X and Y bosons of $SU(5)$ for instance,
are taken into account. Once these criteria are satisfied and heavy particle threshold
effects are included, one can put an error on the unification scale and,
consequently, an error on the proton lifetime.

Fig. 7 shows the evolution, without heavy threshold effects, of $g_3$, $g_2$, 
and $g_1$ of $SU(3) \otimes SU(2) \otimes U(1)$ for the case with three generations. 
Clearly these three couplings do not converge. 

\begin{figure} 

\centerline{ \epsfysize 4.5in \epsfbox{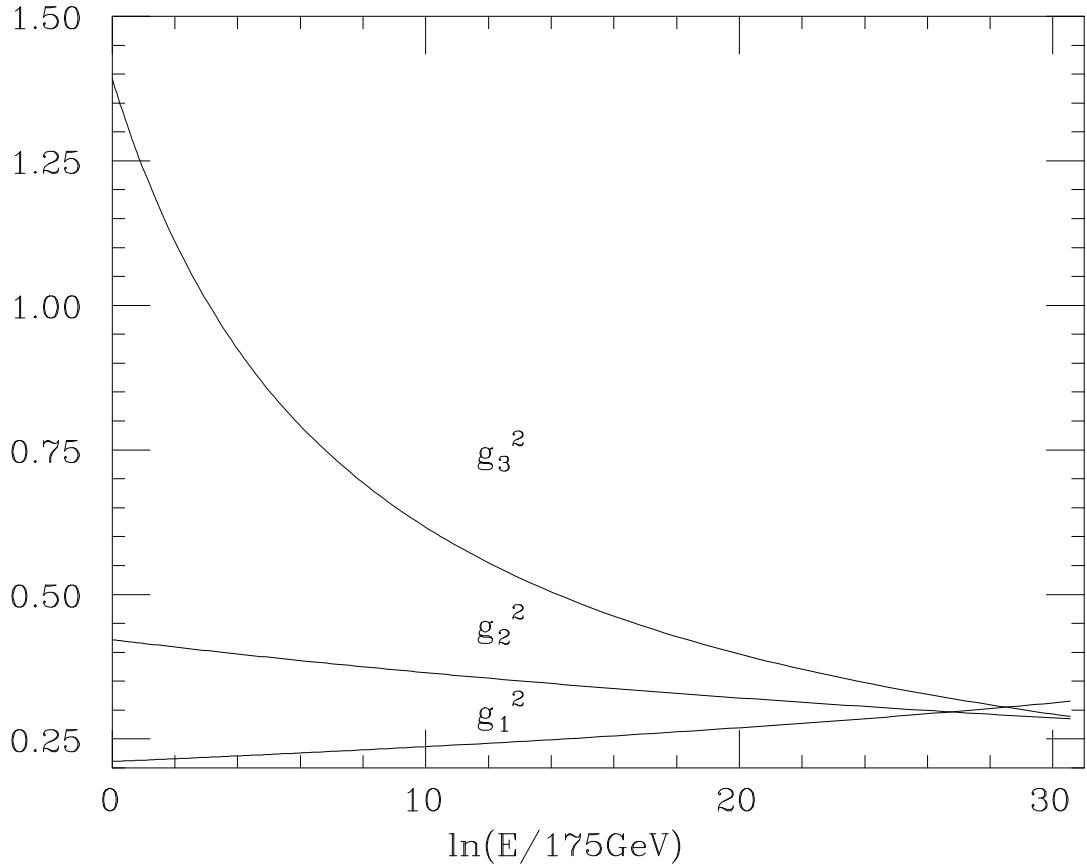}  }

\caption{Evolution of couplings in the three generation case.}

\end{figure}

The question is: How far apart are they 
from each other and at what scales? As far as the scales are concerned, we will be
interested only in those which are above some minimum value implied by the lower
bound on proton decay. A rough estimate of that lower bound is obtained by
noticing that $\tau_{p \rightarrow e^{+} \pi^{0}}(yr) \approx 10^{31}(M_{G}/4.6 \times
10^{14})^{4}$. This gives $M_{G} \gtrsim 1.3 \times 10^{15}$ GeV  (corresponding
to $\ln (E/175) = 29.64$ on the graph) for 
$\tau_{p \rightarrow e^{+} \pi^{0}}(yr) \gtrsim 5.5 \times 10^{32}$. The next question
is the following: Starting from $M_{G}^{min} \sim 1.3 \times 10^{15}$ GeV, how far
apart are the three gauge couplings from each other at a given energy scale? As stated
above, the reason for asking such a question stems from the fact that, if the SM were
to be embedded at $M_G$ in a Grand Unified model such as $SU(5)$ \cite{GUT2} (for instance), the
decoupling of various heavy GUT particles would shift the three couplings from a
common $\alpha_G$ to (possibly) different values. As shown below, for a wide range of
``reasonable'' heavy particle masses, such an effect produces no more than $\sim$ 5\% shift
from the common value and in the same direction. It turns out that the modified 
couplings can differ by no more
than $\sim$ 4\%. From this a reasonable criterion would be to require that, at a scale
$M_{G} \geq m_{G}^{min}$, the three gauge couplings are within 4\% of each other. 

We shall take $SU(5)$ \cite{GUT2} as a prototype of a Grand Unified Theory.
Let us assume the
following heavy particle spectrum:$(X,Y) =(\bar{3}, 2, 5/6) + c.c. $ with
mass $M_V$, real scalars $(8, 1, 0) + (1, 3, 1) + (1,1,0)$ (belonging
to the 24-dimensional Higgs field) with mass $M_{24}$, and
the complex scalars $(3, 1, -1/3)$ (belonging to the 5-dimensional
Higgs field), with mass $M_5$. (The quantum numbers are with respect to $SU(3) \otimes
SU(2) \otimes U(1)$.) The heavy threshold corrections are then \cite{langacker}:
\begin{mathletters}
\begin{equation}
\Delta_1 = \frac{35}{4 \pi} \ln (\frac{M_G}{M_V})-\frac{1}{30 \pi}\ln 
(\frac{M_G}{M_5}) + \Delta^{NRO}_1 ,
\end{equation}
\begin{equation}
\Delta_2 = -\frac{1}{6 \pi} +\frac{21}{4 \pi} \ln (\frac{M_G}{M_V})-\frac{1}{6 \pi}\ln 
(\frac{M_G}{M_{24}}) + \Delta^{NRO}_2 ,
\end{equation}
\begin{equation}
\Delta_3 =-\frac{1}{4 \pi} + \frac{7}{2 \pi} \ln (\frac{M_G}{M_V})-\frac{1}{12 \pi}\ln 
(\frac{M_G}{M_5}) -\frac{1}{4 \pi} \ln(\frac{M_G}{M_{24}})+ \Delta^{NRO}_3 ,
\end{equation}
\end{mathletters}
where
\begin{equation}
\Delta^{NRO}_i = -\eta k_{i} (\frac{2}{25 \pi \alpha_{G}^{3}})^{1/2} \frac{M_G}
{M_{Planck}},
\end{equation}
with $k_{i} =1/2, 3/2, -1$ for $i=1,2,3$, is the correction coming from 
possible dimension 5 operators present between $M_G$ and $M_{Planck}$.
The modified gauge couplings can be expressed in terms of the unified coupling
$\alpha_G$ (at $M_G$) as:
\begin{equation}
\alpha_{i}(M_G) = \frac{\alpha_{G}}{1-\alpha_{G}\,\Delta_{i}},
\end{equation}
where $i=1,2,3$. We then define the fractional difference between the modified gauge couplings as:
\begin{equation}
d_{ij} = \frac{\alpha_{i} - \alpha_{j}}{\alpha_{i}},
\end{equation}
for $i,j = 1,2,3$ and the definition refers to $\alpha_{i}$ as being the larger of the two couplings.
For a wide range of heavy particle masses (in relation with $M_G$) and the parameter $\eta$
appearing in $\Delta_{i}$, and for $\alpha_{G} \sim 0.024 - 0.028$, it is straightforward
to see that $d_{ij}$ can be {\em at most} 4\% \cite{pqh}.
>From this simple analysis, one can reasonably set a criterion for a given scenario
to have a chance of having gauge coupling unification: the fractional difference among
the three gauge couplings at some scale $M_G \geq M_{G}^{min}$ should not exceed 4\%.

For the SM with three generations and taking into account the presence of $M_{G}^{min}
\sim 1.3 \times 10^{15}$ GeV, one finds the following trend: $d_{32}$ decreases from
3\% as one increases the energy scale beyond $M_{G}^{min}$, while $d_{31}$ {\em increases}
from 4\% and $d_{21}$ also {\em increases} from 7\%. (For example, at $M_G \sim 3.3
\times 10^{15}$ GeV, $d_{32} \sim$ 1.4\%, $d_{31} \sim$ 8.4\% and $d_{21} \sim$ 9.7\%.
>From these considerations- and not
from just ``eyeballing'' the curves- one might conclude that the minimal SM with three
generations does indeed have some problem with unification of the gauge couplings. 

There is a drastic change to the whole scenario when one postulates the existence of
a fourth generation of quarks and leptons \cite{pqh}. The main reason is the fact that the
Yukawa contributions to the running of the gauge couplings appear at two loops. In
the three generation case, the top Yukawa coupling actually {\em decreases} sligthly
with energy because its initial value is partially cancelled by the QCD contribution
(at one loop). As a result, the presence of a heavy top quark is insignificant
in the evolution of the gauge couplings at high energies when there are only
three generations. The presence of more than three generations drastically modifies
the evolution of the Yukawa, Higgs quartic self-coupling, and the three gauge
couplings. For example, with a fourth generation which is sufficiently
heavy, all Yukawa couplings {\em grow} with energy, significantly
affecting the evolution of the gauge couplings. 
It turns out, as we shall see below, that 
the Yukawa couplings can develop Landau poles {\em below} the Planck scale. If there
were any possibility of gauge unification, one would like to ensure that it occurs in
an energy region where perturbation theory is still valid. Furthermore, the unification
scale will have to be greater than $M_G^{min}$ (as discussed above). As we shall
see, the validity of perturbation theory plus the lower bound on the proton lifetime
put a severe constraint on the masses of the fourth generation.

The two-loop renormalization group equations applicable to four generations are
given by \cite{RG,pqh}:
\begin{mathletters}
\begin{eqnarray}\label{RG4}
16 \pi^{2} \frac{d\lambda}{dt} =&& 24 \lambda^{2} + 4 \lambda( 3 g_{t}^{2}+
6 g_{q}^{2} + 2 g_{l}^{2}-2.25 g_{2}^{2}-0.45 g_{1}^{2})\nonumber \\
&&-2( 3 g_{t}^{4} + 6 g_{q}^{4} + 2 g_{l}^{4})
+(16 \pi^{2})^{-1}\{30 g_t^{6}\nonumber \\ 
&&+48 g_q^{6}+ 16 g_l^{6} -[3 g_t^{4} + 6 g_q^{4}
+ 2 g_l^{4} - 80 g_3^{2} (g_t^{2}\nonumber \\
&&+ 2 g_q^{2})]\lambda-6\lambda^{2} (24 g_t^{2} + 48 g_q^{2} + 16 g_l^{2})-312
\lambda^{3}\nonumber \\
&&-32 g_3^{2}( g_t^{4} + 2 g_q^{4})\}
\end{eqnarray}
\begin{eqnarray}
16 \pi^{2} \frac{d g_t^{2}}{dt} =&& g_t^{2} \{9 g_t^{2} +12 g_q^{2} + 4 g_l^{2}
-16 g_3^{2}-4.5 g_2^{2}-1.7 g_1^{2}+\nonumber \\
&&(8 \pi^{2})^{-1}  [1.5 g_t^{4}-2.25 g_t^{2}(6 g_q^{2}+ 3 g_t^{2}
+ 2 g_l^{2})\nonumber \\
&&-12 g_q^{4}- (27/4) g_t^{4} - 3 g_l^{4}+ 6 \lambda^{2} +g_t^{2}\nonumber \\
&&(-12 \lambda + 36 g_3^{2})-(892/9) g_3^{4}] \} 
\end{eqnarray}
\begin{eqnarray}
16 \pi^{2} \frac{d g_q^{2}}{dt} =&& g_q^{2} \{6 g_t^{2} +12 g_q^{2} + 4 g_l^{2}
-16 g_3^{2}-4.5 g_2^{2}-1.7 g_1^{2}+\nonumber \\
&&(8 \pi^{2})^{-1}  [3 g_q^{4}-g_q^{2}(6 g_q^{2}+ 3 g_t^{2}
+ 2 g_l^{2})\nonumber \\
&&-12 g_q^{4}- (27/4) g_t^{4} - 3 g_l^{4}+ 6 \lambda^{2} +g_q^{2}\nonumber \\
&&(-16 \lambda + 40 g_3^{2})-(892/9) g_3^{4}] \} 
\end{eqnarray}
\begin{eqnarray}
16 \pi^{2} \frac{d g_l^{2}}{dt} =&& g_l^{2} \{6 g_t^{2} +12 g_q^{2} + 4 g_l^{2}
-4.5 (g_2^{2}+ g_1^{2})+\nonumber \\
&&(8 \pi^{2})^{-1}  [3 g_q^{4}-g_q^{2}(6 g_q^{2}+ 3 g_t^{2}
+ 2 g_l^{2}) -12 g_q^{4}\nonumber \\
&&- (27/4) g_t^{4} - 3 g_l^{4}+ 6 \lambda^{2} -16
\lambda g_l^{2}] \} 
\end{eqnarray}
\begin{eqnarray}
16 \pi^{2} \frac{d g_1^{2}}{dt} =&&g_1^{4} \{ (163/15)+(16 \pi^{2})^{-1}[
(787/75) g_1^{2} + 6.6 g_2^{2}+\nonumber \\
&&(352/15) g_3^{2}-3.4 g_t^{2}-4.4 g_q^{2}-3.6 g_l^{2}] \}
\end{eqnarray}
\begin{eqnarray}
16 \pi^{2} \frac{d g_2^{2}}{dt} =&&g_2^{4} \{ -(11/3)+(16 \pi^{2})^{-1}[
2.2 g_1^{2} + (133/3) g_2^{2}+\nonumber \\
&&32 g_3^{2}-3 g_t^{2}-3 g_q^{2}-2 g_l^{2}] \}
\end{eqnarray}
\begin{eqnarray}
16 \pi^{2} \frac{d g_3^{2}}{dt} =&&g_3^{4} \{ -(34/3)+(16 \pi^{2})^{-1}[
(44/15) g_1^{2} + 12 g_2^{2}\nonumber \\
&&-(4/3) g_3^{2}-4 g_t^{2}-8 g_q^{2}] \} .
\end{eqnarray}
\end{mathletters}
For simplicity, we have made the following assumptions: a Dirac mass for the fourth
neutrino and the quarks and leptons of the fourth generation are degenerate $SU(2)_L$
doublets. The respective Yukawa couplings are denoted by $g_q$ and $g_l$ respectively.
Also, in the evolution of the quartic coupling $\lambda$ and the Yukawa couplings, we
will neglect the contributions of $\tau$ and bottom Yukawa couplings, as well as the
electroweak gauge couplings, $g_1$ and $g_2$, to the two-loop $\beta$ functions since they
are not important. Also, as long as the mixing between the fourth generation and the other
three is small, one can neglect such a mixing.

In the numerical analysis given below we shall fix the mass of the top quark to be
175 GeV. We shall furthermore restrict the range of masses of the fourth generation
so that the Landau poles lie comfortably above $10^{15}$ GeV, in such a way that
unification occurs at a scale which would guarantee the validity of perturbation
theory as well as satisfying the lower bound on the proton lifetime. Concerning the former
requirement, it basically says that one should look at unification scales where the
values of the Higgs quartic and Yukawa coulings are still sufficiently perturbative
that one can neglect contibutions coming from three-loop (and higher) terms to the
$\beta$ functions. 

Fig. 8 shows $g_{1}^2$, $g_{2}^2$ and $g_{3}^2$ as a function of 
energy for a particular set of masses: $m_Q = 151$ GeV, $m_L = 95.3$ GeV, 
where $m_Q$ and $m_L$ denote the fourth generation quark and lepton masses respectively.

\begin{figure} 

\centerline{ \epsfysize 4in \epsfbox{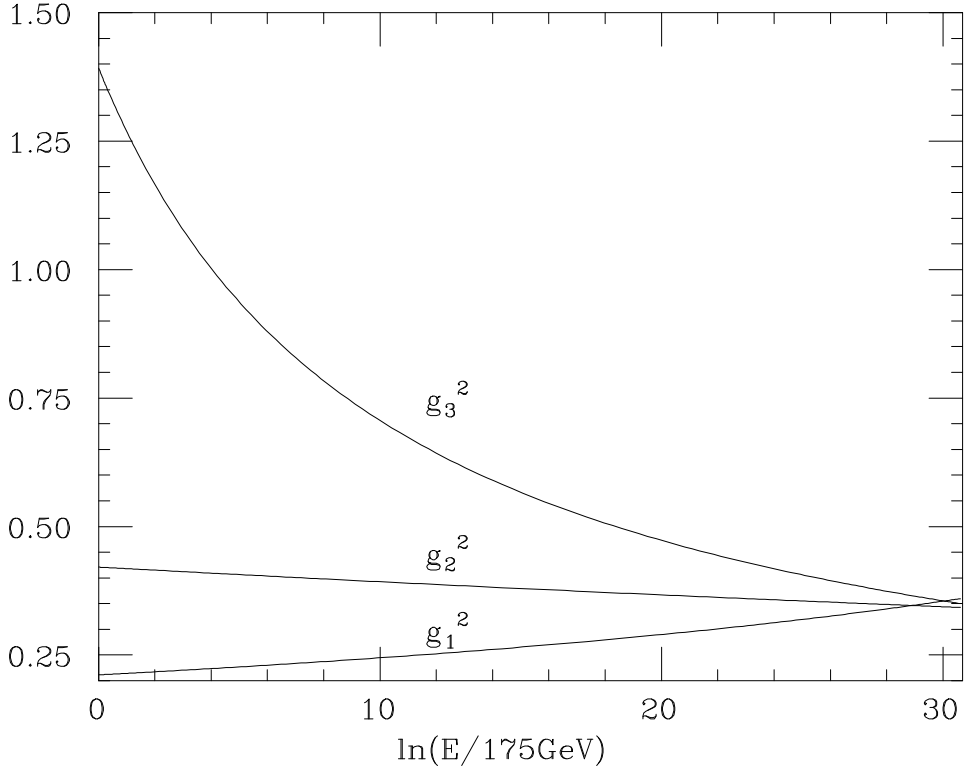}  }

\caption{Couplings as a function of energy for the set of masses given in the text.}

\end{figure}

It is already well known, from the discussion in the previous section, that, by adding more heavy 
fermions, the vacuum will tend to be destabilized unless the Higgs mass is large enough.
As we have seen, the vacuum stability requirement is equivalent to the restriction $\lambda >0$. 
Furthermore, the heavier the Higgs boson is, above a minimum mass that ensures vacuum stability,
the lower (in energy scale) the Landau pole turns out to be. It turns out that this Landau pole
should not be too far from $M_{G}^{min}$ otherwise $g_3$, $g_2$ and $g_1$ would not come
close enough to each other. On the other hand, it should not be too close either because
of the requirement of the validity of perturbation theory.
These considerations combine to give a prediction of the Higgs mass, namely $m_H = 188$ GeV
for the above values of the fourth generation masses \cite{pqh}. The dependance of the Higgs mass
on the fourth generation mass in this analysis is obviously striking.

Following the criteria that we have set for taking into account the heavy threshold
effects, the midified couplings $\tilde{\alpha}_i(M_G)$ expressed in terms of
$\alpha_i (M_G)$ (which can be read off from the graph) and the threshold
correction factors $\Delta_i$ are given by: $1/\tilde{\alpha}_i (M_G)
= 1/\alpha_i (M_G) + \Delta_i$. The choice of the mass scales $M_5$, $M_{24}$,
$M_V$, and the parameter $\eta$ is arbitrary and is only fixed to a certain
extent by the requirement that $\tilde{\alpha}_i (M_G)$'s should be as close to
each other as the precision allows. As an example, the choice $M_5 = M_G$,
$M_{24} = M_G$, $M_V = 0.5 M_G$ and $\eta = 10$ (where we have picked
$M_G \approx 3.5 \times 10^{15}$ GeV) transforms $\alpha_3 (M_G) = 0.0278$,
$\alpha_2 (M_G) = 0.0273$ and $\alpha_1 (M_G) = 0.0285$ (values that can
be read off Fig. (8)) to $\tilde{\alpha}_3 (M_G) = 0.02735$,
$\tilde{\alpha}_2 (M_G) = 0.02662$ and $\tilde{\alpha}_1 (M_G) = 0.02705$.
From these values, one can conclude that the couplings are practically the same
with all three equal to $\alpha_G \approx 0.027$ or $1/\alpha_G \approx 37$.

The above simple exercise simply shows that, with just an additional fourth
generation having a quark mass $m_Q \approx 151$ GeV, a lepton mass
$m_L \approx 95.3$ GeV and a Higgs mass $m_H \approx 188$ GeV, unification
of all three gauge couplings in the nonsupersymmetric SM can be achieved after
one properly takes into account threshold effects from heavy GUT
particles \cite{pqh}. Other combinations of masses are possible for gauge unification
but their values will not be much different from the quoted ones, the reason
being the requirement that the mass range of the fourth generation be restricted 
to one that will have Landau poles only above $10^{15}$ GeV. Do the masses
given above satisfy the requirement of perturbation theory? In fact, at
the unification point $M_G = 3.5 \times 10^{15}$ GeV, one has
(with $\alpha_i \equiv g_i^2/4 \pi$): $\alpha_t = 0.4$, $\alpha_q = 0..16$,
$\alpha_l = 0.48$ and $\lambda / 4 \pi = 0.19$. Although these values are
not ``small'', they nevertheless satisfy the requirements of perturbation
theory, namely $\alpha_{t,q,l} \lesssim 1$ and $\lambda / 4 \pi \lesssim 0.4$.
(The latter requirement comes from lattice calculations which put an upper
bound on the Higgs mass of $\sim$ 750 GeV.) For comparison, $\alpha_t$
in the three-family SM has a value of 0.016 at a comparable scale and this
explains why it is unimportnat in the evolution of the SM gauge couplings.

An important consequence of a fourth generation in bringing about gauge
unification is the value of the unification scale itself. In the example given above,
it is  $M_G = 3.5 \times 10^{15}$ GeV \cite{pqh}. In the nonsupersymmetric SU(5) model,
the dominant decay mode ofthe proton is $p \rightarrow e^{+} \pi^0$ and the
mean partial lifetime is $\tau_{p \rightarrow e^{+} \pi^0} (yr) \approx
10^{31} (M_G/ 4.6 \times 10^{14})^4$. Taking into account various uncertainties such as
heavy threshold effects, hadronic matrix elements, etc., the predicted
lifetime is $\tau_{p \rightarrow e^{+} \pi^0} (yr) \approx 3.3 \times 10^{34\pm 2}$ 
to be compared with $\tau_{p \rightarrow e^{+} \pi^0}^{exp} (yr) > 5.5 \times
10^{32}$ \cite{pqh}. Notice that the central value is within reach of the next generation
of SuperKamiokande proton decay search.

Another hint on the masses of a fourth generation comes from considerations of models
of dynamical symmetry breaking \`{a} la top-condensate\cite{1ai} 
This will be discussed in
Section 5 where one can see how the original idea of using the top quark as 
the sole agent for electroweak symmetry breaking (in the form of $t\bar{t}$ condensates)
led to a prediction for the top quark mass (before its discovery) to be much larger 
than its experimental value. The original form of this attractive idea obviously
has to be modified, most likely by the introduction of new fermions such as
a fourth generation or $SU(2)$-singlet quarks. 

In the above discussion on perturbative gauge unification, as well as in the subsequent related discussion in Section V, the issue of the gauge hierarchy
problem is not considered.  Such an issue is beyond the scope of the largely phenomenological approach that we are taking.   This point was alluded to in our Introduction where we stressed that none of the reasons given for considering quarks and leptons beyond the third generation is fully compelling, but each, including the one on perturbative gauge unification, is suggestive.   It is certainly possible that the ``solution'' of the gauge hierarchy problem will not affect the above arguments; the recently developed alternative to supersymmetry and technicolor, TeV-scale gravity, for example, may not appreciably change results on gauge unification.  A full consideration of the gauge hierarchy problem is beyond the scope of this review.

\newpage

\subsection{Mixing Angles}

In previous sections, we have seen that the masses of quarks and leptons, although
arbitrary, are constrained by phenomenological considerations as well as vacuum
stability and perturbation theory.  The mixing angles of quarks and leptons are
also arbitrary, however there are no constraints from vacuum stability and
perturbation theory (and only weak phenomenological constraints).  Thus, a much
wider range of mixing angles can be accommodated, and one can only be guided by
considering various models for these angles.  In this section, we will discuss
plausible models for mixing angles.   Since we know that the quark sector has
nonzero mixing angles, but that the lepton sector may not, we will first look at
the lepton sector, and then the quark sector.

\subsubsection{Leptons}

The only phenomenological indication of any mixing in the lepton sector comes from
neutrino oscillations.  At the time of this writing, there are three indications of
oscillations:  solar neutrinos\cite{solar}, atmospheric neutrinos\cite{atmospheric}
and LSND\cite{LSND}.  It is difficult, although not quite impossible, to
reconcile all three of these in a three generation model.  If there are four
light neutrinos, in this case, the fourth neutrino must be sterile (an
isosinglet) in order to avoid the bounds from LEP.  Such a neutrino could exist
without requiring the existence of any additional fermions.  It is likely that
the situation will be clarified within a year or so at Superkamiokande and the
Solar Neutrino Observatory.   A detailed discussion of neutrino oscillations
and their phenomenology, including the recent strong evidence for atmospheric
neutrino oscillations at SuperKamiokande, can be found in Ref.
\cite{oscillationreview}.  We will defer to that review in this paper, and will not
discuss the possibility of light isosinglet neutrinos further.

We certainly will, however, discuss the case in which a fourth generation neutrino is very heavy.   This will automatically occur if the fourth generation is vectorlike.  Even if it is chiral, models exist that can give such a mass.  Recently, one of us\cite{pqlast} has considered a model of neutrino masses with four generations where one can obtain dynamically one heavy fourth generation and three light, quasi-degenerate neutrinos.  Ref. \cite{silva} has also considered a scenario with four generations which has similar consequences.

Suppose that the heavy leptons form a standard chiral family, with a right-handed
neutrino.  The bounds from the $Z$ width obtained at LEP force the mass of the $N$ and $E$ to be greater than 45 GeV.  Are
there any phenomenological bounds on the mixing?   In analogy with the quark sector
(as well as the prejudice from most models), one expects the mixing to be the
greatest between the third and fourth generations.  This will affect the
$\tau\nu_\tau W$ vertex, multiplying it by $\cos\theta$, where $\theta$ is the
mixing angle.  Since all $\tau$ decays occur through this vertex, the result will
be a suppression in the overall rates.  For some time, it was believed that the
mass of the $\tau$ was $1782\pm 2$ MeV, and the measured rate was too low; mixing
with a fourth generation was a straightforward
explanation\cite{sharma,rajpoot,li}.  However, the $\tau$ mass has now been
measured to much higher precision at BES\cite{bes} to be $1776.96\pm 0.2\pm 0.2$ MeV,
and the measured rate is now in agreement with theoretical expectations.  This has
been analyzed by Swain and Taylor\cite{swain,taylor}, who find a model-independent
bound on the mixing of $\sin^2\theta < 0.007$.   A similar bound can be obtained
for mixing between the fourth generation and the first two, although one expects
those angles to be smaller.

What values of the mixing might one expect?   There are four plausible (in the view
of the authors, of course) values of the mixing angle between the third and fourth
generations:
$
(a)\sin^2\theta= m_\tau/m_E\quad
(b)\sin^2\theta= m_{\nu_\tau}/m_N\quad
(c)\sin^2\theta= m_W^2/m_{Pl}^2\quad
(d)\sin^2\theta= 0
$

The first of these occurs in typical see-saw models.  The second occurs in models
in which the mixing occurs only in the neutrino mass matrix.  The third occurs in
models with a global or discrete lepton-family symmetry broken by Planck scale
effects, and the fourth occurs when the symmetry is not broken by Planck scale
effects.  We now discuss each of these.

The first relation, $\sin^2\theta= m_\tau/m_E$, occurs in models in which the
$2\times 2$   mass sub-matrices are of the form $\pmatrix{0&A\cr A&B\cr}$.
If the neutrino and charged lepton mass matrices are of that form, then the mixing
angle is given by $\sqrt{m_\tau/m_E}-\sqrt{m_{\nu_\tau}/m_N}$, which gives
$\sin^2\theta = m_\tau/m_E$ for realistic values of the $\nu_\tau$ mass.  Models of
this type were pioneered by Weinberg\cite{weinberg} and Fritzsch\cite{frit}, who
noticed that they will give the successful relation for the Cabibbo angle:
$\sin^2\theta_c = m_d/m_s$. Fritzsch also showed\cite{frita} that there are some very
simple symmetries which automatically give this relation.  When the rate for
leptonic decays was believed to be too low, Fritzsch\cite{fritb} used this relation
to propose that a fourth generation lepton of $100-200$ GeV could account for the
discrepancy.

As noted above, there is a lower bound on the mixing between the $\tau$ and the
$E$ given by $\sin^2\theta < 0.007$.  Using the Fritzsch relation, this becomes
a lower bound on $m_E$, which is given by $m_E > 250$ GeV.  This is very near the
bounds from perturbation theory.  We conclude that a very slight improvement in
the uncertainties in the $\tau$ decay rate will rule out the very general
relationship $\sin\theta=\sqrt{m_\tau/m_E}$ (or discover the effect!).

The second relationship, $\sin^2\theta= m_{\nu_\tau}/m_N$, will occur in models in
which, because of some discrete or global symmetry, the charged lepton mass matrix
is diagonal.  The Fritzsch relationship will then give $\sin^2\theta=
m_{\nu_\tau}/m_N$.  Given the cosmological bound on the $\nu_\tau$ mass, this gives
a value of $\sin^2\theta$ which is less than $10^{-10}$.  The $E$ or $N$ lifetime
(whichever is the lighter) lifetime will then be in the picosecond-nanosecond range,
with extremely interesting phenomenological consequences.

Suppose that one simply assumes that a discrete symmetry forbids any mixing at all
between the $E$ and $N$ and the other three generations.  This is simply an
extension of the familiar electron-number, muon-number and tau-number conservations
laws.  In this case, the mixing angle vanishes and the lighter (the $E$ or the $N$)
is absolutely stable.  As will be seen in the next Section, this would be
cosmologically disastrous if the $E$ is stable, but not if the $N$ is stable.

Finally, one can assume the discrete symmetry which forbids mixing, but note that
Planck mass effects are expected to violate all discrete and global symmetries. 
That means that higher dimension operators, suppressed by the Planck mass, will
violate these symmetries.  Two such examples are given by Kossler et
al.\cite{kossler}.  The mixing angle is then given by $\sin\theta\sim M_W/M_{Pl}$. 
This gives a lifetime for the lighter of the $E$ or $N$ of approximately ten years,
which is very near the bound for charged leptons, discussed in the next Section.

\subsubsection{Quarks}

In the lepton case, one could obtain stringent bounds on mixing with a fourth
generation by considering precise measurements of leptonic decays with theoretical
expectations.  Here, such precision (both theoretical and experimental) is
impossible.  One can still obtain bounds on mixing between the first two
generations and a fourth from the unitarity of the CKM matrix.  As noted in the
Particle Data Group Tables\cite{1aCCCCCC}, the mixing angle between the first and fourth
generations, $V_{uD}$ must be less than $0.08$.  However, other bounds are much
weaker\cite{paschos,rebelo}---the mixing angle between the second and fourth
generations,
$V_{cD}$, is only bounded by $\sin^2\theta < 0.5$.  In the top sector, one can use
constraints\cite{paschos} from $K_L\rightarrow \mu^+\mu^-$ to find that
$Re(V^*_{sU}V_{dU})< 8\times 10^{-4}$.   Since these are all mixings between
the first two and fourth generations, they are expected to be very small--of
greater interest is the bound on the mixing between the third and fourth
generations.  The value of the
$V_{tb}$ element in the CKM matrix is greater than $0.99$ (leaving very little room
for such mixing), however\cite{heinson} this is
determined {\it assuming} only three generations and CKM unitarity.  If one relaxes
this assumption, the value of
$V_{tb}$ could be as small as $0.05$\cite{1aCCCCCC,heinson}.  Thus, the mixing angle
between the third and fourth generations could be extremely large, and there are
effectively no phenomenological constraints on such mixing, if the $D$ is
sufficiently heavy to avoid affecting top quark decays.

Bounds from $K^o-\overline{K}^o$ and $B^o-\overline{B}^o$ mixing are also not
very strong.  The experimental value of 
$B^o-\overline{B}^o$ mixing was the first indication that the top mass might be
heavy, and the observation that it is, in fact, heavy means that only very weak
bounds on fourth generation masses and mixings may be obtained.   Similarly, the
large number of phases and angles in the four-generation CKM matrix (3 and 6,
respectively), implies that only weak constraints can be found from
$\epsilon'/\epsilon$.  

The reason that the bounds for these flavor changing processes are so weak is due
to the GIM mechanism.  This mechanism only applies if the fourth generation is
chiral.  If it is composed of vector-like isosinglets or isodoublets, then the GIM
mechanism will break down and one will have Z-mediated flavor-changing neutral
couplings (FCNC).  In addition, one can have an effect on flavor-diagonal neutral
currents (FDNC), since mixing of a doublet quark with a singlet will reduce its
left-handed coupling.   To be more specific, consider the case of a $Q=-1/3$
isosinglet quark,
$D$.  This case has been analyzed in great detail by Barger, Berger and Phillips
\cite{bargerberger}.
 The mixing between mass and weak eigenstates is given by
\begin{equation}
\pmatrix{d_L'\cr s_L'\cr b_L'\cr D_L'\cr}=\pmatrix{V_{ud}&V_{us}&V_{ub}&V_{uD}\cr
V_{cd}&V_{cs}&V_{cb}&V_{cD}\cr V_{td}&V_{ts}&V_{tb}&V_{tD}\cr V_{0d}&V_{0s}&V_{0b}&
V_{0D}\cr}\pmatrix{d_L\cr s_L\cr b_L\cr D_L\cr}
\end{equation}
In the basis where the $Q=2/3$ mass matrix is diagonal, the first three rows and
column of the $V$ matrix are just the usual CKM matrix.  The fourth row is not
relevant for the weak interactions with the $W$ and $Z$.  Since the entire matrix is
unitary, the CKM matrix will not be, leading to a suppression of flavor-diagonal
couplings.  The $Z$ couplings are given by
\begin{equation}
{\cal L}_{FCNC}={1\over 2}g_Z\sum_{i\neq j}z_{ij}\overline{q}_{iL}\gamma^\mu Z_\mu
q_{jL}
\end {equation}
where, using the unitarity of the $4\times 4$ matrix
\begin{equation}
z_{ij}=\delta_{ij}-V^*_{0i}V_{0j}
\end{equation}
The FDNC couplings are given by
\begin{equation}
{\cal L}_{FDNC}=g_Z\sum_i \overline{q}_i\gamma^\mu Z_\mu \left[ {1\over
4}z_{ii}(1-\gamma_5)-{1\over 3}\sin^2\theta_W\right]q_i
\end{equation}
Thus, mixing with the $D$ quarks reduces the direct left-handed FDNC couplings of the
light quarks, while leaving the right-handed quark couplings unchanged.
For a $Q=2/3$ isosinglet, the results are very similar.

Barger, Berger and Phillips (BBP) then analyze a large number of processes,
constraining elements of the $4\times 4$ matrix.   Their paper (written in early 1995)
has a huge number of references to papers dealing with isosinglet quarks; the reader
is referred to it for earlier references.  Since the top quark was
discovered,  calculation of the effects of isosinglets became much more
precise, since  a major uncertainty in the calculations had disappeared.  This
was the motivation of  BBP for reanalyzing all of the constraints on the
elements of the matrix.  BBP examine Z-decays, meson-antimeson oscillations,
$K_L, B^o, D^o\rightarrow
\mu^+\mu^-$, $B,D\rightarrow X{l}^+{l}^-$, and radiative $B$ decays, for
both the $Q=-1/3$ and $Q=2/3$ cases.   They list bounds on the off-diagonal matrix
elements.  

The above argument that $V_{tb}$ could be very small, while $V_{tD}$ could be very
large, does not work in the isosinglet $Q=-1/3$ case.  The reason is that the mixing,
by reducing the left-handed FDNC couplings of the left-handed quarks, can be
constrained from high-precision SLC and LEP results.  In the case where one just adds
a chiral family, the FDNC couplings are not reduced.  This is discussed in detail by
Nardi, et al.\cite{nardi}, who show that the precision data gives $|V_{0d}|^2<0.0023,
 |V_{0s}|^2<0.0036, |V_{0b}|^2<0.0020$.  Unitarity of the matrix then gives
$|V_{0D}|>0.996$, which in terms gives $|V_{qD}|<0.09$ for $q=u,c,t$.  Thus, the
mixing between the top quark and the $D$ cannot be very large.

In the $Q=-1/3$ case, BBP find that the off-diagonal matrix elements (in the fourth
row or column) have upper bounds ranging from $0.045$ to $0.09$.  Tighter bounds on
the geometric mean of two couplings are also found.   In the $Q=2/3$ case, the bounds
are weaker.  In that case, in fact, the $|V_{t0}|$ and $|V_{Ub}|$ mixings are
completely unconstrained, thus one could have large mixings between the third and
fourth generations.  The bounds on $|V_{Ud}|$ and $|V_{Us}|$ are very weak, given by
$0.15$ and $0.56$ respectively, while the bounds on $|V_{u0}|$ and $|V_{c0}|$ are as
strong as the corresponding terms in the $Q=-1/3$ case.

Other papers discuss certain particular processes in more detail, and look at
constraints in more specific models.    Let us first consider the case in which the
isosinglet quark has
$Q=2/3$.  This possibility was discussed in detail by Branco, Parada and
Rebelo\cite{rebelo}.  They showed that if one has an isosinglet
$Q=2/3$ quark, then the strongest signal will come from $D^o-\overline{D}^o$ mixing. 
Suppose the
$Uu,Uc$ and $Ut$ elements of the $Q=2/3$ mass matrix are given by $J_u/M_U,J_c/M_U$ and
$J_t/M_U$ respectively.  They show that if one assumes that the $J_i$ are equal,
then the current bound on $D^o-\overline{D}^o$ mixing gives an upper bound on $J/M_U$
of $0.033$.  This bound could easily be saturated in realistic models, and thus
$D^o-\overline{D}^o$ mixing gives the strongest constraint.  If one were to take the
$J_i$ to be hierarchical, with a value of $k m_i$, then the current limits give an
upper  bound on $k$ of $2$, and thus the well-motivated $k\sim 1$ is well within
reach of the next round of experiments. 

The $Q=-1/3$ case is more strongly motivated (since such states appear naturally
in representations of $E_6$).  These models have been analyzed in great detail in
papers by Silverman and collaborators.  The bounds on the $Zds$ vertex coming from
$K_L\rightarrow \mu^+\mu^-$, $\epsilon$, and the $K_L-K_S$ mass difference were
considered in early papers by Shin, Bander and Silverman\cite{shin} and by Nir and
Silverman\cite{nir1,nir2}; bounds on the $Zbd$ and $Zbs$ vertices arising from
$B_q-\overline{B}_q$ mixing and rare $B$ decays were discussed in Refs. \cite{nir1}
and
\cite{nir2}, and followed up in Refs.
\cite{silv1} and \cite{choong}. Bhattacharya et al.\cite{bhatt} analyzed radiative
B-decays in detail, and in another paper, Bhattacharya\cite{bhatta} analyzed the
bounds from Z-decays.   Although these processes are also discussed  by BBP, they are
described in much more detail in the above papers.    More recently,
Silverman\cite{silv3} has analyzed the mixing constraints using the latest data from
$B$ physics, and looks at the constraints that will be reached in upcoming $B$
factories.

 Lavoura and Silva\cite{lavoura} extended the
analysis to the case of multiple isosinglets.  They also pointed out that  a very
strong bound comes from
$K^+\rightarrow \pi^+\nu\overline{\nu}$, a process which was not considered by BBP.  
As noted by Branco, Parada and Rebelo\cite{rebelo}, using the realistic assumptions on
the mixing mentioned above, the bound from $K^+\rightarrow
\pi^+\nu\overline{\nu}$ is the most stringent for the $Q=-1/3$ case.  The resulting
bounds on the
$J/M$ are more stringent than in the $Q=2/3$ case, $J/M<.008$.  It is interesting
that the rate for $K^+\rightarrow \pi^+\nu\overline{\nu}$, which gives the strongest
bound, has been measured and may be high (one event seen and a quarter of an event
expected), but drawing conclusions on the basis of a single event is certainly
premature.

The results  of above two paragraphs apply to the case of an isosinglet
vector-like fourth generation.  The results are quite different in the case of a
vector-like isodoublet.  As shown in the analysis of the Aspon Model by Frampton and
Ng\cite{framptonng}, the flavor changing $Z\overline{d}_id_j$ vertex will be
suppressed relative to the isosinglet case by a factor of $m_im_j/m_D^2$.  This is
because, in the isodoublet case, the mismatch with light quarks occurs in the right
handed sector, forcing a helicity flip of the usual quarks. (In the isosinglet case,
the mismatch is in the left handed sector, thus no helicity flip is required.)   This
extra factor eliminates any significant constraints from flavor changing neutral
currents.  What about more exotic states?  Recently, del Aguila, Aguilar-Saavedra and
Miquel\cite{paco} looked at the constraints on anomalous top quark couplings in
models with exotic quarks.  They look at chiral and non-chiral singlets and doublets,
including mirror quarks, and find some very general inequalities which allow one to
go from LEP bounds on diagonal $Z$ couplings to stringent bounds on the off-diagonal
couplings.

Thus, the bounds on mixing of a chiral fourth generation with the third are
virtually non-existent, as are bounds on an isodoublet fourth generation, but the
bounds on an isosinglet vector-like fourth generation are getting near the
``interesting" range--and may be  improved significantly with more measurements on
$K^+\rightarrow
\pi^+\nu\overline{\nu}$ and on 
$D^o-\overline{D}^o$ mixing.

It should also be pointed out that the four-generation model does have many
additional phases.  A detailed analysis of CP violation in the isosinglet $Q=-1/3$
case has been carried out by Silverman\cite{silv3,silv2}. This is also discussed in
BBP.  The entire ``unitarity quadrangle" is analyzed.  We will discuss CP violation in
more detail in Section VI.

What are the theoretical expectations for the mixings?  The Fritzsch ansatz (for
the $3\times 3$ quark mass matrices) fails\cite{gilman,ma1,kfp} for a $174$ GeV
top quark, although the generic expressions $\sin\theta=\sqrt{m_b/M_D}$ or
$\sqrt{m_t/M_U}$ could easily be accommodated in other models. As noted above in
the lepton case, many models with flavor symmetries will  have  the
$2\times 2$   third-fourth generation mass sub-matrices  of the form
$\pmatrix{0&A\cr A&B\cr}$.  These will have a 3-4 mixing angle of
$O(\sqrt{m_t/M_U})$, which is of order unity.  Thus, one should keep in mind that
the mixing angle between the third and fourth generation could be very large.
As in the lepton case, one could imagine a symmetry in which the $Q=2/3$ quarks
are diagonal, and then the 3-4 mixing angle would be of order $\sqrt{m_b/M_D}$.

The possibility that there is a symmetry prohibiting mixing
altogether  cannot be excluded.  With such a symmetry, the lighter
of the $U$ or $D$ would be stable, leading to a cosmological disaster; however
one could assume that Planck mass effects violate the symmetry, giving a long
(but possibly acceptable) lifetime of $O(10-100)$ years.

In the vector-like case, the mixing angles are also related to the $J_i$ discussed
above, and the expectations are not very different. It should be noted that in the
Aspon Model, the mixing angles are typically
$10^{-5}-10^{-3}$ in order to account for the appropriate amount of CP 
violation.

\newpage
\section{Lifetime and Decay Modes.}

In the previous Section, we discussed the masses and mixing angles of additional
quarks and leptons.  Now, we consider the lifetime and decay modes of such
fermions.  In the standard model, the lifetime and decay modes of the most
recently discovered fermion, the top quark, were not particularly interesting--it
was known that the top quark would decay very quickly (quickly enough that the
width is large enough to obscure any structure in the toponium system) and that
it would decay almost entirely into a $b$ and a $W$.    

However, there are several interesting possibilities for the case of additional
fermions.  In the chiral case, the $N$ could be heavier than the $E$, forcing the
$E$ to decay only via mixing; if the mixing angles are small (as discussed in the
last Section), the lifetime could be very long.  In the quark case, the mass of
the $D$ is likely less than the sum of the masses of the top quark and $W$, and
thus the $D$ will decay only via the doubly-Cabibbo suppressed $c+W$ mode or the
one-loop $b+Z$ mode; either could give a long lifetime, especially if mixing
angles are very small.  In the non-chiral case, the GIM mechanism will not be
operative, leading to tree-level flavor-changing neutral decays, such as
$E\rightarrow \tau Z$, $N\rightarrow \nu_\tau Z$ and $D\rightarrow bZ$; these
decays could give very unique and interesting phenomenological signatures.  In
addition, we will see that the mass-splitting of the leptons and quarks in the
non-chiral doublet case is calculable, and gives lifetimes with potentially
observable decay lengths.

We will begin by discussing the lepton sector, first for the chiral case and then
for the non-chiral case, and then turn to quarks.
\newpage
\subsection{Leptons}

\subsubsection{Chiral Leptons}

Much of the early work on the phenomenology of heavy
leptons\cite{barger}  assumed that $m_N$ is less than $m_E$, however, as
discussed in the last Section, there is no particular reason for that
assumption.  Let us first assume the opposite--that $m_E<m_N$ (and both
greater than $M_Z/2$.   This case has been discussed in detail by Hou and
Wong\cite{houwong}.  The $E$ will only be able to decay via mixing,
$E\rightarrow \nu_\tau W^*$, where $W^*$ is a real or virtual $W$.  The $N$
will decay via either $N\rightarrow EW^*$ or $N\rightarrow \tau W^*$.  The
decay rates are
\begin{eqnarray}\label{phase}
\Gamma(N\rightarrow EW^*)&=& 9\cos^2\theta_{34}{G^2_Fm^5_N\over
192\pi^3}f\left(m_N^2/m_W^2,m^2_E/m^2_N)\right),\cr\cr
\Gamma(N\rightarrow \tau W^*)&=& 9\sin^2\theta_{34}{G^2_Fm^5_N\over
192\pi^3}f\left(m_N^2/m_W^2,0)\right),
\end{eqnarray}
where $f(\alpha,\beta)$ is given by\cite{bigi}

\begin{equation}
f(\alpha,\beta)=2\int_0^{(1-\sqrt{\beta})^2}
{dx
[(1-\beta)^2+(1-\beta)x-2x^2](1+\beta^2+x^2-2(\beta+\beta x+x))^{1/2}
\over [(1-x\alpha)^2+\Gamma_W^4/M_W^4]^2]}
\end{equation}  
This function accounts for both real and virtual $W$'s.  The rate for
$E\rightarrow \nu_\tau W^*$ is identical to the second of these equations with
$m_N\rightarrow m_E$.  Since the angle is expected to be small, one might
expect that $N\rightarrow EW^*$ would be favored over $N\rightarrow \tau W^*$,
however, one must recall that the $S$ and $T$ bounds discussed above imply
that the $N$ and $E$ must be fairly close in mass, and thus the decay might be
significantly phase space suppressed.  These rates are plotted in Figure 9.

\begin{figure} 

\centerline{ \epsfysize 4in \epsfbox{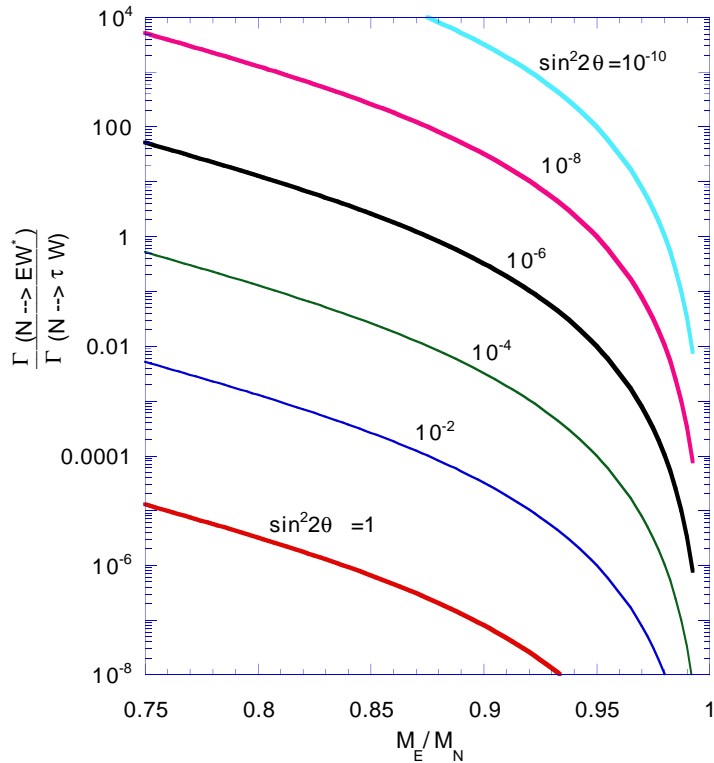}  }
\vskip .25in
\caption{Relative branching ratios of the $N$ into $EW^*$ vs. $\tau W^*$ for various values of
the mixing angle and the $E$ to $N$ mass ratio.  One sees that the decay into $\tau W^*$ will
dominate unless the mixing angle is very small. $\sin^2 2\theta$ is the mixing
between the third and fourth generations.}

\end{figure}


We see that the decay of $N$ into $\tau W^*$ tends to dominate, unless the
mixing angle is extremely small.  This leads to interesting phenomenology, as
will be discussed in Section VII.  If the
$E$ is heavier than the $N$, the results are the same with $N\leftrightarrow
E$ and
$\nu_\tau\leftrightarrow \tau$.

Note that the decay rate of $E\rightarrow \nu_\tau W^*$ is proportional to
$\sin^2\theta_{34}$.  In the previous Section, we noted that this angle could
be very small.  Simplifying the expression for the decay, and assuming that
the mass of the $E$ is greater than the $W$ (thus the $W$ is real), the
width of the $E$ is given by
\begin{equation}
\Gamma(E\rightarrow \nu_\tau W)=(180\ {\rm MeV})\sin^2\theta_{34}\left({m_E\over
m_W}\right)^3
\end{equation}

Consider the four plausible values of $\sin^2\theta_{34}$ discussed in the
previous Section.  If $\sin^2\theta_{34}=m_\tau/m_E$, then the decay is very
rapid and would occur at the vertex.  If $\sin^2\theta_{34}=m_{\nu_\tau}/m_N$,
then the lifetime is of the order of a few picoseconds, which has very interesting
phenomenological consequences---it might be possible to see the charged track.  If
$\sin^2\theta_{34}=0$, then the
$E$ is stable; this would be cosmologically disastrous, since they would bind with
protons to form anomalously heavy hydrogen.  If $\sin^2\theta_{34}=m_W^2/m_{Pl}^2$,
then the lifetime is approximately $10-100$ years.   We now address whether such
a lifetime would pass cosmological muster.

The bounds on the lifetime of a fourth generation charged lepton were first
discussed several years ago\cite{sheryao}.  They considered charged lepton
masses ranging from $50$ GeV to $50$ TeV.  Stable leptons are ruled out by
searches for heavy hydrogen.  Since any decay of the $E$ will result in photon
emission, failure to observe such emission in the diffuse photon background
implies that the lifetime must be less than $10^{13}$ seconds (time of the
cosmic background radiation (CMBR) production).  Using COBE data on the CMBR,
and requiring that the radiation in the decay not distort the CMBR, they found
a bound on the lifetime which ranges from $10^9-10^{11}$ seconds as the mass
ranges up to $1$ TeV.   Very recently, an analysis by Holtmann, Kawasaki,
Kohri and Moroi\cite{hkkm} looked at the radiative decay of a long-lived
particle, $X$, and the effects on big-bang nucleosynthesis (the $E\rightarrow
\nu_\tau W$ is not ``radiative", however, a significant fraction of the energy
will eventually turn into photons, and the results are the same).  The photons
emitted in the decay may photodissociate deuterium (lowering its abundance) and
helium (which raises the deuterium abundance), destroying the agreement between
theory and observation.  They give bounds on the lifetime as a function of
$m_XY_X$, where $Y_X\equiv n_X/n_\gamma$ is the relative abundance of the $X$.
For heavy leptons, the abundance as a function of mass was calculated in Ref.
\cite{sheryao}.  For heavy lepton masses between $100$ and $500$ GeV, the
contribution to $\Omega h^2$ varies from $0.05$ to $0.01$, leading to a value of
$m_XY_X$ which varies from $6\times 10^{-10}$ to $1.2\times 10^{-10}$ GeV (for a
Hubble constant of $65$ km/sec/Mpc).  From Tables 3-5 of Holtmann, Kawasaki, Kohri
and Moroi\cite{hkkm}, one can see that this correponds to an upper bound on the
lifetime of between $10^7$ and $10^8$ seconds.  Given the uncertainties in the
abundance calculation, nucleosynthesis calculation, deuterium and helium
abundance observations, etc., this is not inconsistent with a lifetime of $10-100$
years.  Thus, the possibility
that $\sin^2\theta_{34}=m_W^2/m_{Pl}^2$ is marginally allowed. 

To summarize, if the mixing angle $\theta_{34}$ is not very small, then both
the $N$ and $E$ will decay via Cabibbo-suppressed decays:  $N\rightarrow \tau
W^*$ and $E\rightarrow \nu_\tau W^*$.   If the angle is very small (of the
order of $m_{\nu_\tau}/m_N$ or less), then the heavier of the two will
decay into the lighter, while the lighter decays via the Cabibbo-suppressed
decay.  In this case, the latter lifetime could be quite long, as long as
a few picoseconds for $\sin^2\theta_{34}=m_{\nu_\tau}/m_N$ and as long as $10$ years
for $\sin^2\theta_{34}=m_W^2/m_{Pl}^2$.

\subsubsection{Nonchiral Leptons}

As noted in the previous Sections, an interesting feature of models with a vector-like
doublet (with small mixing with light generations) is that the two members of the
doublet are degenerate in mass, at tree level.  This
degeneracy will be split by radiative corrections, and the size of this splitting
is crucial in understanding the lifetimes and decay modes of the heavy leptons.

\begin{figure} 

\centerline{ \epsfysize 3in \epsfbox{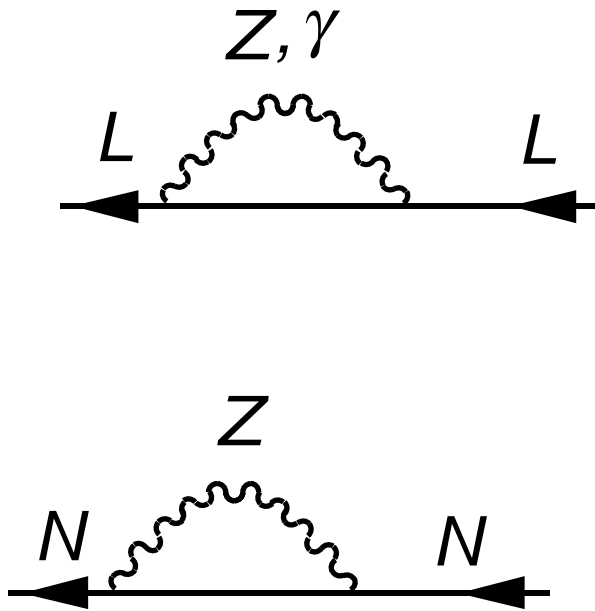}  }

\caption{Diagrams contributing to the $E$-$N$ mass difference.}

\end{figure}

The splitting is due to the diagrams in Figure 10.   The size of the splitting
was first calculated by Dimopoulos, Tetradis, Esmailzadeh and Hall\cite{dteh}
(DTEH), and later  calculated by Sher\cite{sherlept} (S) and even later by
Thomas and Wells\cite{thomas} (TW).  The result is that the charged lepton is
heavier than the neutrino, with a mass splitting of
\begin{equation}
\delta m={\alpha\over 2}m_Zf(m^2_E/m^2_Z)
\end{equation}
where
\begin{equation}
f(x)={\sqrt{x}\over\pi}\int_0^1 dx\ (2-x)\ln\left(1+{x\over r(1-x)^2}\right).
\end{equation}
For small $x$, $f(x)\rightarrow 0$, but for large $x$, $f(x)\rightarrow 1$, and thus
the splitting reaches an asymptotic value of ${1\over 2}\alpha m_Z\simeq 350$ MeV
for $m_E>>m_Z$.  The splitting is plotted in Fig. 11.

\begin{figure} 

\centerline{ \epsfysize 4in \epsfbox{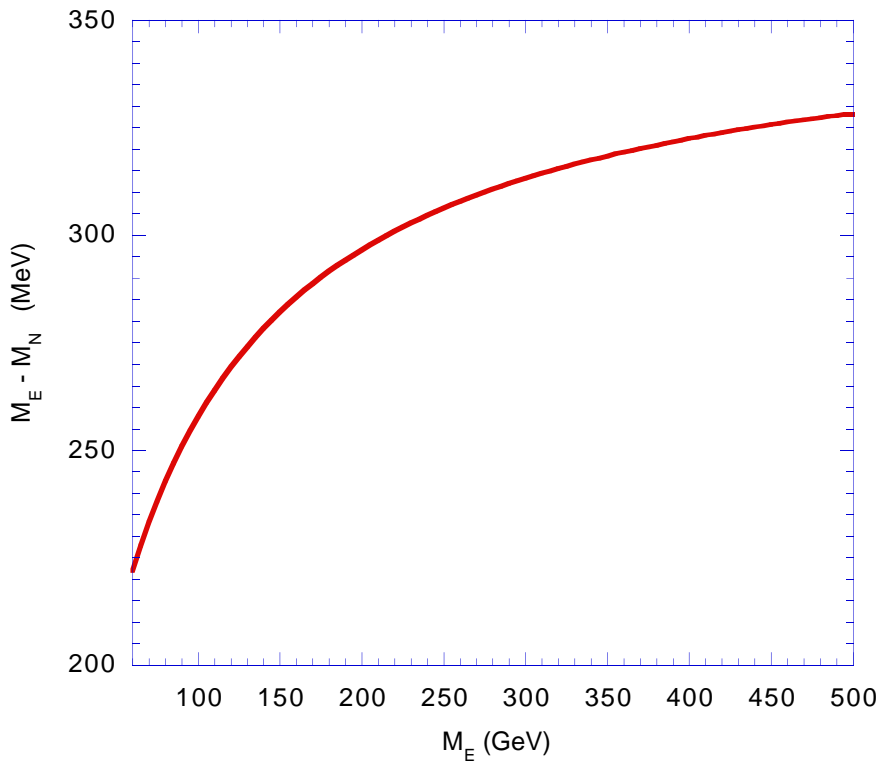}  }

\caption{Mass difference between the $E$ and the $N$ as a function of the $E$ mass.}

\end{figure}

In DTEH\cite{dteh}, the authors considered very heavy leptons (of the order of
several TeV) and looked at the question of whether such leptons could constitute
the dark matter (they can't).  In S\cite{sherlept}, the decay width of
$E\rightarrow Ne\overline{\nu}$ and $E\rightarrow N\mu\overline{\nu}$ was
calculated--the inverse of these widths corresponded to a lifetime of $1-2$
nanoseconds, which is obviously of great phenomenological interest.  In S, it was
also pointed out that the value of the splitting is very robust, and that
supersymmetric contributions to the splitting turn out to be very small,
since for most of parameter-space, the contributions turn out to be proportional
to
(${1\over 4}-\sin^2\theta_W$).  In TW\cite{thomas}, it was pointed out that the
dominant decay of the $E$ will be into $N\pi$ (a decay neglected in S),
resulting in a much shorter (by roughly a factor of $10$) lifetime.  The decay
length then is of the order of centimeters, rather than tens of centimeters,
greatly complicating detection.  In Fig. 12, we plot the decay distance in the
lab frame as a function of $m_E$ for a variety of center of mass energies.  TW
do propose an interesting signature involving triggering on an associated hard
radiated photon; this will be discussed in Section VII.   

Thus, in the vector-like doublet case, without significant mixing with the lighter
generations, the charged member of the doublet is heavier and decays primarily into
$N\pi$ (with a VERY soft pion) with a decay length given in Fig. 12.

\begin{figure} 

\centerline{ \epsfysize 4in \epsfbox{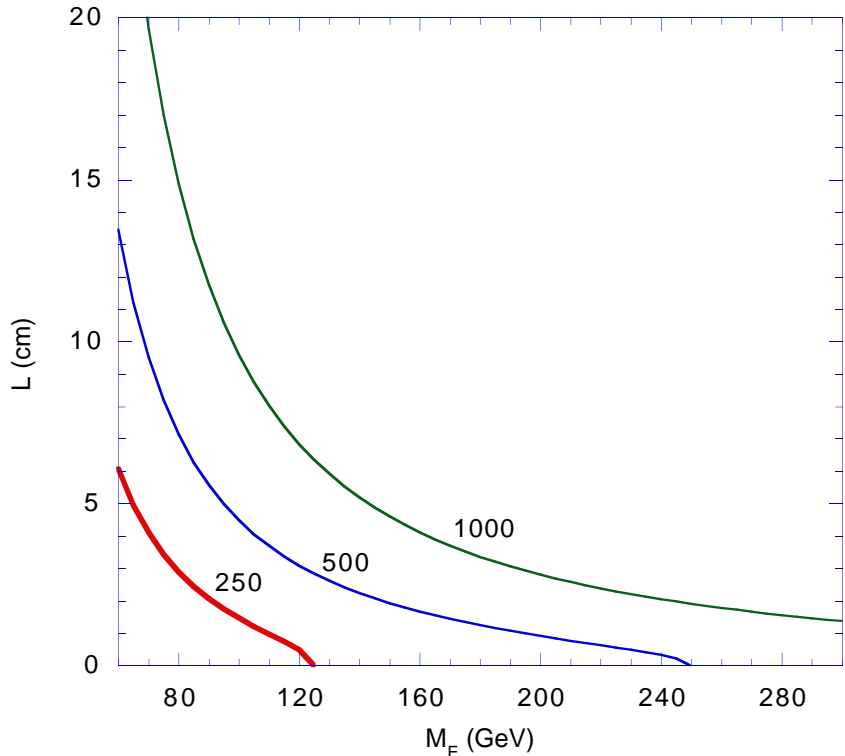}  }

\caption{Decay length for the $E$ as a function of its mass.  The lines  are labelled with the
center of mass energy of the collider in GeV.}

\end{figure}

  The fully
leptonic decays, even though they have branching ratios of only a few percent, might
be easier to detect, although even in that case, the very soft electron or muon would
be difficult to separate from backgrounds.   If there is significant mixing, then
both the $E$ and the $N$ will decay  into the lighter generations.  We
now consider this possibility (which also applies to vector-like singlets).

We now consider the case in which there is significant mixing of vector-like
leptons.  By ``significant", we mean that the mixing angle, $\sin^2\theta_{34}$, is
greater than about $10^{-11}$, so that the decay (see the expression for the
lifetime above) occurs at the vertex.  The results will be very similar to the
chiral case, with one extremely important difference.  Due to the breakdown of the
GIM mechanism, the decays $E\rightarrow\tau Z$ and $N\rightarrow \nu_\tau Z$ will
occur.

For $m_E>M_Z$, the branching ratios of the $E$ are given by\cite{fram}
\begin{equation}
{\Gamma(E\rightarrow\tau Z)\over\Gamma(E\rightarrow \nu_\tau W)}=
{|U_{E\tau}|^2\over 2\cos^2\theta_W|U_{E\nu_\tau}|^2}
{(m_E^2-2m^2_Z+m_E^4/m_Z^2)(m_E^2-m_Z^2)\over
(m_E^2-2m^2_W+m_E^4/m_W^2)(m_E^2-m_W^2)}
\end{equation}
There is no particular reason to believe that this branching ratio should be
small.  One might expect 
$|U_{E\nu_\tau}|^2$ to be of the order of $m_\tau/m_E$, as discussed in detail in
the last Section.  As discussed in the last Section, in the isosinglet heavy lepton
case, one finds that $|U_{E\tau}|$ is  of
order of the ratio of $M_{34}$ to $M_{44}$ in the leptonic mass matrix which would
be similar to $|U_{E\nu_\tau}|$; the resulting branching ratio would be very large.
In the isodoublet heavy lepton case, there is an additional suppression of
$m_\tau/m_E$, which results in a branching ratio of about $0.1\%$.  Thus, one
expects a branching ratio for
$E\rightarrow
\tau Z$ to be a fraction of a percent in the isodoublet case and very large in the
isosinglet case.

It is important to note that even if the branching ratio is as low as a
fraction of a percent, the background for a particle decaying into $\tau Z$ would be
extremely small.  A major problem with conventional heavy-lepton detection has been
backgrounds; so the
$E\rightarrow\tau Z$ signature, even with a branching ratio of a fraction of a
percent or so, might very well be easiest to detect.  This will be discussed in more
detail in Section VII.
\newpage
\subsection{Quarks}

The discussion of the lifetime and decay modes of the quarks follows the general
discussion of those of leptons, with a few crucial differences.  While the $E$ and
$N$ are certainly much heavier than the $\tau$, it is not necessarily the case that
the $U$ and $D$ quarks are much heavier than the top.  In addition, the mass
splitting in the vector-like quark case is a factor of three smaller than that for
vector-like leptons, which can drastically affect the decay modes and lifetime.

\subsubsection{Chiral Quarks}

Due to constraints from the $\rho$ parameter, the $U$ and $D$ quarks cannot
have masses which are too different, and thus the decay $U\rightarrow D+W$ or
$D\rightarrow U+W$ cannot occur into real $W$'s.  Suppose, for the moment, that
$m_U>m_D$.  Then the allowed decays of the $U$ will be $U\rightarrow (D\ {\rm or}\ q)
+ W^*$,
$U\rightarrow q+W$, where $W^*$
refers to a virtual $W$. The allowed decays of the $D$ are $D\rightarrow (t\  {\rm
or}\  q) + (W^*\ {\rm or}\  W)$.  In addition, one can have a flavor-changing
neutral current decay $D\rightarrow b+Z$, which (in the chiral case) can occur
through one loop.  The fact that the flavor-changing neutral decay of a
fourth generation quark could be significant was first pointed out by Barger,
Phillips and Soni\cite{bargerps}, and followed up by Hou and
Stuart\cite{hou1,hou2}.  In the latter works, they noted that the decay
$D\rightarrow b+Z$ dominates over other flavor-changing decays, such as
$D\rightarrow b+\gamma$ and $D\rightarrow b+g$. The possible $D\rightarrow b+Z$
mode, like the $E\rightarrow \tau+Z$ mode, is very important phenomenologically due
to the very clear signatures\cite{carls,gunion}. Precise analytical formulae for the
various decays, and a discussion of the decays of the
$D$ (from which much of the discussion below is taken), can be found in the recent
work of Frampton and Hung\cite{framptonhung}.

 The two-body and three-body decay widths are given by the first of Eq. \ref{phase},
with the obvious substitutions of $|V_{Dq}|$ or $|V_{Uq}|$ for $\cos^2\theta_{34}$.
The flavor-changing neutral current decay is given by
\begin{equation}
\Gamma(D\rightarrow bZ)=|V_{Dt}|^2{G_F(m_D)^3\over 4\pi\sqrt{2}\cos^2\theta_W}
\left( ({g^2\over 16\pi^2})^2\Delta(m_U,m_t)\right)I_2(m_b/m_D,m_Z/m_D)
\end{equation}
where we have assumed, for simplicity, that $V_{Ub}=-V_{Dt}$ so there will be a GIM
suppression when $m_U=m_t$ (this occurs in most models).  We have also assumed
that $V_{UD}$ is approximately unity.  Here,
\begin{equation}
\Delta(m_U,m_t)=\left( {M_U^2-m_t^2\over m^2_W}(\ln({m_W^2\over
m^2_{heavy}})-1)\right)^2
\end{equation}
and $m_{heavy}$ refers to the heavier of $U$ and $t$.  The factor $I_2$ is the
standard two-body phase space factor, which is unity in the limit $m_D>>m_Z$.

The $U$ decays very rapidly, at the vertex (unless the masses are unusually close
together, which is unlikely in the chiral quark case, since they arise from
different terms); however the $D$ can be long-lived.  The decay modes of the
$D$ depend crucially on the mass.

First, suppose the $D$ is lighter than the top quark.  The decay can only occur
into $c + W$ or $b+Z$ (one expects the decay into $u+W$ to be highly
suppressed).     Two important issues arise:  which of these has a larger branching
ratio (since their phenomenological signatures are very different), and is it
possible that the decay length might be large enough to be detected?  We now
address both of these  questions. 

\begin{figure} 

\centerline{ \epsfysize 4in \epsfbox{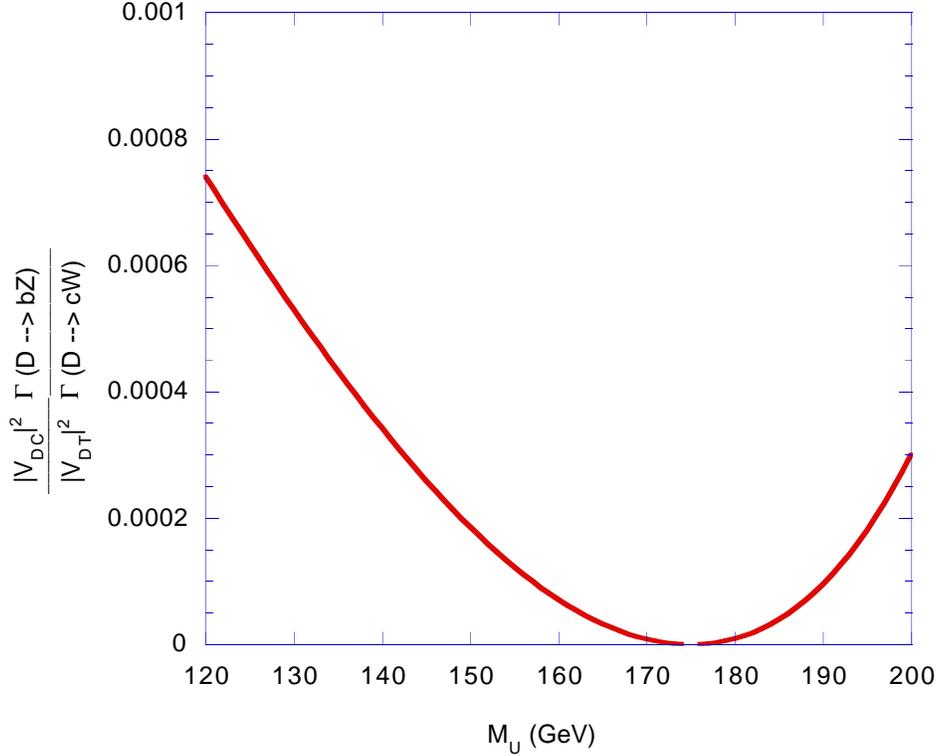}  }
\vskip .25in
\caption{Ratio of width of  $D\rightarrow b+Z$ to that of $D\rightarrow c+W$.  Note the GIM
suppression of the FCNC decay when the $U$ mass equals the top mass.}

\end{figure}

From the 
above formulae, the ratio of the widths can be calculated.  The only dependence of the D-mass
in this ratio is in the ratio of phase spaces--unless the $D$ is fairly close in mass to the
$Z$ this ratio is near unity.   The only mass-dependence is then in the $U$ mass-dependence in
$\Delta(m_U,m_t)$ above.  The result is given in Figure 13, for various values of the
$U$ mass (recall the fact that the
$U$ and
$D$, in this range, cannot be different in mass by more than $20$ GeV due to
$\rho$ parameter constraints).  The results
depend on the ratio of
$|V_{Dc}|$ to
$|V_{Dt}|$.  Since the former is ``doubly-Cabbibo-suppressed" (crossing two
generations), one expects the ratio to be small.  We see that the $D\rightarrow
bZ$ decay mode dominates if the ratio is small, whereas $D\rightarrow cW$ dominates if
it is large (unless the $U$ mass is very near the top mass.  Since there is little theoretical
guidance as to the size of this ratio, both signatures should be looked for.

 Could a vertex be seen?  From the expression for $\Gamma(D\rightarrow bZ)$, one
can see that the width of this mode\cite{hou1} is between $10^{-4}|V_{Dt}|^2$ MeV and
$10^{-2}|V_{Dt}|^2$ MeV over the mass range of interest.  This implies that the
lifetime will be at most $6\times 10^{-18}/|V_{Dt}|^2$ seconds.  This would mean
that one must have $|V_{Dt}|<< 10^{-3}$ in order to detect a vertex.  Although such
a small angle is not expected in the chiral sequential quark case (the Aspon
Model involves non-chiral quarks), it is not excluded, if one has a flavor
symmetry with an almost unbroken $3+1$ structure.

For $m_D$ between $177$ and $256$ GeV, the $D$ can decay into the top quark via the
three-body decay into a virtual $W$.  In this case, $D\rightarrow t+W^*$ will be
competetive  with $D\rightarrow c+W$ and $D\rightarrow b+Z$---the latter are
suppressed by the doubly-Cabibbo suppressed mixing angle or the extra loop, the
former is suppressed by three body phase space.  If we compare the rate for
$D\rightarrow t+W^*$ with $D\rightarrow b+Z$, the mixing angle cancels out. 
In Figure 14, we have plotted the ratio of these two decays.
\begin{figure}

\centerline{ \epsfysize 4in \epsfbox{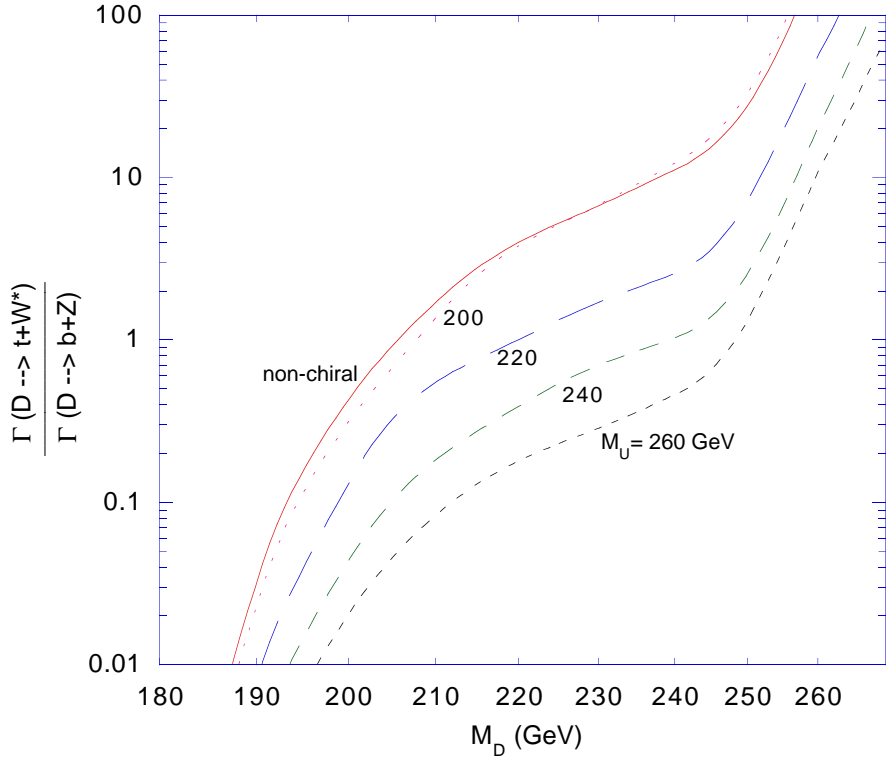}  }
\vskip .25in
\caption{Ratio of width of  $D\rightarrow t+W^*$ to that of $D\rightarrow b+Z$
 for various values of $M_U$.  The non-chiral line corresponds to the vectorlike doublet case, which is independent of $M_U$.  For the non-chiral isosinglet case, the ratio is too small to be seen on the graph.  }

\end{figure}
  Thus, as pointed out in Ref.
\cite{framptonhung}, the three-body decay of the $D$ is irrelevant for $D$ masses
below about $230$ GeV (depending on  $M_U$), and thus the arguments of the previous two paragraphs are
unchanged.

As the mass increases beyond $230$ GeV, the three body decay becomes more
important, and as soon as the mass exceeds $256$ GeV, the two-body decay
$D\rightarrow t+W$ becomes accessible.  At that point, $D\rightarrow t+W$ becomes
the dominant decay.  Again, very small mixing would be needed to see the vertex,
$|V_{Dt}|<10^{-5}$.

What about the decay of the $U$ quark (again, assuming that $m_U>m_D$)?  The U can
decay into $D$ and a virtual $W$, $U\rightarrow D+W^*$, or into a light quark
(most likely a $b$) and a real W, $U \rightarrow b+W$.  Which decay mode dominates
depends on how close the $U$ and $D$ are in mass and on the $|V_{Ub}|$ mixing angle.
This was discussed in Ref. \cite{framptonhung}.  As $m_U$ varies from $180$ to
$250$ GeV, the width for $U\rightarrow b+W$ varies from $1.75|V_{Ub}|^2$ GeV to
$4.7|V_{Ub}|^2$ GeV.  For $U\rightarrow D+W^*$, the width (assuming $|V_{UD}|$ is
nearly unity) is $5.2\times 10^{-5}$ GeV for $m_U/m_D=1.1$, and drops to $3\times
10^{-8}$ GeV for $m_U/m_D=1.02$.  For moderate mixing angles, $|V_{Ub}|>0.003$, the
$U\rightarrow b+W$ decay mode will always dominate.   For mixing angles in the
range between $10^{-3}$ and $10^{-4}$, which dominates dependes sensitively on the
mass ratio between the $U$ and the $D$.  For mixing angles below $10^{-4}$, the
$U\rightarrow D+W^*$ will dominate.   In no cases will the width be small enough so
that a vertex can be seen:  the $U$ will decay at the vertex.

If $m_D>m_U$, everything we have said above will carry through, exchanging $t$ with
$b$, etc.   A principal difference is that one can consider lighter long-lived $U$
quarks (since one can have $m_U<m_t$) than for $D$ quarks.

To summarize, if $m_U>m_D$, then the primary decays modes of the $D$ will be
$D\rightarrow c+W$ and $D\rightarrow b+Z$ if the $D$ mass is below about $230$ GeV.
The former will dominate if $|V_{cD}|/|V_{Dt}|$ is greater than $0.01$, the latter
will dominate if it is less than $0.001$.  As the $D$ mass increases above $230$
GeV, the decay $D\rightarrow t+W^*$ begins to dominate.  In all cases, one must
have very small mixing angles ($|V_{Dt}|<10^{-3}$ or $10^{-5}$, depending on the
mode) in order for the decay to occur a measureable distance from the vertex.  The
$U$ will always decay at the vertex, into either $D+W^*$ or $b+W$, depending on the
precise masses and mixing angles.

\subsubsection{Nonchiral Quarks}

For vector-like quarks, either isosinglet or isodoublet, many of the above results
are unchanged.  As noted in Ref. \cite{framptonhung}, the decay widths into real or
virtual $W$'s will not be significantly changed.  As in the nonchiral lepton case,
there are two major differences:  the mass difference between the $U$ and the $D$
in the isodoublet case is calculable (if mixing is small), and the flavor-changing
neutral decay
$D\rightarrow bZ$ can occur at tree level.

In the isodoublet case, the mass difference between the $U$ and the $D$ can be
shown\cite{sherlept} to be $1/3$ that of the lepton case, and will thus be between
$70$ and $110$ MeV, if the mixing with lighter generations is small.  This means
that the hadronic decay is forbidden, and the only decay would be
$U\rightarrow De\nu$, with a lifetime of the order of microseconds.  Thus, in the
absence of mixing, both the $U$ and the $D$ would be absolutely stable, as far as
accelerators are concerned, and the mass difference is irrelevant.  We thus must
consider both $U$ and $D$ decays.

The $D$ can decay, as in the chiral
case, into either $c+W$, $t+W^*$, $t+W$ or $b+Z$.  The only difference in the above
discussion is the rate for $D\rightarrow b+Z$.  In the expression for
$\Gamma(D\rightarrow b+Z)$ in the last section, one must replace the factor of
$g^4\Delta(m_U,m_t)/64\pi^4$ from the loop with $m_b^2/m_D^2$, in the isodoublet
case, and with unity in the isosinglet case (the factor of $|V_{Dt}|^2$ is the same
in each case).  In the isodoublet case, this does not change the result for the width
by more than an order of magnitude, but in the isosinglet case, it increases the
width by roughly three orders of magnitude.

First, if the $D$ is lighter than the top quark, then the decay modes in the
isodoublet case are very similar to the chiral case---the results depend
sensitively on the ratio of $|V_{Dc}$ to $|V_{Dt}|$, and either the $D\rightarrow
c+W$ and $D\rightarrow b+Z$ will be dominant.  A displaced vertex can only be seen
if 
$|V_{Dt}|<10^{-3}$.  This is expected in the Aspon Model case, and thus one can
expect a significantly displaced vertex in the model (as emphasized in Ref.
\cite{framptonhung}).  In the isosinglet case, the absence of the $m_b^2/m_D^2$
suppression makes the $D\rightarrow b+Z$ mode dominant, and also requires that
$|V_{Dt}|<<10^{-5}$ in order for a displaced vertex to be seen (which is not
expected in the Aspon Model).

If the $D$ is heavier than the top quark, then the argument in the previous
section carries through without significant modification in the isodoublet case;
the cross-over where the three body decay becomes relevant is closer to $210$ GeV
than to $230$ GeV.  In the isosinglet case, however, the $D\rightarrow b+Z$ mode
dominates, even over the two-body top quark decay, until $m_D$ is well over $300$
GeV, and remains significant even to much higher masses.   

Thus, we see that in either the chiral or non-chiral case, the decay $D\rightarrow
b+Z$ is very important.  It was  pointed out in Ref.
\cite{framptonhung} that, if this mode were detected, then one could look at the
chirality of the $Z$ to determine which of the two cases applies.  This might be
the quickest way to determine the chirality of the heavy quarks.

Unless the $U$ is very heavy, above $310$ GeV, the flavor-changing neutral decay
$U\rightarrow t+Z$ is forbidden or highly suppressed by phase space.  Since
$U\rightarrow c+Z$ is suppressed by small mixing angles, one has the $U\rightarrow
b+W$ decay mode dominating.  Again, one can only detect the vertex if the angle is
very small, less than $10^{-5}$.

\newpage

\section{Dynamical Symmetry Breaking}

It is fair to say that perhaps one of the most important discoveries that can be made in
the future will be that of the Higgs boson. In the absence of any alternative plausible
explanation for particle masses, the concept of spontaneous breakdown of a gauge symmetry
via the Higgs field as the origin of all masses is universally accepted. The only problem
is that it
has not been found. Not only does one not know its mass, but one also does not know in 
what shape or form it should be. One fact that we do know however , regardless
of how massive and in what form the Higgs boson may be, is the scale of electroweak
symmetry breaking: $v = 246$ GeV. All masses in the SM are expressed in terms of that
scale, e.g. $M_W = (1/2) g v$ and $m_i = g_{Y_i} v/\sqrt{2}$, where $g_{Y_i}$ are Yukawa
couplings. There is a rather intriguing fact: With $v/\sqrt{2} \sim 174$ GeV, it follows
that the top quark Yukawa coupling, $g_t$, is of order unity, unlike all other fermions.
That the top quark is so heavy and its mass is so close to the electroweak 
breaking scale is cause to wonder about any relationship that it might have 
with the mechanism of symmetry breaking itself. 

If the top quark mass is so close to the electroweak scale, is it possible that
it itself is reponsible for the breaking of the electroweak symmetry? This fascinating 
possibility was first entertained by Refs.\cite{1af}-\cite{1ai}. It is now 
commonly refered to as
``top-condensate models''. In these models, the Higgs boson is generally viewed as a
$t\bar{t}$ composite field generated by some unknown dynamics at a scale 
$\Lambda \gg 246$ GeV. In other words, the electroweak symmetry breaking is
dynamical in these scenarios, i.e. it is broken by a 
$t\bar{t}$ condensate or something similar. One interesting feature of these models is
the prediction of a heavy fermion mass (e.g. the top quark mass or a fourth
generation mass) as a function of the Higgs boson mass through the
so-called compositeness conditions.
Before describing these models and their variants,
let us first summarize the results of the simplest version, that of Bardeen, Hill
and Lindner \cite{1ai}. In Ref. \cite{1ai}, the first scenario including only a
heavy top quark predicted a top mass of order 230 GeV. This is now excluded. Ref.
\cite{1ai} also presented results involving a heavy, degenerate fourth generation quark
doublet, assuming that the top quark and other fermions are much lighter.
This result combined with the known top quark mass indicates that
something else other than the top quark alone must exist if this picture has any chance
of being correct. 

The subject of dynamical symmetry breaking \`{a} la top-condensate deserves
a whole review; a book on the topic  already exists
\cite{miransky}. The best we could do here is to review salient points and results,
especially those  pertaining to the subject of this review. We first state the
so-called compositeness conditions used in the top-condensate type of models. As
we shall see below, these conditions allow us to relate the masses of
the heavy fermions to that of the Higgs boson. Simply speaking, it is
the requirement that the Higgs quartic coupling $\lambda$ and the relevant
Yukawa couplings diverge at the {\em same} scale $\Lambda_C$ (the Landau
poles) while the ratio of $\lambda$ to the square of the Yukawa coupling(s)
remains finite, namely
\begin{eqnarray}
\label{boundary}
&& \lambda(\mu),~g_t(\mu)  \stackrel{\mu\to \Lambda_C}{\longrightarrow}   \infty  \\
&& \lambda(\mu)/g^2_t(\mu)  \stackrel{\mu\to \Lambda_C}{\longrightarrow}  \mbox{const.}
\end{eqnarray}
where $g_t$ can be either the top or just a generic Yukawa coupling. The
above boundary conditions modify the structure of the SM Lagrangian at
the scale $\Lambda_C$ in the following way. Let us consider a simple toy
model where there is a degenerate heavy quark doublet $Q= (U, D)$ 
coupled to the Higgs field $\Phi$ (light fermions will be ignored in this 
discussion). Let us rescale $\Phi$ as follows
\begin{equation}
\Phi \longrightarrow \Phi_0/g_f ,
\end{equation}
where $g_f$ is the Yukawa coupling of the degenerate quark doublet. The SM 
Lagrangian (with only that degenerate doublet present) becomes
\begin{equation}
{\cal L}= {\cal L}_{kinetic}(U,D) +
 Z_{\Phi} D_{\mu}\Phi^{\dagger}_{0} D^{\mu}\Phi_{0} 
+{\widetilde{m}}^2
\Phi^{\dagger}_{0} \Phi_{0} -\frac{\widetilde{\lambda}}{4}
(\Phi^{\dagger}_{0} \Phi_{0})^2
+ {\bar Q}_{L}\Phi_{0} D_R +{\bar Q}_{L}{\Phi^C_{0}} U_R + \mbox{h.c.}, \label{SM}
\end{equation}
where
\begin{equation}
\Phi^C_0 = i\sigma_2 \Phi_0^{\ast} , \qquad
Z_{\Phi} = 1/ g^2_f,  \qquad
\widetilde{m}^2 = Z_{\Phi} m^2 , \qquad\mbox{and}\qquad  
\widetilde{\lambda} = Z_{\Phi}^2\lambda.
\end{equation}

The first remark one can make when one looks at the above expressions is the
compositeness condition itself: the vanishing of the wave function renormalization 
constant. Indeed, as can be seen from the expression for $Z_{\Phi}$, one
immediately notices that, if $g_f$ has a Landau singularity at some scale
$\Lambda_C$, the wave function renormalization constant for the Higgs field,
$Z_{|Phi}$, vanishes at that same scale. In other words, the Landau pole
is identified with the compositeness scale. Furthermore, if $\lambda$ and
$g_f$ develop a singularity at the {\em same} scale in such a way that
conditions \ (\ref{boundary}) are satisfied, the Lagrangian at the compositeness scale becomes
\begin{equation}
 {\cal L}= {\cal L}_{kinetic}(U,D) + 
{\bar Q}_{L}\Phi_{0} D_R +{\bar Q}_{L} \Phi^C_{0} U_R 
+{\widetilde{m}}^2 \Phi^{\dagger}_{0} \Phi_{0} +\mbox{h.c.} 
\end{equation}
$\Phi_0$ is now just an auxiliary field and can be integrated out, resulting in
a Nambu-Jona-Lasinio form for the Lagrangian, namely
\begin{equation}
{\cal L} = {\cal L}_{kinetic}(U,D) + G_0 {\bar Q}_L( U_R {\bar U}_R + 
D_R {\bar D}_R) Q_L , \label{composite}
\end{equation}
where $G_0 = -1/\widetilde{m}^2.$ In this picture, the Higgs boson becomes
a fermion-antifermion composite particle below the scale $\Lambda_C$.

What might be even more interesting is the relationship between the Higgs
boson mass and that of the heavy fermions in the top-condensate type of
scenario. In particular, the Higgs mass, $m_H$, can be seen to be bounded
from above by $2 m_f$ and from below by $m_f$, where $m_f$ is a heavy
fermion mass. The search for a heavy fermion is intimately tied to
the search for the Higgs boson-a feature already seen in the discussion
of gauge coupling unification \cite{pqh}. To see this in the context
of the top-condensate type of model, let us look at the RG equations
for $\lambda$ and $g_f$, the heavy fermion Yukawa coupling, at one loop
level (the one-loop terms of Eq. \ (\ref{RG4})) in the toy model
with one doublet of heavy quarks. Let us define
$x = \lambda/g_f^2$. The one-loop RG for $x$ is then
\begin{equation}
16 \pi^2 \frac{dx}{dt} =24 g^2_f (x - x_{+})(x - x_{-}) ,
\label{rat}
\end{equation}
where $x_{\pm} = (-1 \pm 3)/4$, if both members of the 
quark doublet are degenerate in mass, or
$x_{\pm} = \frac{1}{16} (-1 \pm \sqrt{65})$,
if one member is much heavier than the other one 
(e.g. the 3rd generation case). The boundary conditions \ (\ref{boundary})
are satisfied if $x$ is one of the two fixed points, $x_{\pm}$.
The solution $x_{-}$ (always negative) is ruled out by vacuum stability.
This leaves us with $x_{+}$. From the definition of $m_f$ and $m_H$,
one can then obtain a relationship between these two masses as follows:
\begin{equation}
m_H^2 = 4 m_f^2\, x_{+}.
\end{equation}
In the large $N_c$ limit, the right hand side of Eq. \ (\ref{rat}) becomes
$4 N_c (x-1)$ with the fixed point being $x=1$ which implies
$m_H = 2 m_f$, a familiar result found in the Nambu-Jona-Lasinio model. Since 
$x_{+}$ is always greater than 1/4, one can easily see that, for finite $N_c$,
one obtains the bound
\begin{equation}
m_f < m_H < 2 m_f.
\end{equation}

As we have mentioned above, a necessary condition for this scenario to
work is the boundary \ (\ref{boundary}). The minimal SM with three
generations unfortunately does not satisfy these boundary conditions:
the top quark with a mass of 175 GeV is simply too ``light''. One 
needs eiher a heavier quark (216-230 GeV), which cannot be the
case with the minimal SM, or more heavy quarks or leptons such as in the
four generation scenario. The four generation case was studied by Hill,
Luty and Paschos \cite{hill3} in the context of Majorana neutrinos.
Later, Hung and Isidori \cite{HuIs} also examined the four generation 
scenario as part of an overall ``anatomy'' of the Higgs mass spectrum,
starting form $m_H$ of 65 GeV to $m_H \geq 2 m_Z$. It was in this last
mass range, $m_H \geq 2 m_Z$, that the focus on top-condensate types
of models was diverted to.
In their analysis, Ref. \cite{HuIs} assume a Dirac mass
for the fourth neutrino and degenerate quarks and degenerate leptons
for the fourth generation. It turns out that the inclusion of such
a fourth generation drastically modifies the evolution of the couplings
even if these fermions were lighter than the top quark. In the
analysis of Ref. \cite{HuIs}, Eqs. \ (\ref{RG4}) were used at the
one loop level. The results are shown in Fig.15 below, where the
mass of the top quark is fixed at 175 GeV and that of the fourth
lepton doublet is fixed at 90 GeV.

\begin{figure} 

\centerline{ \epsfysize 4in \epsfbox{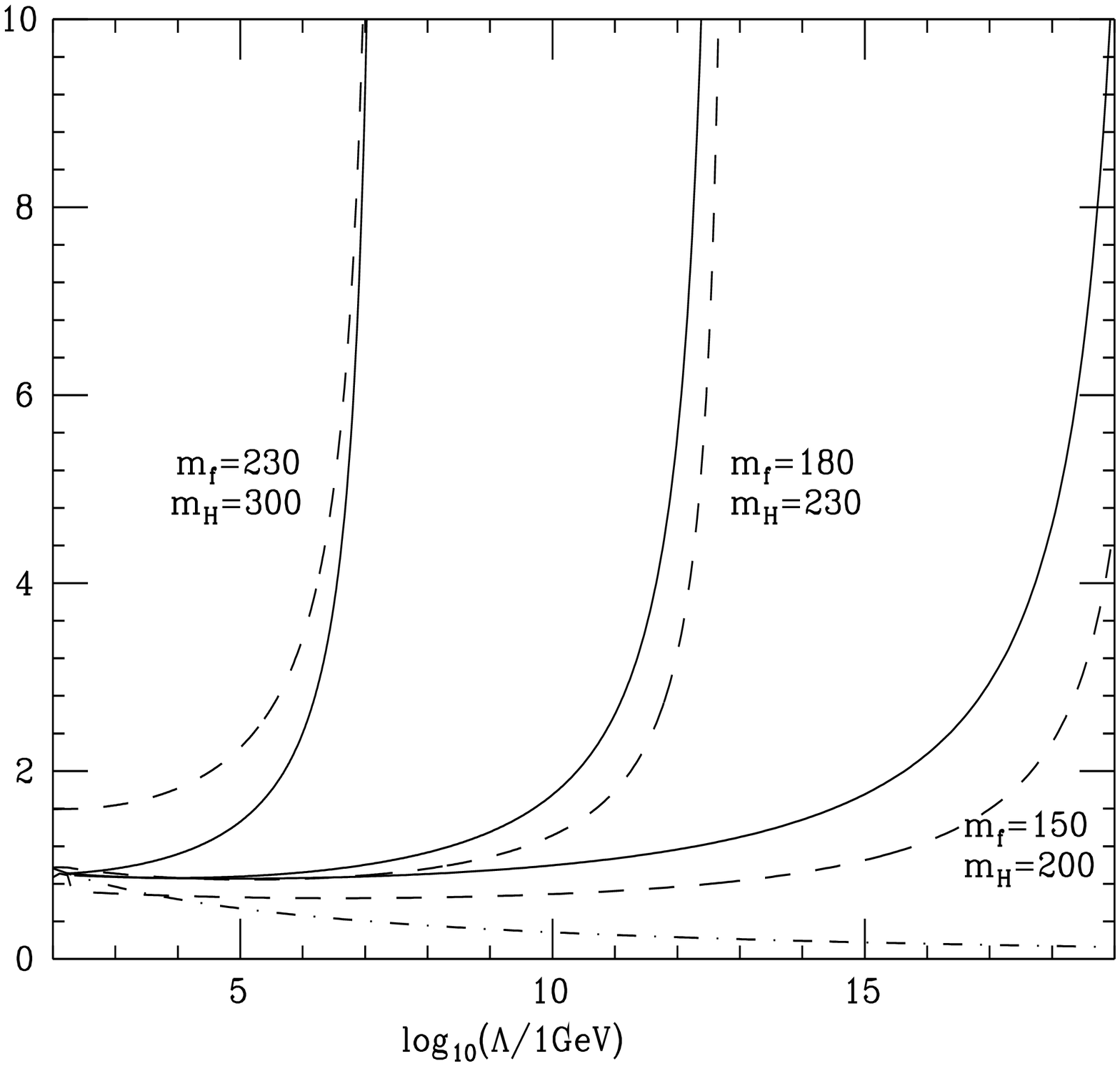}  }
\caption{RG evolution of the Yukawa couplings $g^2_t$ (full and dash-dotted 
lines) and $g^2_q$
(dashed lines).  The top mass is always fixed to be 175 GeV and the heavy-lepton mass is assumed
to be 90 GeV.  The dash-dotted line is the evolution of $g^2_t$ without extra fermions.  Near
each dashed line is indicated the value of $m_q$ and the corresponding value of $m_H$ obtained
by the requirement $\Lambda_L=\Lambda_q$ (the error on both $m_H$ and 
$m_q$ is about $\pm 10$
GeV.}

\end{figure}

A few remarks are in order concerning the above figure. As the figure caption
already indicates, the dash-dotted line represents the top Yukawa
coupling squared, $g_t^2$, as a function of energy for the minimal SM.
One can see that $g_t^2$ actually decreases in value with energy and
remains finite at the Planck scale. This is so because 
a mass of 175 GeV is ``too small'' to provide a large enough
initial value plus the fact that the contribution from the
QCD coupling to the $\beta$ function for $g_t^2$ occurs with the opposite
sign. When a fourth generation is added, the evolution of the couplings
change drastically, partly due to the fact that there are more
degrees of freedom than the minimal case. So, as long as the initial
Yukawa couplings of the fourth generation are not too small, there
will be Landau poles that appear below the Planck scale. As one can
see from Fig.15, this occurs when the fourth generation quark mass 
$m_Q \gtrsim 150$ GeV. Also an interesting phenomenon occurs: all
Yukawa couplings ``drag'' each other in such a way that they all 
``blow up'' at the same point. The top quark Yukawa coupling,
which by itself {\em decreases} with energy, is now ``dragged''
in such a way as to develop a Landau pole at the same time as
the fourth generation quarks. These results- the existence of
Landau poles below the Planck scale in the presence of a fourth
generation- encourage a reconsideration of the top-condensate
type of model. 

Since the quartic coupling $\lambda$ is a free
parameter, one can now choose its initial value, i.e. its mass,
in such a way that the boundary conditions \ (\ref{boundary}) are
satisfied. This procedure results in a relationship between the
fourth generation quark mass and the Higgs mass. This is shown
in Fig. 15 ($m_f$ is the notation used in the figure to refer
to the fourth generation quark mass instead of $m_Q$ which is used
here). The relationship for low 4th generation quark mass (e.g. 150 GeV)
is strikingly similar to the one obtained in the discussion of
the gauge coupling unification of Ref. \cite{pqh}. The range of
mass used in Ref. \cite{HuIs} for $m_Q$ is however considerably larger
than that of Ref. \cite{pqh} because, in that study of a
top-condensate type of model, no constraint on where the Landau poles
should be has been used. In fact, for masses larger than 150 GeV,
the Landau poles move down in energy. For example, if $m_Q = 230$
GeV which corresponds to $m_H = 300$ GeV, the Landau pole is 
situated at approximately 100 TeV. As remarked by Ref. \cite{HuIs},
the relationship between the Higgs mass and the fourth generation
quark mass approaches more and more the fixed point value for
a degenerate quark doublet, $m_H = \sqrt{2} m_Q$, as the Landau
pole ``approaches'' the electroweak scale. This can be seen in
the figure shown above.

The scenario described above intimately links the search for the
fourth generation quarks and leptons to that of the Higgs boson,
and vice versa. It is not clear at this point how the higher mass
values ($> 160$ GeV) for the fourth generation quarks would affect
the evolution of the gauge couplings, except for the fact that 
perturbation theory breaks down above the Landau poles and
it is not legitimate to evolve those couplings beyond that point.
For the purpose of this report, we shall however leave open
the possibility of such top-quark condensate type of models as
a possible mechanism for electroweak breaking. It is partly for
this reason that the range of masses considered in the search for
long lived quarks by Frampton and Hung \cite{framptonhung}
is larger than the mass range used in the study of gauge coupling
unification of Hung \cite{pqh}.

Recently \cite{pirogov}, there was an analysis of the two-loop
RG equations in the SM with three and four generations. Although
many results presented there \cite{pirogov} were already discussed
in \cite{pqh} and \cite{HuIs}, there were statements that are not
correct. In order to clarify the issues, we repeat the statements
of Ref. \cite{pirogov} and explain why they are misleading.
Ref. \cite{pirogov} chose the masses of the leptons and the quarks
of the fourth generation as follows: $m_L/m_Q = 1/2$ and 1, and
restricted $m_Q$ to be greater than 180 GeV which was referred to 
by the authors as the direct experimental constraint. Furthermore,
an upper bound of 200 GeV for $m_H$ was used. Using these constraints,
Ref. \cite{pirogov} claimed that a fourth generation is ruled out
by plotting the allowed regions in the $m_H-m_Q$ plane (Fig. 11
of \cite{pirogov}). First, $m_Q$ less than 180 GeV is not ruled
out by direct experiment if a long-lived quark decays in the
detector at a distance between 100 $\mu$m and 1 m, a subject
discussed by Frampton and Hung \cite{framptonhung}. As
we shall discuss in the section on experiments, there is and
will be such a search at the Tevatron. Second, the constraints
on $m_H$ at the present time from precision experiments are
rather loose at best. A much larger bound than 200 GeV is
possible \cite{PDG}. For example Refs. \cite{ddd,ddd1,ddd2}
gave an upper bound as high as $\sim$ 280 GeV, while Ref.
\cite{chanowitz} gave an upper bound (within 95 \% CL) from
340 GeV to 1 TeV. In summary, the mass ranges used by
\cite{pirogov} to rule out a fourth generation are not 
warranted. 
In fact, the values used by Hung \cite{pqh},
$m_Q = 151$ GeV and $m_H = 188$ GeV, were shown to lead to
a better unification of the SM gauge couplings, and
higher masses (such as 180 GeV) for $m_Q$ were not used because 
precisely of the fact that the Landau poles were much too low
to trust perturbation theory in the evolution of the gauge
couplings.

As the above discussion and earlier ones made it crystal 
clear, there are several theoretical reasons for looking at 
quarks and leptons beyond the third generation. In particular,
our primary motivation in this review is to examine
those reasons which ``predict'' fermion masses which are within reach either
of present experiments such as the ones performed at the Tevatron
or at future machines such as the LHC, NLC, etc. By ``within
reach'', we mean masses which are close to the top quark mass
( $\sim$ 150 -230 GeV) for the quarks. It goes without saying
that there exists also plenty of reasons for considering fermions
which are much heavier than the top quark, some of which are
even primary focus of the topics discussed in this section: the
techniquarks and leptons of Technicolor models
\cite{techcol,techcol1,techcol2,techcol3}. It is  outside of our goal to review such
topics but for completeness, we shall give a brief description
of one of such scenarios: the topcolor assisted Technicolor model.

Our discussion of top-condensate type of models above relies
on one crucial assumption: The entire electroweak symmetry
breaking is due to the condensate of the top quark and/or
that of the fourth generation quark. The topcolor assisted Technicolor
model of Hill \cite{hill4} did away with that assumption. 
But then how does one explain the fact that the top quark mass
is so much larger than all other fermion masses and so close
to the electroweak breaking scale?
The salient points of the topcolor model are basically the assumptions 
that the electroweak symmetry is still broken by some form of Technicolor,
there exists an extra (``topcolor'') group, $SU(3)_{1} \otimes
U(1)_{Y_1}$, which couples preferentially to the third family
and which triggers a dynamical condensate for the top quark,
giving rise to  a large top mass. This new ``topcolor'' group
is assumed to be spontaneously broken by Technicolor at a
scale $\sim$ 1 TeV. An extended Technicolor interaction (ETC) is
also assumed so that quarks and leptons can obtain some mass
which is much smaller than the top mass. The top quark itself 
obtains most of its mass from the condensate with a small
contribution coming from ETC. Variations of the model include
cases in which there are $SU(2)$ singlet quarks which are
however very massive ($\sim$ 1 TeV). In any case, all fermions
beyond the 3rd generation is these models have masses around 1 TeV.
Although there is a rich phenomenology involving objects such as
top-pions, etc., it is beyond the scope of this review to
discuss it here. 

We end this section by mentioning that there are interesting
recent developments on the subject of electroweak symmetry
breaking involving the so-called seesaw mechanism of quark
condensation \cite{hill5,hill5a}. This is yet another variant of the topcolor 
model. In this particular scheme, the top quark mass obtains
its ``observed'' value dynamically by a mass mixing with a
{\em new SU(2) singlet} quark $\chi$ of the form
$\mu_{\chi t} \bar{\chi}_L t_R + m_{t\chi} \bar{t}_L \chi_R + 
\mu_{\chi \chi} \bar{\chi}_L \chi_R$ with $m_{t \chi} \sim 0.6$ TeV,
$\mu_{\chi t} \sim 0.9$ TeV and $\mu_{\chi \chi} \sim 2$ TeV.
The physical top mass is found by diagonalizing that $2\times 2$ matrix
giving $m_t \approx \mu_{\chi t} m_{t \chi}/ \mu_{\chi \chi}$.
The other mass eigenstate would be $\sim 2$ TeV and it might not
be easy to be observed directly. The model predicts a number
of pseudo Nambu-Goldstone bosons, some of which are $\chi$-
bound states. How this new degree of freedom manifest itself
experimentally is a subject which is under active investigation.

One last comment which is worth mentioning here is the mass of
the Higgs boson in the different variants of the topcolor model.
Generically, the physical Higgs scalar would have a mass
of the order of a TeV just like standard Technicolor models. However,
in the seesaw version of the topcolor model, there appears an
extra ``light'' Higgs scalar (in addition to the charged ones)
as one would have in a two-Higgs doublet model. Depending on a
delicate cancellation of some parameters of the model, one
could have a ``truly light'' scalar with mass of O(100) GeV.
These remarks are meant to emphasize the importance of the
search of Higgs scalar(s) in addition to that of new quarks and
leptons.

\section{CP Violation.}

\subsection{CP Violation in the Standard Model.}

At first sight, there may appear to be no connection
between additional quarks and leptons and the violation
of CP symmetry. The object of this subsection is therefore to
convince the reader that the better understanding of
CP violation may necessitate the incorporation of
additional fermions. This is not an inevitable conclusion
but it is a suggestive one.

To set the scene, we need to describe the status 
of CP symmetry and its violation in the context of the unadorned 
standard model. This discussion will be in two separate parts:
weak CP violation (the KM mechanism), and the strong CP
problem. The former is not really a problem for
the standard model, merely one that is not yet verified unambiguously
by experiment. The latter, the strong CP problem, is definitely
a difficulty, a shortcoming, for the standard model and
one whose solution (we shall mention the axion 
possibility) is still unknown.

The gauge group of the standard model is 
$SU(3)_C \times SU(2)_L \times U(1)_Y$, broken at the weak scale
to $SU(3)_C \times U(1)_Y$. Under 
the standard group the first generation transforms as:
\begin{equation}
Q_L = \left( \begin{array}{c} u \\ d \end{array} 
\right)_L , \bar{u}_L , \bar{d}_L;\hspace{0.2in}
L_L = 
\left( \begin{array}{c} \nu_e \\ e^- \end{array} \right)_L, 
e^+_L       \label{gen}
\end{equation}
and the second ($c, s, \nu_{\mu}, \mu $) and third
($t, b, \nu_{\tau}, \tau $) generations are assigned
similarly. 

The quarks acquire mass from the vacuum expectation value (VEV)
of a complex $SU(2)_L$ doublet of scalars $\phi = \left( \begin{array}{c}
\phi^+ \\ \phi^0 \end{array} \right)$ giving rise
to up and down quark mass matrices:
\begin{equation}
M(U) = \lambda_{ij}^U<\phi^0> ; M(D) = \lambda_{ij}^D<\phi^0>  \label{M}
\end{equation}
which are arbitrary matrices that may, without loss of generality,
be chosen to be hermitian. The matrices $M(U), M(D)$
of Eq. (\ref{M}) are defined so that the Yukawa terms give {\it e.g.}
$Q_LM(U)u_R + h.c. $ and
can be diagonalized by a bi-unitary transformation:
\begin{equation}
K(U)_L M(U) K(U)_R^{-1} = diag (m_u, m_c, m_t )      \label{MU}
\end{equation}
\begin{equation}
K(D)_L M(D) K(D)_R^{-1} = diag (m_d, m_s, m_b )      \label{MD}
\end{equation}
These mass eigenstates do not coincide with the gauge eigenstates
of Eq.(\ref{gen}) and hence the charged $W$ couple to the
left-handed mass eigenstates through the $3 \times 3$ CKM matrix
$V_{CKM}$ defined by:
\begin{equation}
V_{CKM} = K(U)_L K(D)_L^{-1}     \label{ckm}
\end{equation}
This is a $3 \times 3$ unitary matrix which would in general have
9 real parameters. However, the five relative phases of the 6 quark
flavors can be removed to leave just 4 parameters comprising
3 mixing angles and a phase. This KM phase underlies the KM mechanism
of CP violation.

With N generations and hence a $N \times N$ mixing matrix
there are $N(N - 1)/2$ mixing angles and $(N - 1)^2$ parameters
in the
generalized CKM matrix. The number of CP violating phases is therefore
$(N-1)^2 - N(N-1)/2 = (N - 1)(N - 2)/2$. This is zero for $N = 2$,
one for $N = 3$, three for $N = 4$, and so on. In particular, as
Kobayashi and Maskawa \cite{1x} pointed out, with three generations
there is automatically this source of CP violation arising
from the $3 \times 3$ mixing matrix. This is the most conservative
approach to CP violation. This source of CP violation is necessarily
present in the standard model; the only question is whether it is the {\it only}
source of CP violation. Since the only observation of CP violation
remains in the neutral kaon system, there is not yet sufficient
experimental data to answer this question definitively.
 
There are various equivalent ways of parametrizing the
CKM matrix. That proposed\cite{1x} by KM involved writing:
\begin{equation}
V_{CKM} = \left( \begin{array}{ccc} cos\theta_1 & -sin\theta_1cos\theta_3 
& -sin\theta_1sin\theta_3 \\ sin\theta_1cos\theta_2  
& cos\theta_1cos\theta_2cos\theta_3 
- sin\theta_2sin\theta_3 e^{i\delta}  & cos\theta_1cos\theta_2sin\theta_3 
+ sin\theta_2cos\theta_3 e^{i\delta}  \\    
sin\theta_1sin\theta_2 & cos\theta_1sin\theta_2cos\theta_3 
+ cos\theta_2sin\theta_3 e^{i\delta}   & cos\theta_1sin\theta_2sin\theta_3 - cos\theta_2cos\theta_3 e^{i\delta}  \\ \end{array} \right)   \label{CKM}
\end{equation}
Another useful parametrization\cite{6b} writes:
\begin{equation}
V_{CKM} = \left( \begin{array}{ccc} 1 - \frac{1}{2}\lambda^2 & 
\lambda & \lambda^3 A (\rho - i\eta) \\
-\lambda & 1 - \frac{1}{2}\lambda^2 & \lambda^2 A \\
\lambda^3 A (1 - \rho - \eta ) & -\lambda^2 A & 1 \\
\end{array} \right)      \label{wolf}
\end{equation}
In Eq.(\ref{wolf}), $\lambda$ is the sine of the 
Cabibbo angle $sin\theta_1$ in Eq.(\ref{CKM}) and
CP violation is proportional to
$\eta$. If we write the CKM matrix  a third time as:
\begin{equation}
V_{CKM} = \left( \begin{array}{ccc} V_{ud} & V_{us} & V_{ub} \\
V_{cd} & V_{cs} & V_{cb} \\
V_{td} & V_{ts} & V_{tb} \\
\end{array} \right) \label{ckmelements}
\end{equation}
then the unitarity equation $(V_{CKM})^{\dagger}V_{CKM} = 1$ dictates, for
example, that
\begin{equation}
V_{ub}^*V_{ud} + V_{cb}^*V_{cd} + V_{tb}^*V_{td} = 0
\label{triangle}
\end{equation}
This relation is conveniently represented as the addition
of three Argand vectors to zero in a {\it unitarity triangle.}
Dividing out the 
middle term of Eq.(\ref{triangle}) and using the parametrization
of Eq.(\ref{wolf}) leads to the prediction of the standard model 
with KM mechanism that the vertices of the unitarity triangle
in a  $\rho-\eta$ plot are at the origin $(0, 0)$, at $(1, 0)$
and $(\rho, \eta)$. Thus, the area of the unitarity triangle
is proportional to $\eta$ and hence to the amount of CP violation. 

The measurement of the angles and sides of this unitarity
triangle are the principal goals of the B Factories (see
{\it e.g.} \cite{6c} for a review)\footnote{A recent prelimary report from
CDF\cite{CDFoneta} gives a value of $0.8\pm 0.4$ for $\sin 2\beta$.}.  As we shall see, alternative
models will give quite different predictions and hence be easily distinguishable
from the KM mechanism when accurate measurements on B-meson decays are made in
B Factories. \\

\subsection{Strong CP and the Standard Model.}

Next we turn to a brief outline of the strong CP problem
in the standard  model. (More detailed reviews are available in 
\cite{6d,6e,6f}). The starting observation is that
one may add to the QCD lagrangian an extra term:
\begin{equation}
L = \sum_k \bar{q}_k (i\gamma_{\mu} D_{\mu} - m) q_k - 
\Theta G_{\mu\nu}\tilde{G}_{\mu\nu}
\end{equation} 
where the sum over $k$ is for the quark flavors and $D_{\mu}$
is the partial derivative for gauged color $SU(3)_C$.
The additional term proportional to $\Theta$ violates $P$
and $CP$ symmetries. This term is a total divergence of a
gauge non-invariant current but can contribute
because of the existence of classical instanton solutions.
It turns out that chiral transformations can change the
value of $\Theta$ via the color anomaly but cannot change
the combination:
\begin{equation}
\bar{\Theta} = \Theta - arg det M(U) - arg det M(D)   \label{barr}
\end{equation} where det$M(U, D)$ are the determinants of the 
up, down quark mass matrices respectively.
Thus $\bar{\Theta}$
which is an invariant under chiral transformations measures
the violation of CP symmetry by strong interactions.
A severe constraint on $\bar{\Theta}$ arises from the neutron
electric dipole moment $d_n$ which has been measured to
obey $d_n \leq 10^{-25} e.-cm.$\cite{6ff,6fff}.
A calculation of $d_n$\cite{6ffff,6fffff} leads to an estimate that
$\bar{\Theta} \leq 10^{-10}$. This fine-tuning of
$\bar{\Theta}$ is unexplained by the unadorned standard model
and raises a serious difficulty thereto.

A popular approach (which does {\it not} necessitate
additional fermions) involves the axion mechanism which
we briefly describe, although since only a relatively
narrow window remains for the axion mass, and since the
mechanism is non-unique, it is well worth looking for
alternatives to the axion for solving the strong CP problem.

In the axion approach, one introduces a color-anomalous
global $U(1)$ Peccei-Quinn symmetry\cite{6g,6h} such
that different Higgs doublets couple to the up- and down-
type quarks. The effective potenetial now becomes a function
of the two Higgs fields and $\bar{\Theta}(x)$ regarded as
a dynamical variable. An analysis then shows that
the potential acquires the form:
\begin{equation}
V = V(H_1, H_2) - cos \bar{\Theta}
\end{equation}
and hence the minimum energy condition relaxes
$\bar{\Theta}$ to zero.

Because a continuous global symmetry is spontaneously
broken, there is a pseudo-Goldstone boson\cite{6i,6j},
the axion, which acquires a mass through the color
anomaly and instanton effects. The simplest model
predicts an axion with mass of a few times $100keV$,
but this particle was ruled out phenomenologically.
Extensions of the axion model\cite{6k,6l,6m,6n} lead
to an axion mass which becomes a free parameter.
Empirics constrain the mass to lie between about
a micro-electronVolt and a milli-electronVolt, and searches are underway
for such an axion.

In one variant of the axion approach, new heavy quarks
are necessary\cite{6k}; in fact, this was the 
first-proposed model to avoid the experimentally-excluded
``visible'' axion in favor of an ``invisible'' axion.
Alternative versions of the invisible axion \cite{6m,6n}
do not involve extra quarks.

A second solution of the strong CP problem is to
assume that $m_u = 0$, but this seems to be at
variance with the successes of chiral perturbation
theory\cite{6nn,6nnn,6nnnn,6nnnnn}.

\subsection{Strong CP and Extra Quarks.}

For present purposes, let us believe {\it neither}
the axion {\it nor} the massless up quark.
In this case we are inevitably led to
the existence of further fermions beyond the standard
model. Such a conclusion can have the additional
bonus of connecting the strong CP solution to
the occurrence of weak CP violation. 

The appearance of new fermions in this context
was first suggested by Nelson\cite{6o} and Barr\cite{6p,6q}
who looked within the framework of GUTs. Their
additional fermions have gigantic masses $\sim 10^{12}$GeV,
well beyond accessible energy, but their basic idea involving
the quark mass matrix texture is one that will reappear
in the models where the new fermions are at accessible
masses.

Nelson\cite{6o} invented a GUT based on $SU(5)_{gauge}
\times (SO(3)\times U(1))_{global}$. The fermions
are assigned to the representations:
\begin{equation}
[10. 3. 0] + [\bar{5}, 3, 0] + [10, 1, 1] + [\bar{10}, 1, -1]
+ [5, 1, 1] + [\bar{5}, 1, -1]
\label{nelson}
\end{equation}      
Note, in particular, the additional fermion representations
in the last four terms of Eq.(\ref{nelson}).

The scalars are the $(5, 1, 0)$ which is complex and
contains the standard Higgs doublet together with a
superheavy color triplet; then there are $(r_i, 3, 0)$
where $r_i$ are $SU(5)$ representations containing singlets
of $SU(3) \times SU(2) \times U(1)$ such as {\bf 1, 24, 75}.

If we write the most general Yukawa terms we find that the
couplings of the quarks to the light Higgs is complex but has a real
determinant. Thus, if the lagrangian respects CP symmetry,
the value of $\bar{\Theta}$ vanishes at tree level.

Looking at the loop corrections to $\bar{\Theta}$, it turns
out that the additional fermions must be lighter than
the GUT scale by three or four orders of magnitude to
suppress $\bar{\Theta}$ adequately. 

The secret of the Nelson model lies in the arrangement of additional
fermions such that a basis may be chosen where
complex entries in the quark mass matrix are always multiplied
by zero in the evaluation of the determinant. Barr\cite{6p,6q} examined what
are the general circumstances under which this suppression
of $\bar{\Theta}$ happens.

Barr was led to the following rules for a GUT in which
$\bar{\Theta} = \Theta_{QCD} + \Theta_{QFD} = 0$ at tree
level. Let the gauge group be $G$ and let CP be a symmetry
of the lagrangian. Then $\Theta_{QCD} = 0$ and the couplings
have no CP violating phases. Let the fermion representation
be divided into two sets F and R where F contains the fermions
of the three families and R is a real non-chiral set. Let
R be composed of a set $C$ and its conjugate set $\bar{C}$.
Then the following two conditions are sufficient to ensure
that $\bar{\Theta} = 0$ at tree level:

\begin{itemize}

\item At tree level there are no Yukawa mass terms coupling
F fermions to $\bar{C}$ fermions, or $C$ fermions to $\bar{C}$
fermions.

\item The CP violating phases appear at tree level only in those 
Yukawa terms which couple $F$ fermions to $R = C + \bar{C}$ fermions.

\end{itemize}

In such Nelson-Barr GUT models, strong CP is solved by
physics (the additional fermions) close to the GUT scale. The
unrelated (in this approach) physics of weak CP violation
arises from the usual KM mechanism.

Non-GUT models which adopt the Nelson-Barr mechanism have 
been discussed especially in the papers\cite{6r,6s,6t,6u,6v}.
The model proposed in \cite{6r} by Bento, Branco and Parada
introduces a non-chiral charge $-1/3$ quark together with
a complex singlet scalar $S$. The field $S$ develops a
complex VEV $<S> = Ve^{i\alpha}$ while the standard Higgs
doublet has VEV $<\phi> = v$ which is real. The KM phase
$\delta_{KM}$ is generated from $\alpha$ in an unsuppressed
manner. $\bar{\Theta}$ is zero at tree level and its loop 
corrections are suppressed by powers of $(<\phi>/<S>) = (v/V)$.    
Consequently $\bar{\Theta}$ can be naturally sufficiently
small; {\it e.g.} if $V > 100$TeV, the Yukawa can be $\sim 10^{-1}$
while even if $V > 1$TeV, the Yukawa coupling can still be as large as
$\sim 10^{-2}$. Further papers examine other consequences of
such a model. In \cite{6s} the impact for $B-\bar{B}$ mixing
is found, while in \cite{6t} $D-\bar{D}$ is also found to differ
from the standard model predictions.
One general feature of this type of model is that
$\bar{\Theta}$ is suppressed by the Nelson-Barr type
mechanism and, as in Nelson-Barr, the CP violation arises
from $\delta_{KM}$. In the Aspon Model described below, the
CP violation necessarily arises from an additional mechanism
because there the generation of $\delta_{KM}$ is highly suppressed.

\subsection{CP and a Fourth Generation.}

Before moving to that, let us mention a number of valuable papers which
discuss the parametrisation of the CKM matrix when the number of
generations of quarks and leptons is increased from three to
four or more\cite{6w,6x,6y,6z,hepakvasa,6zzz,6zzzz}. In particular,  
Harari and Leurer\cite{6zzz} claim to
have an optimal parametrisation for general generation number.
Let us here merely quote\cite{hepakvasa} an example of a parametrisation
for $V_{CKM}^{(4)}$ in terms of six mixing angles
and three CP-violating phases:
\begin{equation}
V_{CKM}^{(4)} = \left( \begin{array}{cccc}
c_1 & s_1c_3 & s_1s_3c_5 & s_1s_3s_5 \\
-s_1c_2  & c_1c_2c_3 + s_2s_3c_6e^{i\delta_1} & 
c_1c_2s_3s_5 - s_2c_3s_5c_6e^{i\delta_1} & 
c_1c_2s_3s_5 - s_2c_3s_5c_6e^{i\delta_1} \\ 
 & & 
+s_2s_5s_6e^{i(\delta_1 + \delta_3)} & 
-s_2s_5s_6e^{i(\delta_1 + \delta_3)} \\
-s_1s_2c_4 & c_1s_2c_3c_4 - c_2s_3c_4c_6e^{i\delta_1} &
c_1s_2s_3c_4c_5 + c_2c_3c_4c_5c_6e^{i\delta_1} & 
c_1s_2s_3c_4c_5 + c_2c_3c_4c_5c_6e^{i\delta_1} \\ 
 & -s_3s_4s_6e^{i\delta_2} &
-c_2c_4s_5s_6e^{i(\delta_1 + \delta_3)} &
+c_2c_4s_5s_6e^{i(\delta_1 + \delta_3)} \\
 & &
+c_3s_4c_5s_6e^{i\delta_2} &
+c_3s_4c_5s_6e^{i\delta_2} \\
 & &
+s_4s_5c_6e^{i(\delta_2 + \delta_3)} & 
-s_4s_5c_6e^{i(\delta_2 + \delta_3)} \\
-s_1s_2s_4 & c_1s_2c_3s_4 - c_2s_3s_4c_6e^{i\delta_1} &
c_1s_2s_3s_4c_5 + c_2c_3s_4c_6e^{i\delta_1} & 
c_1s_2s_3s_4c_5 + c_2c_3s_4c_6e^{i\delta_1} \\
 & +s_3c_4s_6e^{i\delta_2} &
-c_2s_4s_5s_6e^{i(\delta_1 + \delta_3)}  &  
+ c_2s_4s_5s_6e^{i(\delta_1 + \delta_3)} \\
 & & 
-c_3c_4c_5s_6e^{i\delta_2} & 
-c_3c_4c_5s_6e^{i\delta_2} \\   
 & &
-c_4s_5c_6e^{i(\delta_2 + \delta_3)} & 
+ c_4s_5c_6e^{i(\delta_2 + \delta_3)}

\end{array} \right)
\end{equation}

The large number of phases and mixing angles makes a thorough analysis of CP
violation in the model with four chiral generations impractical.  However, if one
considers the case of an additional isosinglet quark, then the discussion of CP
violation becomes tractable, and this case has all of the essential features of the
more general case.   An extremely extensive and detailed analysis of CP violation in
models with a
$Q=-1/3$ isosinglet quark has been given in a series of papers by
Silverman\cite{silv3,silv2,nir2,silv1}; the $Q=2/3$ case was discussed briefly by
Barger et al.\cite{bargerberger}.   Using the notation of Barger et al. (see the last
subsection of Section III), one can write the unitarity relations of the $4\times 4$
matrix as
\begin{equation}
V^*_{ui}V_{uj}+V^*_{ci}V_{cj}+V^*_{ti}V_{tj}+V^*_{0i}V_{0j}=\delta_{ij}
\end{equation}
or
\begin{equation}
V^*_{id}V_{jd}+V^*_{is}V_{js}+V^*_{ib}V_{jb}+V^*_{iD}V_{jD}=\delta_{ij}
\end{equation}
For $i\neq j$, these can be expressed as closure of a unitarity quadrangle.  The
first three terms are the three sides of the unitarity triangle (which would be
closed if the $3\times 3$ submatrix were unitary).  Note that in the first relation,
the fourth side of the quadrangle is $V_{0i}^*V_{0j}=-z_{ij}$, which is the
coefficient governing the size of the flavor-changing neutral current interactions.
There are two major consequences for CP-violation in the B sector.  The CP violating
angles $\alpha$ and $\beta$, which are directly measureable in $B_d$ decays, no
longer have the same values as they do in the standard model; and tree-level
Z-mediated graphs give a contribution to $B_d-overline{B}_d$ mixing.
The contribution of $z_{db}$ to the unitarity quadrangle has a fixed magnitude (since
the quadrangle must close), but has any phase.  Barger et al. give the CP asymmetry for
$B_d\rightarrow \psi K_S$ as a function of two parameters:  the phase of $z_{bd}$ and
$|z_{bd}/(V_{td}V^*_{tb}|$.  For some values of the parameters, the CP asymmetry is
the same as the standard model; for some values it is different, and can even have
the opposite sign.

The most extensive analysis of the model is the recent paper of
Silverman\cite{silv3}.  He first finds the presently allowed ranges for the
unitarity triangle angles, the mixing and asymmetry in $B_s-\overline{B}_s$ mixing. 
Then, the model with an isosinglet $Q=-1/3$ is presented.   He analyzes all of the
constraints in this mdoel, including experiments to determine the CKM submatrix
elements, $|\epsilon|$, $K_L\rightarrow\mu^+\mu^-$, $B_d-\overline{B}_d$ mixing,
$B_s-\overline{B}_s$ mixing, $B\rightarrow\mu^+\mu^-X$, $R_b$ in $Z\rightarrow
b\overline{b}$, and the recent event in $K^+\rightarrow\pi^+\nu\overline{\nu}$.
The results are presented as plots (one for the standard model and one for the
isosinglet model) in the $(\rho,\eta)$ plane, in the $(\sin(2\alpha),
\sin(2\beta))$ plane, and in the $(x_s,\sin\gamma)$ plane.   In each case, he shows
the current bounds (with $1\sigma,2\sigma,3\sigma$ contours), and then shows
projected results from the upcoming B-factories.   There are several interesting
features of the isosinglet model case:  the presently observed CP violation in
$\epsilon$ can come entirely from new phases, and the value of $\sin\gamma$ can take
on any value whatsoever.  The range of $A_{B_s}$ is just as large for its negative
values as for its positive values.  It is clear from the plots how the upcoming
B-factories will drastically narrow the allowed parameter space, can could easily
rule out standard model CP-violation. 

\subsection{CP in the Aspon Model.}

Here we shall concentrate specifically on
the Aspon Model\cite{6aa,6ab,6ac,6ad,6ae,6af,6ag,6ah,framptonhung}
and its requirement of additional quarks to solve strong CP.
The first generation of the standard model contains quarks with 
the following $(T_3, Y)$ values:
\begin{equation}
\left(-\frac{1}{2}, \frac{1}{6} \right) =  d_L;
\left(0, \frac{1}{3} \right) = \bar{d}_L;
\end{equation}
\begin{equation}
\left(\frac{1}{2}, \frac{1}{6} \right) = u_L;
\left(0, -\frac{2}{3} \right) = \bar{u}_L.
\end{equation}
We introduce\cite{6aa} a $U(1)_{new}$ symmetry
and assign $Q_{new} = 0$ to all of the above quark states
and to the leptons, although the latter do not play
a significant role in solving strong CP. The second and third families
have parallel assignments under the same $U(1)_{new}$.

In the model there is also a real representration of 
exotic heavy quarks corresponding to a complex representation $C$
and its conjugate $\bar{C}$. In $\bar{C}$ the exotic heavy quarks 
have quantum numbers exactly like some of the usual quarks; for 
example, in $\bar{C}$ there may be one doublet
\begin{equation}
\left( -\frac{1}{2}, \frac{1}{6} \right) = D_L; 
\left( \frac{1}{2}, \frac{1}{6} \right) = U_L
\end{equation}
These have charges $Q_{new} = +h$. In representation $C$ we shall then have
\begin{equation}
\left( \frac{1}{2}, -\frac{1}{6} \right) = D^C_L; 
\left( -\frac{1}{2}, -\frac{1}{6}) \right)  = U^C_L
\end{equation}
These have $Q_{new} = -h$.

The Higgs sector has one complex doublet
\begin{equation}
\phi \left( +\frac{1}{2}, -\frac{1}{2} \right), Q_{new} = 0,
\end{equation}
and two complex singlets
\begin{equation}
\chi_{1,2} (0, 0), Q_{new} = +h.
\end{equation}
The gauge group is $SU(3)_C \times SU(2)_L \times U(1)_Y \times U(1)_{new}$,
where we gauge $U(1)_{new}$ to avoid an unwanted Goldstone boson when it is
spontaneously broken.

In breaking the symmetry, we give a real vacuum expectation value to
$\phi$ and {\it complex} VEVs to $\chi_{1,2}$ with a nonvanishing 
relative phase. 

The lagrangian contains bare mass terms $M(U^C_LU_L + D^C_LD_L)$ for
the extra quarks. The allowed Yukawa coupling include 
$\bar{u}^i_L u^j_L \phi$, 
$\bar{d}^i_L d^j_L \phi$ 
for the families and
$U^C_L u^i_L \chi_{\alpha}$, 
$D^C_L d^j_L \chi_{\alpha}$
coupling light quarks to $C$ heavy quarks 
($\alpha = 1,2$ and $i = 1,2,3$).
Because the families have no couplings to $\bar{C}$ exotics, the quark mass
matrix determinant arising from spontaneous symmetry
breaking is real at lowest order; it has the required texture.

We do not allow terms in the Higgs potential
which explicitly break $U(1)_{new}$. 
Disallowed terms include $\bar{\phi} \phi \chi^2$, 
$\chi^2$, $\chi^3$ and $\chi^4$. If any of these terms
are present, $U(1)_{new}$ is explicitly broken and the model can have
$\Theta = 0$ at tree level only in very special cases where
{\it e.g.} we choose particular representations of a grand unified group
such that the quark matrix is real.

Without explicit breaking of the $U(1)_{new}$,
there is correct texture at tree level; the
mass matrix has the tree-level texture (F = family)
\begin{equation}
(F C \bar{C}) \left( \begin{array}{ccc} real & 0 & complex \\
complex & real & 0 \\
0 & 0 & real \end{array} \right)
\left( \begin{array}{c} F \\ C \\ \bar{C} \end{array} \right)
\end{equation}

Thus $\Theta_{QCD} = 0$ at tree level. If we assume CP
symmetry of the Lagrangian then $\Theta_{QFD} = 0$ also.
In this case $\bar{\Theta} = 0$ at tree level and will be nonzero by
a small amount through radiative corrections; this can be consistent
with experiment if the Yukawa couplings are within a certain window,
as will be shown below.    

Because it is anomaly-free, we may gauge the $U(1)_{new}$ symmetry
and the Higgs mechanism will lead to a massive gauge boson, {\it the aspon},
which couples only to the exotic quarks and indirectly
via nondiagonal mixing to quarks and leptons.

The Yukawa interactions of the model are given by
\begin{equation}
-L_Y = 
\bar{q}_L {\bf m}_d d_R \left[\frac{\sqrt{2}}{v}\Phi\right]
+ 
\bar{q}_L {\bf m}_u  u_R \left[\frac{\sqrt{2}}{v}\tilde{\Phi} \right]
+
\bar{l}_L {\bf m}_e e_R \left[\frac{\sqrt{2}}{v}\Phi\right]
+
{\bf h}^{\alpha} \bar{q}_L Q_R \chi_{\alpha}
+ 
h.c.
\end{equation}
where $v/\sqrt{2}$ is defined as the VEV of $\phi^0$
and $\tilde{\Phi}$ as $(\bar{\phi}^0,-\phi^-)^T$.
The generation indices are implicit. Usual quarks and leptons
acquire their masses through spontaneous symmetry breaking
(SSB) induced by the VEV of the doublet Higgs scalar.
The new quarks acquire their mass through a gauge invariant mass
term of the form $M\bar{Q}_LQ_R$. Hence, $U$ and $D$ quarks 
are degenerate in mass. ${\bf m}_d, {\bf m}_u, {\bf m}_e, v,
{\bf h}^{1,2},$ and $M$ are real by the assumption of CP invariance.
The VEVs of $\chi_1$ and $\chi_2$ are chosen to be
\begin{equation}
<\chi_1> = \frac{1}{\sqrt{2}}\kappa_1e^{i\theta};\hspace{0.2in}
<\chi_2> = \frac
{1}{\sqrt{2}}\kappa_2
\end{equation}
Hence CP is broken spontaneously. [CP can be
broken softly by $i(\chi^*_1\chi_2 - \chi^*_2\chi_1)$.]

The up- and down- quark mass matrices linking the 
right-handed sector to the left-handed sector are of the form
\begin{equation}
M_{up} = \left[ \begin{array}{cc} {\bf m}_u & {\bf F} \\
0 & M \end{array} \right]; \hspace{0.2in} 
M_{down} = \left[ \begin{array}{cc} {\bf m}_d & {\bf F} \\
0 & M \end{array} \right]  \label{MUD}
\end{equation}
where
\begin{equation}
{\bf F} = {\bf h}^1 <\chi_1> + {\bf h}^2 <\chi_2>
\end{equation}
The Kobayashi-Maskawa (KM) matrices will be generalized to $4 \times 4$.
From the constraint $|V_{ud}|^2 + |V_{us}|^2 + |V_{ub}|^2 =
0.9979 \pm 0.0021$, we find $|F_1|/M$ and $|F_2|/M$ to be less than
$10^{-2}$ and $10^{-1}$, respectively. 
Although {\bf F} is a complex column matrix, the determinants of
$M_{up}$ and $M_{down}$ are real.
All entries become complex and, therefore a nonvanishing 
value of $\bar{\Theta}$
arises through radiative corrections.
The calculation of $\bar{\Theta}$ at one loop level
will be done below.

First, we discuss how the flavor-changing neutral currents (FCNC)
in the presence of new quarks are suppressed at the tree level.
After introducing the new vector-like quark doublet, we
find that there are FCNC's induced by $Z$ coupling because of the
mismatch of the new and usual quarks in the right-handed sector.
Therefore, the flavor-changing $Z$ couplings are
induced by the terms
\begin{equation}
L^{FCNC}_Z = \left( -\frac{1}{2} \right) 
\frac{g_2}{cos\theta_W} \bar{D}_R 
\gamma_{\mu} D_R Z^{\mu} + ( U_R \hspace{0.2in} contributions ),
\label{FCNC}
\end{equation}
where the factor $-\frac{1}{2}$ is the isospin of $D_R$ and
$g_2$ is the $SU(2)$ gauge coupling constant.
Let us consider the down sector first.
Without losing any generaity, we assume the down-quark mass matrix is 
in the partially diagonalized form
\begin{equation}
M_{down} = \left[ \begin{array}{cccc} m_d & 0 & 0 & F_1 \\
0 & m_s & 0 & F_2 \\
0 & 0 & m_b & F_3 \\
0 & 0 & 0 & M \end{array}
\right]
\end{equation}
This mass matrix can be diagonalized by a biunitary transformation,
${\bf K}^{\dagger}_L M_{down} {\bf K}_R$. The
transformation matrices are given in Ref. \cite{6ab}.
 Thus Eq.(\ref{FCNC}) can be written in
terms of mass eigenstates $d^{'i}$ as
\begin{equation}
L_Z^{FCNC} (down) = \beta_{ij} \bar{d}_R^{'i} \gamma_{\mu} d_R^{'j} Z_{\mu}
\end{equation}
for $i \neq j$, where
\begin{eqnarray}
\beta_{ij} & = & \left(\frac{1}{2}\right) \frac{g_2}{cos\theta_W} (K_R)^*_{4i}
(K_R)_{4j} \cr
& = & \left( \frac{1}{2} \right) \frac{g_2}{cos\theta_W}\frac{m_{d_i}m_{d_j}}
{M^2} x_i x_j^*
\end{eqnarray}
where $x_i\equiv F_i/M$.
Therefore, the FCNC induced by $Z$ coupling is highly
suppressed by the small mass ratio of usual to new quarks.
It is because the mixings of right-handed quarks
require a helicity flip of the usual quarks.
For example, $\beta_{ij} \simeq 544.8\times 10^{-8}x_1x_2^* < 10^{-10}$
for $M = 100$GeV, while the experimental limits on FCNC's require only
that $\beta_{ij} < 10^{-6}$.

FCNC can also be induced by aspon (A) couplings which are
given by
\begin{equation}
L_A^{FCNC} (down) = \alpha_{ij} \bar{D}_L^{'i} \gamma_{\mu} d_L^{'j} A^{\mu}
\end{equation}
for $i \neq j$, where
\begin{equation}
\alpha_{ij} = - g_A x_i x_J^*
\end{equation}
Therefore, FCNC's induced by $A$ will be important if $A$ is
not too heavy compared to $Z$. Consider the $K^0 - \bar{K}^0$
mixing matrix element $M_{12}$.
e($M_{12}$) is expected to be
dominated by standard $2W$-exchange box diagrams,
while Im($M_{12}$) receives its largest contribution
from the $A$ exchange shown in Figure 16.

\begin{figure} 

\centerline{ \epsfysize 2in \epsfbox{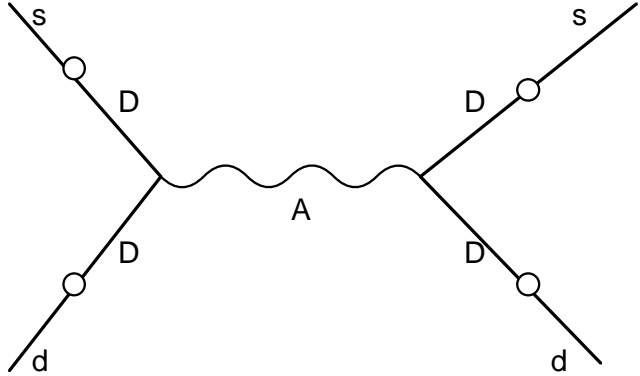}  } 

\caption{Contributions to Im($\Delta M_{12}$) by aspon exchange (Circles mean mixings.) }

\end{figure}

\noindent
We obtain
\begin{equation}
Im(M_{12}) = \frac{f_K^2m_K}{6}\frac{1}{\kappa^2}Im(x_1x_2^*)^2
\label{IM}
\end{equation}
where $\kappa^2 = \kappa_1^2 + \kappa_2^2$. The color factor has been taken into
account in Eq.(\ref{IM}). Im($M_{12}$) receives contributions from the new
$2W$-exchange box diagrams shown in Figure 17,

\begin{figure} 

\centerline{ \epsfysize 4in \epsfbox{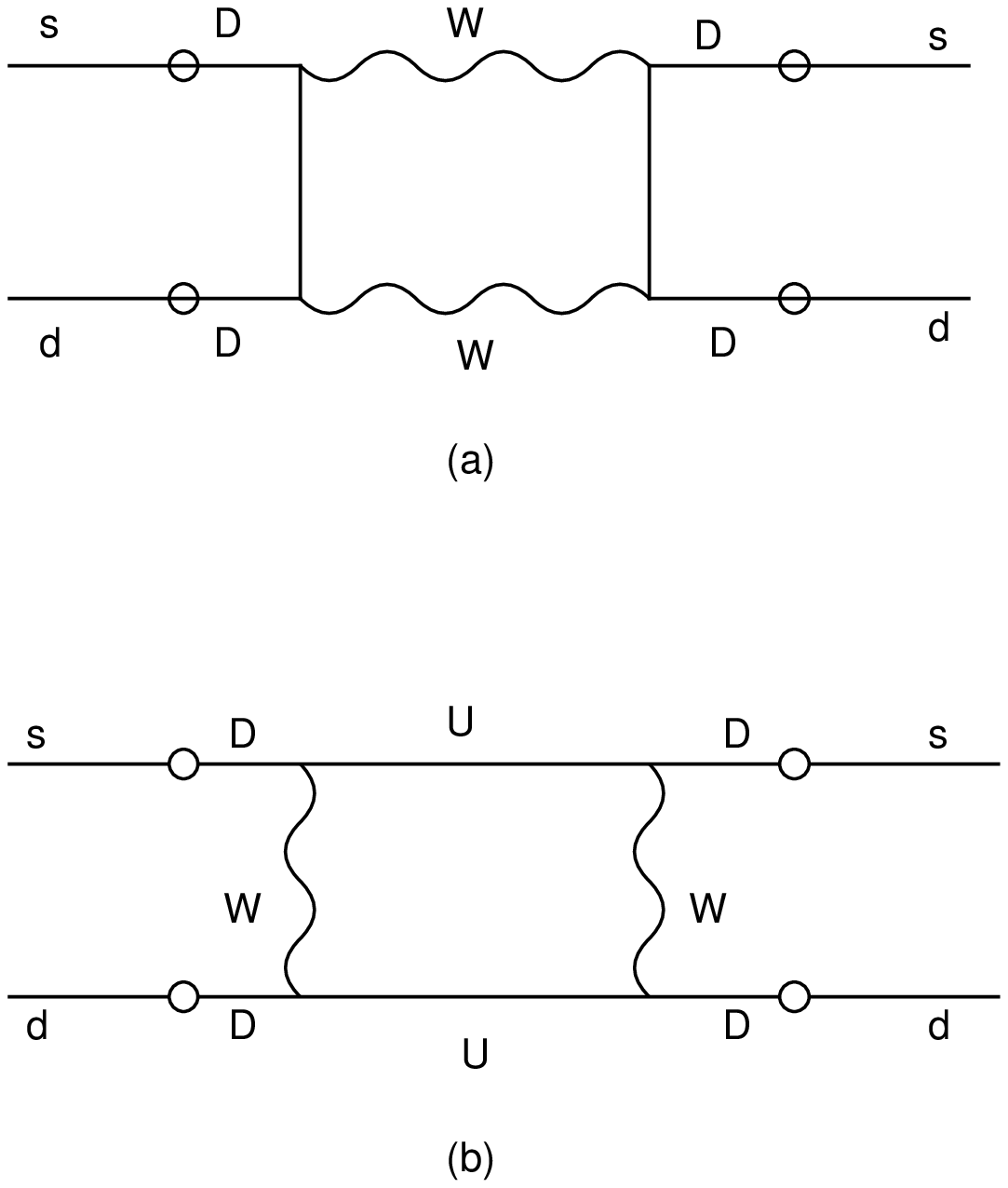}  }
\vskip .25in
\caption{Contributions to Im($\Delta M_{12}$) by new quarks and two-W box diagrams. (Circles mean
mixings.)}

\end{figure}

\noindent 
but these contributions are negligible.
We will consider the CP-violating parameters, such
as $Im(M_{12})/\Delta M_K$ in more detail below.

Next we consider the up sector for
completeness. We can also choose states such that $M_{up}$
in Eq.(\ref{MUD}) is replaced by
\begin{equation}
M_{up} = \left( \begin{array}{cccc}
m_u & 0 & 0 & \tilde{F}_1 \\
0 & m_c & 0 & \tilde{F}_2 \\
0 & 0 & m_t & \tilde{F}_3 \\
0 & 0 & 0 & M \end{array} \right)
\label{Mup}
\end{equation}
with
\begin{equation}
\tilde{F}_i = C_{ij}F_j
\end{equation}
where {\bf C} is the real standard $3 \times 3$ KM matrix.
The transformation matrices ${\bf J}_L$ and ${\bf J}_R$ that
diagonalize $M_{up}$
in Eq.(\ref{Mup}) can be related to ${\bf K}_L$
and ${\bf K}_R$ by changing $x_i$ into $\tilde{x}_i (= \tilde{F}_i/M)$
and $m_d, m_s$ and $m_b$ into
$m_u, m_c$ and $m_t$. The generalized $4 \times 4$ KM matrix
is given by
\begin{equation}
{\bf V}_{KM}^{(4)} = {\bf J}_L^{\dagger} \left[ \begin{array}{cc}
{\bf C} & 0 \\
0 & 1 \end{array} \right] {\bf K}_L
\end{equation}
\noindent
Before proceeding, we note that flavor-changing Z-coupling can be induced by the
one-loop diagram shown in Figure 18.

\begin{figure} 

\centerline{ \epsfysize 2in \epsfbox{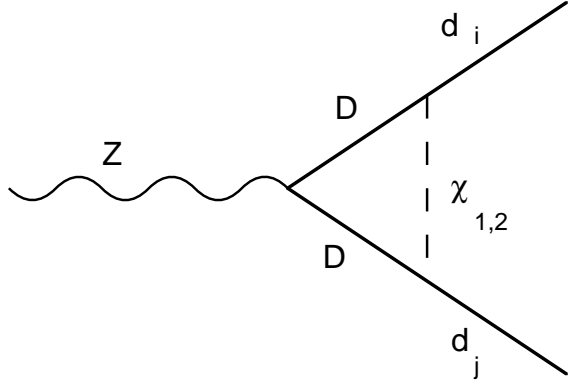}  }

\caption{Flavor-changing Z coupling induced by new quarks at one loop.}

\end{figure}

\noindent
By naive dimensional arguments, the effective coupling constant $\beta_{ij}$
is
\begin{equation} \beta_{ij}=
\left( -\frac{1}{2} \right) \frac{g_2}{cos \theta_W} 
\frac{h_ih_j}{16 \pi^2} \left( \frac{M}{m_{\chi}} \right)^2
\end{equation}
Using $h_{1,2,3} \simeq 0.01$ and $(M/m_{\chi})^2 \simeq 0.1$, we
conservatively estimate $\beta_{ij}$
to be less than $10^{-7}$. Therefore, we expect these FCNC's to be smaller
than those in the standard model.

Consider now the one-loop corrections to $\bar{\Theta}$. Although
the mass matrices in Eq.(\ref{MUD}) are complex, their
determinants are real. Therefore, $\bar{\Theta}$ defined
in Eq.(\ref{barr}) is zero at tree level. 
$\bar{\Theta}$ will be nonzero when the mass matrices
receive radiative corrections. For example, the contributions to
$\bar{\theta}$ from the up sector are given by
\begin{eqnarray}
\bar{\Theta} (up) & = & Arg[ det(M_{up} + \delta M_{up})]  \cr
& = & Im Tr ln [ M_{up} (1 + M_{up}^{-1} \delta M_{up})] \cr
& \simeq & Im Tr ( M_{up}^{-1} \delta M_{up}),   
\label{thet}
\end{eqnarray}
where we have used the fact that $M_{up}$ is real and that 
the corresponding radiative corrections $\delta M_{up}$ are small.
The last line in Eq.(\ref{thet}) is valid at one-loop order. Defining
the one-loop corrections $\delta M_{up}$ by
\begin{equation}
\delta M_{up} = \left( \begin{array}{cc}
\delta {\bf m}_u & \delta {\bf F} \\
\delta {\bf m}_{Uu} & \delta M \end{array} \right)
\end{equation}
and combining with Eq.(\ref{thet}), we obtain
\begin{equation}
\bar{\Theta} (up) = Im Tr ( {\bf m}_u^{-1} \delta {\bf m}_u 
- {\bf m}_u^{-1} {\bf F} M^{-1} \delta {\bf m}_{Uu} +
M^{-1} \delta M).
\label{THE}
\end{equation}
Notice that $\delta F$ will not contribute to $\bar{\Theta}$ at
one-loop order because of the structure of $M_{up}$. The expression
for $\bar{\Theta} (down)$ is strictly analogous to Eq.(\ref{THE}).

By studying all possible one-loop diagrams\cite{6ab,6ah} we find
that the only contribution to $\bar{\Theta}$ comes through $\delta {\bf m}_u$
in Eq.(\ref{THE}) and, in particular, from the diagram depicted in Figure 19.
The imaginary part gives a contribution
\begin{equation}
\bar{\Theta} (up) = \frac{1}{\sqrt{2}} \frac{1}{(4 \pi)^2}
\sum_{\alpha=1, l=1}^{\alpha =2, l=3} h_l^{\alpha} Im[x_l^*] 
\lambda_{\alpha} \frac{\kappa}{M} 
\end{equation}

\begin{figure} 

\centerline{ \epsfysize 3in \epsfbox{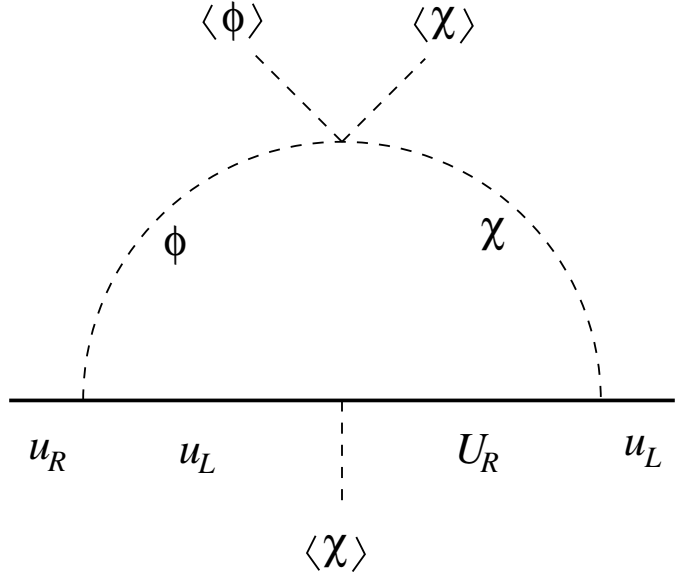}  }

\caption{One loop diagram of which the imaginary part contributes to $\overline{\Theta}$.}

\end{figure}

\noindent
which gives the estimate (for convenience we shall write x to denote
$x = |x_i|$, the modulus, and taken to be generation independent since the limits
on $x$ are not sensitive to the generation considered)
\begin{equation}
\bar{\Theta} = \frac{\lambda x^2}{16 \pi^2}
\label{oneloop}
\end{equation}
\noindent
Here $\lambda_{\alpha}$ is the coefficient of the quartic interaction
between the two types of Higgs $|\phi|^2|\chi_{\alpha}|^2$,
and $\lambda$ with no subscript is an average value. (Actually there are
three independent $\lambda$ corresponding to indices 11, 12+21, 22 
but our estimates will not distinguish these).

As mentioned earlier, the neutron electric dipole moment $d_n$
has been calculated\cite{6fffff,6g} in terms of $\bar{\Theta}$
long ago with the result that
\begin{equation}
d_n \simeq 10^{-15} \bar{\Theta} e.cm.
\label{NEDM}
\end{equation}
and so we know from $d_n \leq 10^{-25}$ e.cm. empirically that
$\bar{\Theta} \leq 10^{-10}$, from which it
follows by Eq.(\ref{oneloop}) that $\lambda x^2$ is less
than $10^{-8}$.

In the kaon system, the CP violation parameter $|\epsilon_K|$ is
given\cite{6ab,6ag} by
\begin{equation}
|\epsilon_K| = \frac{1}{\sqrt{2} \Delta m_K} m_K \frac{f_k^2}{3} 
\frac{2}{\kappa^2} x^4     \label{epsilon}
\end{equation}
Using $(\Delta m_K/m_K) = 7.0 \times 10^{-15}$, $f_K = 0.16$GeV
gives the relationship between $x^2$ and the $U(1)_{new}$ breaking
scale
\begin{equation}
\kappa / x^2 = 2.9 \times 10^7 GeV
\end{equation}
Thus, if we insist that the $U(1)_{new}$ is broken above the electroweak
breaking scale ($\sim 250GeV$) then
\begin{equation}
x^2 \gtrsim 10^{-5}
\end{equation}
From Eq.(\ref{oneloop}), this means that $\lambda < 10^{-3}$.

In \cite{6af}, it was argued plausibly that $\lambda > 10^{-5}$
on the basis of naturalness; this would imply that $\bar{\Theta}
> 10^{-12}$ and hence $d_n > 10^{-27}e.cm.$

But we can find a yet more solid and conservative lower bound
on $d_n$. It follows from the fact that the $|\phi|^2 |\chi_{\alpha}|^2$
interaction receives a one-loop correction from the quark loop box
diagram where three sides are the top quark and the fourth
is the heavy $U$-quark (see Figure 20).

\begin{figure} 

\centerline{ \epsfysize 3in \epsfbox{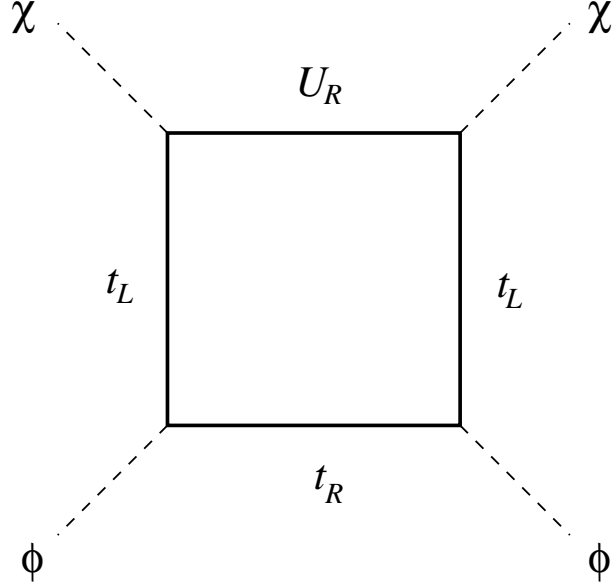}  }

\caption{One loop $|\phi|^2|\chi|^2$ counterterm for the coupling $\lambda$.}

\end{figure}

\noindent
The full $\lambda$ is given by
\begin{equation}
\lambda = \lambda_{tree(bare)} + 
\lambda_{1-loop} (\mbox{including counterterm}) + \mbox{higher loops}
\end{equation}
and the one-loop finite contribution, for the dominant diagram, Figure (20),
neglecting quark masses and taking $h_3^{\alpha}, g_t$ as the 
respective Yukawa copuplings to $\chi_{\alpha}$ and 
$\phi$ of the third generation by
\begin{equation}
\lambda_{1-loop} \simeq \int \frac{d^4k}{(2\pi)^4} |h_3^{\alpha}|^2 |g_t|^2 \frac{1}{k^4}  + \mbox{counterterm}
\end{equation}
which imples that the lowest value for $\lambda$ (without accidental
cancellations) is:
\begin{equation}
\lambda \gtrsim \frac{x^2}{16 \pi^2}
\label{counter}
\end{equation}
Combining Eqs.(\ref{oneloop}) and (\ref{counter}) then gives the 
estimate for $\bar{\Theta}$ of
\begin{equation}
\bar{\Theta} \gtrsim \frac{x^4}{16 \pi^4}
\end{equation}
which implies that $x^2 \leq 10^{-3}$ (incidentally in full agreement
with \cite{6af}) and that $10^{-10} \geq \bar{\Theta} \geq 10^{-14}$.

From Eq.(\ref{NEDM}), this then gives a lower limit on the neutron
electric dipole moment of
\begin{equation}
d_n \geq 10^{-29} e.cm.
\end{equation}
This is more than two orders of magnitude greater than the prediction of the
KM mechanism and thus provides another distinguishing
feature of spontaneous CP violation.

Before addressing other consequences of the Aspon Model, let
us point out that the production of the predicted heavy quarks and
the aspon in a hadron collider is discussed in Ref. \cite{6ac}.

A promising approach to finding new quark flavors is through searching for
heavy quark bound states. This fails for the top quark since the single $t$
quark decay to $bW$ is more rapid than the formation time of toponia.
However, the $Q$ of the Aspon Model have a very much longer lifetime
because of the small mixing with the three light generations. Such
production of the exotic quarkonia is dicussed in \cite{1ac}.

The experimental information regarding CP violation still comes primarily from
the neutral kaon system and is inadequate to determine whether the KM 
mechanism is the correct underpinning of CP violation. In dedicated
$B$ studies, with more than $10^8$ samples of $B^0(\bar{B}^0)$ decay,
it will be possible\cite{6aj,6ak,6al,6am,6an} to test this assumption
stringently by measuring the angles of the well-known unitarity
triangle whose sides correspond to the complex terms of 
Eq.(\ref{triangle}). If CP is spontaneously broken, as in
the Aspon Model, the outcome of these measurements will be different from the predictions
of the standard model.

The three angles of the unitarity trangle are conventionally
defined as $\alpha, \beta$, and $\gamma$ between the first
and second, second and third, third and first sides in Eq.(\ref{triangle})
respectively. These angles can be separately measured 
for the standard model by the time-dependent
CP asymmetry
\begin{equation}
a_f(t) = \frac
{\Gamma(B^0(t) \rightarrow f) - \Gamma( \bar{B}^0 \rightarrow f)}
{\Gamma(B^0(t) \rightarrow f) + \Gamma( \bar{B}^0 \rightarrow f)}
\end{equation}
where the final state $f$ is a CP eigenstate.
We define $q, p$ in $B^0 - \bar{B}^0$ mixing by the mass eigenstates
$B_{1,2}$:
\begin{equation}
|B_{1,2}> = p|B^0> \pm q|\bar{B}^0>
\end{equation}
and similarly for $K_{1,2}$ in the kaon system. Also, $A, \bar{A}$ are
the decay amplitudes
\begin{equation}
A, \bar{A} = <f|H|B^0, \bar{B}^0>
\end{equation}.
Let us consider the specific cases of $f = \pi^+\pi^-, \psi K_S$
from $B_d$ decay and $f = \rho K_S$ from $B_s$ decay.
We define $\lambda(f)$ by

\begin{equation}
\lambda(\pi^+\pi^-) = \left( \frac{q}{p} \right)_{B_d} \left( \frac{\bar{A}}{A} 
\right)_{B_d\rightarrow\pi^+\pi^-} ,
\label{pi}
\end{equation}
\begin{equation}
\lambda(\psi K_S) = \left( \frac{q}{p} \right)_{B_d} \left( \frac{\bar{A}}{A} 
\right)_{B_d\rightarrow\psi K} \left(\frac{q}{p} \right)_K , 
\label{psi}
\end{equation}
and
\begin{equation}
\lambda(\rho K_S) = \left( \frac{q}{p} \right)_{B_s} \left( \frac{\bar{A}}{A} 
\right)_{B_s\rightarrow \rho K} \left( \frac{q}{p} \right)_K^* .
\label{rho}
\end{equation}
The complex conjugate appears in Eq.(\ref{rho}) because $B_s^0 \rightarrow \bar{K}^0$
while $B_d^0 \rightarrow K^0$. If to a sufficiently good approximation
$|q/p| = 1$ and $|\bar{A}/A| = 1$, as we shall show for the Aspon Model below,
then $\lambda(f)$ is related to the CP asymmetry through the $B_1-B_2$ mass difference
$\Delta M$ by
\begin{equation}
a_f(t) = - \mbox{Im} \lambda(f) \mbox{sin}(\delta M t).
\label{aft}
\end{equation} 
In the standard model the angles of the unitarity triangle
are related to the $\lambda(f)$ by:
\begin{equation}
\mbox{sin} 2\alpha = \mbox{Im} \lambda(\pi^+\pi^-);
\hspace{0.2in}\mbox{sin} 2\beta = \mbox{Im} \lambda(\psi K_S);
\hspace{0.2in}\mbox{sin} 2\gamma = \mbox{Im} \lambda(\rho K_S).
\end{equation}
Such relations are no longer valid in the Aspon Model because
$Im (q/p)_{B_d}$ has a major contribution from
aspon exchange and $Im (q/p)_K$ is dominated by aspon exchange.

To evaluate the CP asymmetries in B decays for the Aspon Model
one needs to evaluate the different factors in the $\lambda(f)$
given in Eqs.(\ref{pi}) - (\ref{rho}) above. More precisely
we need, from Eq.(\ref{aft}), the imaginary part of the $\lambda(f)$.
The Aspon Model adds new Feynman diagrams involving aspon exchange to those already
present in the standard model. Because CP is only spontaneously
broken, the W-exchange amplitudes are predominantly real
amd have very small phases while the aspon exchange has a much smaller magnitude but an 
unpredicted arbitrary phase. As a result, the $|\mbox{Im}\lambda(f)|$
appearing in Eq.(\ref{aft}) are of order $0.002$ or less,
compared to the standard model expectation that $|\mbox{Im}\lambda(f)|$
be of order of, although less than, unity.
Thus CP asymmetries in B decays are predicted to be correspondingly smaller 
than in the standard model.  The technical details can be found in Ref.
\cite{6ae}.

It is found,  setting
$M_A = 300$ GeV, that
\begin{equation}
|\mbox{Im} \lambda (\pi^+\pi^-)| \leq 1 \times 10^{-5},
\end{equation} 
\begin{equation}
|\mbox{Im} \lambda(\psi K_S)| \leq 2 \times 10^{-3},
\end{equation} 
\begin{equation}
|\mbox{Im} \lambda (\rho K_S)| \leq 2 \times 10^{-3}.
\end{equation} 
The resulting asymmetries $a_f(t)$ are much smaller than those
predicted by the standard model where these imaginary parts are all of order unity.

It is also interesting to observe from
Eqs.(\ref{pi}) - (\ref{rho}) that
\begin{eqnarray}
\frac{\lambda(\psi K_S) \lambda(\rho K_S)}{\lambda(\pi^+\pi^-)} & = &
\left( \frac{q}{p} \right)_{B_s} \left( \frac{\bar{A}}{A} \right)_{B_d\rightarrow\psi K}
\cr
& = & \left( \frac{q}{p} \right)_{B_s} \left( \frac{\bar{A}}{A} \right)_{B_s \rightarrow
D_s^+D_s^-}       \cr
& = & \lambda(D_s^+D_s^-).
\end{eqnarray}
In the Aspon Model, where the $\lambda(f)$ have unit moduli and
$|\mbox{Im} \lambda(f)| \ll 1$, this relation implies a linear relation for the imaginary
parts:
\begin{equation}
\mbox{Im} \lambda(\psi K_S) + \mbox{Im} \lambda(\rho K_S) - \mbox{Im} \lambda (\pi^+\pi^-) 
- \mbox{Im} \lambda(D_s^+ D_s^-) = 0,
\label{test}
\end{equation}
which provides an additional test of the Aspon Model.

In conclusion, our result is that, if the Aspon Model is correct, CP asymmetries in B decays
would be much smaller than predicted by the standard model and the relation (\ref{test})
would be satisfied.

One may ask\cite{6af} whether the present experimental situation of B decay
is compatible with the Aspon Model?

In the model, the quark mixing matrix is a complex unitary $4 \times 4$ matrix
$C_{\mu\nu}$. Mixings of the conventional six quarks with one another are specified
by the $3 \times 3$ matrix $C_{ij}$, whose indices run from 1 to 3.
$C_{ij}$ is neither real nor unitary because of $\chi$-induced mixings
to the undiscovered quark doublet. However, these terms are $\sim x^2$. Thus,
since $x^2 \leq 10^{-3}$, $C_{ij}$ is, to a precision of at least $0.1\%$,
a real orthogonal matrix. It
is a generalized Cabibbo matrix rather than a Kobayashi-Makawa matrix. This is
unfortunate for the search for CP violation in the beauty sector, but has 
other observable consequences.

In the standard model, the Kobayashi-Maskawa matrix $V$ is complex and unitary.
The sides of the unitarity trangle are unity and
\begin{equation}
R_b = \left| \frac{V_{ud}V_{ub}}{v_{cd}V_{cb}} \right|,
\hspace{0.3in} 
R_t = \left| \frac{V_{td}V_{tb}}{v_{cd}V_{cb}} \right|.
\end{equation}
The value of $R_b$ has been measured.
According to \cite{6an}, $R_b = 0.35 \pm 0.09$. The
value of $R_t$ cannot be extracted from experimental data alone.
Appeal must be made to a theoretical evaluation of the neutral B-meson
mass difference using the standard model. The analysis in \cite{6an}
yields $R_t=0.99\pm 0.22$. These results
suggest a rather large value of the CP-violating angle $\beta$,
namely, $0.34 \leq \beta \leq 0.75$.

In the Aspon Model, the matrix $C_{ij}$ is orthogonal up to terms of order
$x^2$ arising from mixings with unobserved quarks. Thus, we anticipate
no readily observable manifestations of CP violation in the beauty sector.
Furthermore, the unitarity triangle
must degenerate into a straight line: $|R_b \pm R_t| = 1$.
In this case, we cannot appeal to a theoretical calculation of the neutral B-meson mass difference since it depends
on unknown parameters. On the other
hand, the matrix $C_{ij}$ with neglect of terms $\sim x^2$ involves only three parameters.
>From data at hand, in this context, we obtain
$R_t = 1 - \rho$ in
the Wolfenstein\cite{6ap} parametrization, whence $R_t = 0.637 \pm 0.09$.
We note that present data yields $R_b + R_t = 0.99\pm 0.13$.
This result is compatible with an approximately orthogonal mixing matrix and hence with the 
Aspon Model.

As a final CP violation parameter in the Aspon Model, we shall discuss the value predicted
for 
Re$\left( \frac{\epsilon^{'}}{\epsilon} \right)$
 - a measure of direct CP violation in K decay.
To evaluate 
Re$\left( \frac{\epsilon^{'}}{\epsilon} \right)$
requires the study of several Feynman diagrams\cite{6aq,6ar,6as,6at,6au,6av},
and their comparison to the standard model.
Recall that the most recent evaluations completed at CERN (NA31) \cite{6aw}
and FNAL (E371) \cite{6ax,6ay} give results
Re$\left( \frac{\epsilon^{'}}{\epsilon} \right) = (23\pm3.6\pm5.4) \times 10^{-4}$
and
Re$\left( \frac{\epsilon^{'}}{\epsilon} \right) = (7.4\pm5.2\pm2.9) \times 10^{-4}$
respectively, where the first error is statistical and the second is systematic.
These results are consistent within two standard deviations; the error is
expected to be reduced to $1 \times 10^{-4}$ in foreseeable future
experiments\footnote{A new result from KTeV\cite{newktev} at Fermilab gives
$Re \left({\epsilon^\prime\over\epsilon}\right)= (28 \pm 4.1)\times 10^{-4}$; it
will be interesting to see whether CERN confirms this result.}.

A detailed analysis of $\epsilon^\prime/\epsilon$ in the Aspon model was
carried out by Frampton and Harada\cite{6ag,6ah}.  They showed that penguin
diagrams involving the additional quarks give the dominant contribution.
The outcome of these considerations is that
Re$\left( \frac{\epsilon^{'}}{\epsilon} \right)$
is not larger than $10^{-5}$ in the Aspon Model. While it does not vanish exactly,
it does correspond closely to a superweak model prediction\cite{6bb}.

This discussion of the Aspon Model is presented as a motivation for additional quarks
beyond the six discovered flavors. Because of the smallness of $x^2$ which
characterizes the mixing of the additional quarks with the known ones, the
new quarks have a long lifetime. This is important in their experimental
detection, as discussed the next Section.

\vfill

\section{Experimental Searches.}

\subsection{Search for long-lived quarks}
\subsubsection{Present Searches}
The search for long-lived quarks is an ongoing process at the Fermilab
Tevatron.   The first search at the Tevatron was made by the D0 collaboration
\cite{D0ref} which looked for signals $b^\prime\rightarrow b\gamma$, where $b^\prime$ is a charge -1/3 quark, setting a limit $m_{b^\prime}> M_Z + m_b$.  
The second and most recent search was made by the CDF collaboration\cite{abe1}
which looked for a displaced vertex for $Z\rightarrow e^+e^-$ coming from $b^{\prime}\rightarrow bZ$ resulting in $m_{b^\prime}> 148$ GeV (for $c\tau=1$ cm.).
The latter search will be described below.
Eventually, such a search will  be carried out at the LHC which
has a much greater C.M. energy, with a much larger production
cross section. In this section, we will concentrate on the current
limits coming out of two operating facilities: LEP2 and Fermilab. In the
next section, we will discuss the prospects for future searches, both
at the Fermilab Tevatron and at the LHC. We will only briefly discuss
prospects for such a search at facilities which are under discussion,
such as the NLC, etc.  Also in this section, we will focus primarily on
hadron colliders such as the Tevatron since these are the machines
which can explore the mass range that was discussed earlier.

In the search for a new particle, there are two principal activities to attend 
to: how to produce the particle and how to detect it. For a particle which 
is somewhat ``exotic'', such as supersymmetric particles, the production process
would be highly model-dependent. Fortunately, for a heavy quark, this
is rather standard: it proceeds through the $q \bar{q}$
and $gg$ channels. For the range of heavy quark masses considered
in this Report, the $q \bar{q}$ process via the electroweak channels
$W, \gamma, Z$ is completely negligible compared with the QCD process
with gluons. The production cross section, at the Tevatron
and at the LHC, for the top quark as a function of its mass has been
computed up to the next-to-leading order in QCD \cite{berger}-\cite{berger5}. 
(The $q \bar{q}$ channel is dominant at the Tevatron 
while the $gg$ channel is dominant at the LHC.)
This can be directly applied
to the present search for long-lived quarks whose production mechanism
should be similar to that of the top quark. The production cross sections
as a function of the heavy quark mass are shown in Figs. 21 and 22
for the Tevatron at $\sqrt{s} = 1.8$ TeV and 2 TeV respectively,
and in Fig. 23 for the LHC at $\sqrt{s} = 10, 14$ TeV. Here
``$m_t$'' will stand for a generic heavy quark mass, for both the
Tevatron and the LHC.

\begin{figure} 
\centerline{ \epsfysize 6in \epsfbox{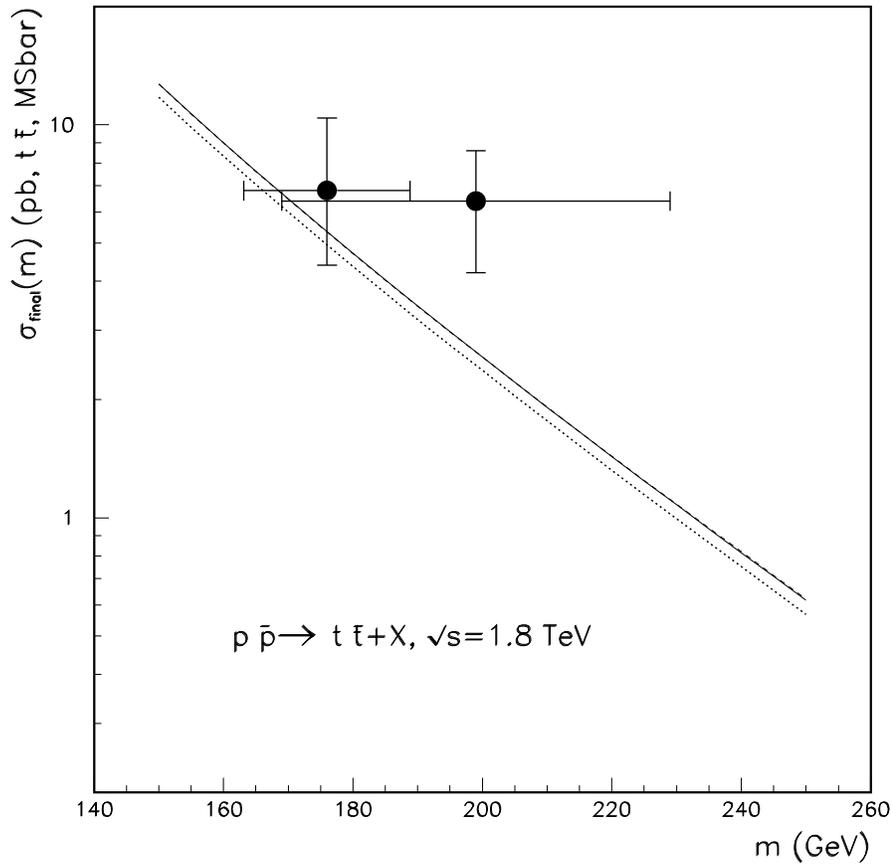}  }
\vskip .25in
\caption{Physical cross section for $p\protect\overline{p}\protect\rightarrow
t\protect\overline{t}X$  at $\protect\sqrt{s}=1.8$ TeV as a function of the top
mass.  This cross section applies to a heavy quark $Q$ as well with $t$
changed to
$Q$.  The two data points are from CDF and D0 respectively for the top quark.}
\vskip .25in
\end{figure}

\begin{figure} 
\centerline{ \epsfysize 6in \epsfbox{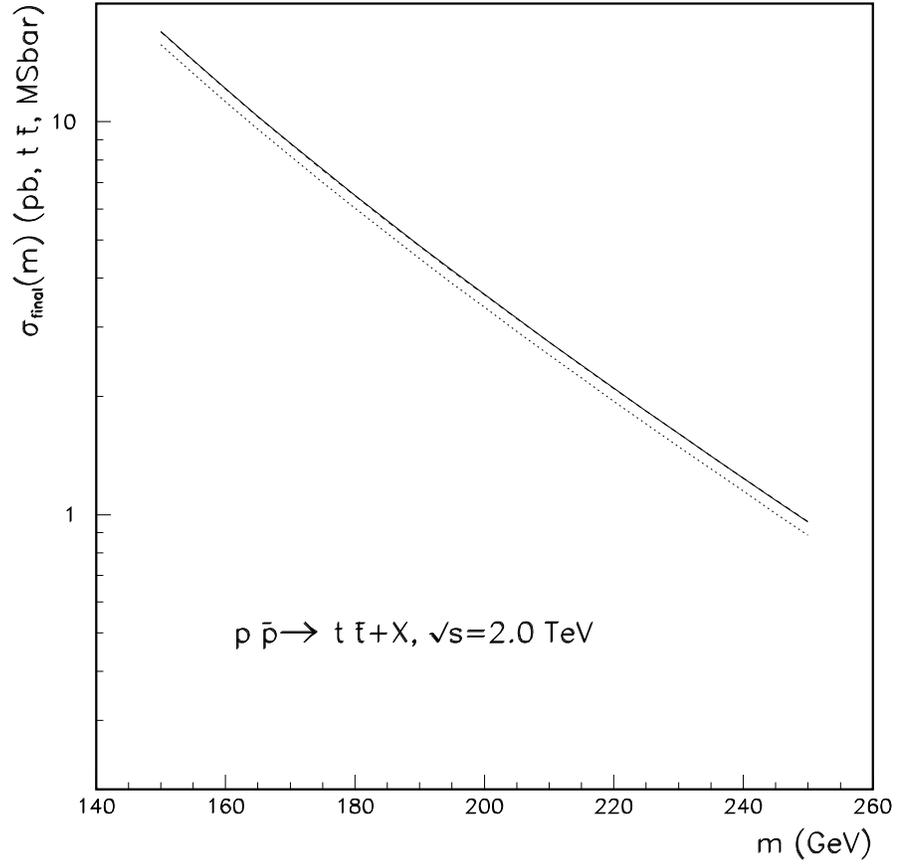}  }
\vskip .25in
\caption{Physical cross section for $p\protect\overline{p}\protect\rightarrow
t\protect\overline{t}X$  at $\protect\sqrt{s}=2.0$ TeV as a function of the top
mass.  This cross section applies to a heavy quark $Q$ as well with  $t$
replaced by
$Q$.}
\vskip .25in
\end{figure}
\vspace{0.5in}
\begin{figure} 
\centerline{ \epsfysize 6in \epsfbox{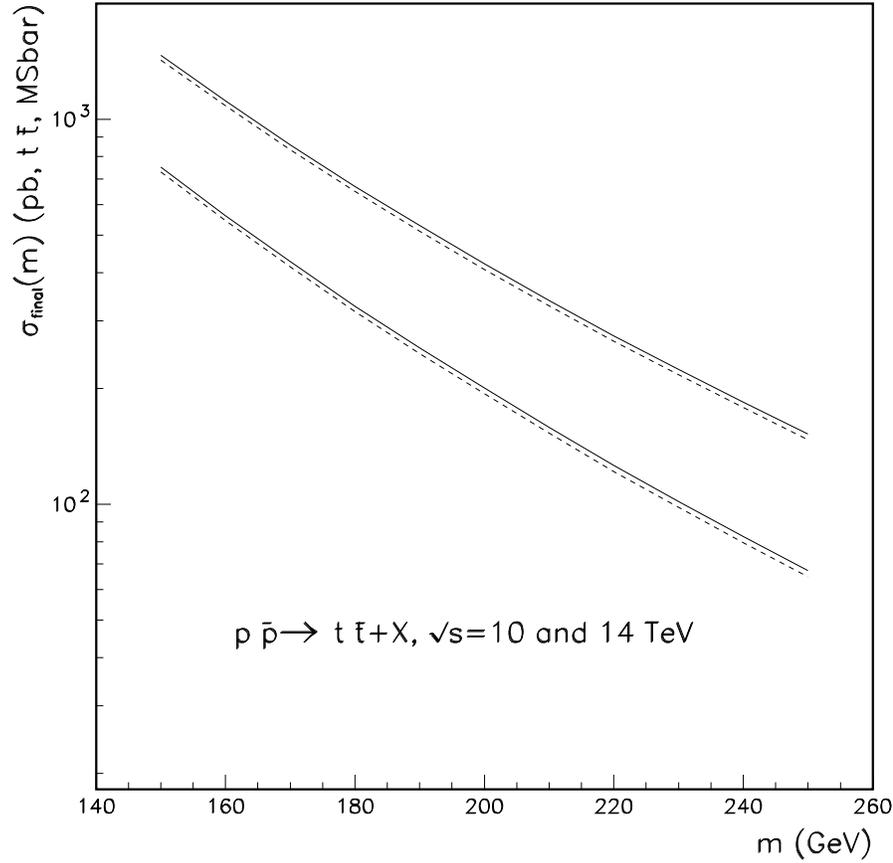}  }
\vskip .25in
\caption{Cross section for $pp\protect\rightarrow t\protect\overline{t}X$ at
$\protect\sqrt{s}=10$ and
$14$ TeV as a function of the top mass.  Notice that $p$ has been mislabeled
as $\protect\overline{p}$ in the figure.  The same prediction applies to a heavy
quark
$Q$.}
\vskip .25in
\end{figure}

As can be seen above, the predicted
cross section at the LHC, for a given heavy quark mass, exceeds that at
the Tevatron by more than two orders of magnitude, which will facilitate the
search for such an object.

The next task is to define the detection capability of various 
detectors. Since the latest constraint on long-lived quarks
come from CDF \cite{abe1}, we shall use it as a prototype of detectors dedicated
to such a purpose. Other detectors such as D0 are very similar in layout. 
Needless to say, the CDF detector is a complicated,
multipurpose one whose specifications can be found in \cite{abe2}. A generic
detector for hadron colliders generally consists of a (silicon) vertex detector
immediately surrounding the beam pipe for high precision
determination of the location of the tracks. Next comes a central
tracking chamber which measures charged tracks and momenta of
charged particles. Surrounding these two units are generally
hadron calorimeters which measure the energy deposited by hadrons.
Next comes the muon chambers which detect the location of the
particles which penetrate the calorimeters. As described below,
the first two parts (vertex detector and central tracking chamber)
were used to search for displaced vertices coming from the decay
of a long-lived particle. For stable or very long lived particles,
the muon chambers are used in conjunction with the ionization
energy loss in the tracking chambers to make such a search.

The current search at CDF can be
divided into two categories: the search for those quarks whose
decay lengths, $l = \gamma \beta c \tau$ with $\tau$ being the
proper decay time, is 1) less than 1 meter, and 2) greater than
1 meter.

Let us first concentrate on the first category 
($l< 1$ meter) \cite{abe1}.
The parts of the detector which are relevant here 
consist of two components: a silicon vertex detector immediately
surrounding the beam pipe for precision tracking and a central
tracking chamber embedded in a 1.4 T solenoid magnetic field
which measures the momenta and trajectories of charged particles.
In the search for a new particle, a crucial task would be the 
identification of a characteristic signature which would
distinguish it from background. In the present case, that
characteristic signature is the decay $Z \rightarrow e^{+}
e^{-}$ with the $e^{+} e^{-}$ vertex displaced from the 
$p\overline{p}$ interaction point. This
$Z$ boson could come from the decay of a charge
-1/3 quark (denoted by $b^{\prime}$ in Ref.\cite{abe1} and
by $D$ in Ref. \cite{framptonhung} in the process
$D \rightarrow b + Z$  with a subsequent decay
$Z \rightarrow e^{+} e^{-}$. In this case, $D$
would be the long-lived parent of the $Z$ boson. The search at
CDF concentrated on events containing an electron-positron
pair whose invariant mass is consistent with the $Z$ mass and
whose vertex is displaced from the $p\overline{p}$ interaction point.
The data used was from the 1993-1995 Tevatron run with an
integrated luminosity of $90 \,pb^{-1}$ of $p\overline{p}$ collisions
at $\sqrt{s}$ = 1.8 TeV.

In the search for a long-lived parent of the $Z$, the CDF
collaboration focused on the measurement of $L_{xy}$ which
is the distance in the transverse ($r-\phi$) plane between
the $p\bar{p}$ interaction point and the $e^{+} e^{-}$ vertex.
Notice that $L_{xy} = \gamma \beta_{xy} c \tau$ with 
$\beta_{xy}$ being the transverse component of the parent
particle didived by $c$. As defined by the CDF collaboration,
$L_{xy}$ can be either positive or negative. For prompt 
$Z$'s coming from the SM process $q \overline{q} \rightarrow Z
\rightarrow e^{+} e^{-}$, one would expect $L_{xy} \approx
0$ because of the short lifetime of the $Z$. The $L_{xy}$
distribution with appropriate cuts taken into account
is shown in Fig. 24 below

\vspace{0.5in}

\begin{figure}
\begin{center}
\ \hskip -4cm \epsfysize 4in \epsfbox{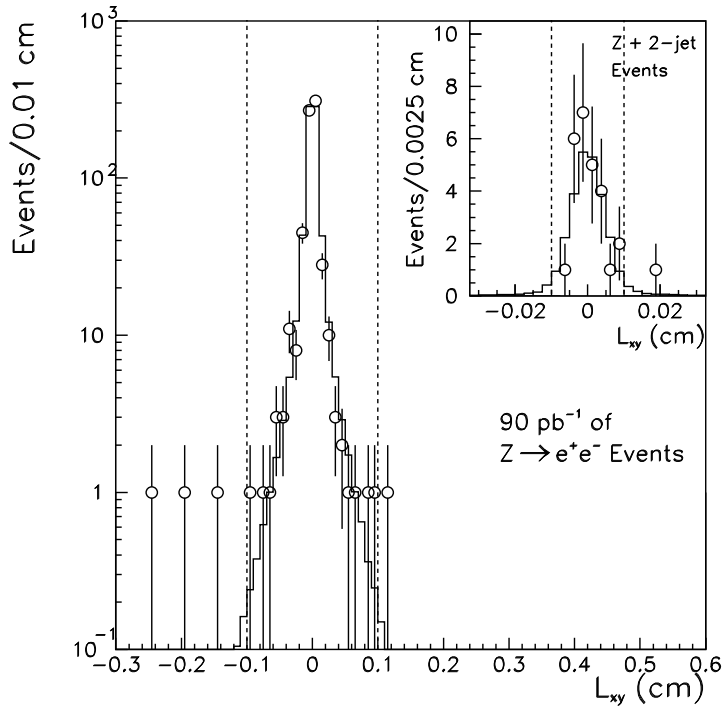}

\caption[]{The $L_{xy}$ distribution of the Z's after applying all cuts.  The 
data are represented by the circles.  The histogram is the expected $L_{xy}$
distribution for prompt Z's based on the measured $L_{xy}$ uncertainty in the
event sample.  The inset shows the distribution after the 2 jet requirement is
applied.  The vertical dashed lines separate the prompt and non-prompt regions.}

\end{center}
\end{figure}
\vskip .25in

As emphasized by \cite{abe1}, this distribution is consistent
with that for prompt $Z$'s where one would expect less than
one event for $|L_{xy}| < 0.1$ cm. \cite{abe1} also pointed
out that the number of events with $L_{xy}$ significantly less
than zero is an effective measure of the background. The CDF
collaboration observed one event for $L_{xy} > 0.1$ cm and
3 events for $L_{xy} < -0.1$ cm. As stated, there is no evidence
for a long-lived parent of the Z. This is shown in Fig. 25 where
the constraint is expressed in terms of the 95 
level upper limit on the product of the
production cross section for the long-lived parent, $\sigma_X$,
its branching ratio, Br($X \rightarrow Z$), the branching ratio,
Br($Z \rightarrow e^{+} e^{-}$), and the $e^{+} e^{-}$ acceptance
for pseudorapidity $|\eta| < 1$.
\vspace{0.5in}

\begin{figure}
\begin{center}
\ \hskip -4cm \epsfysize 4in \epsfbox{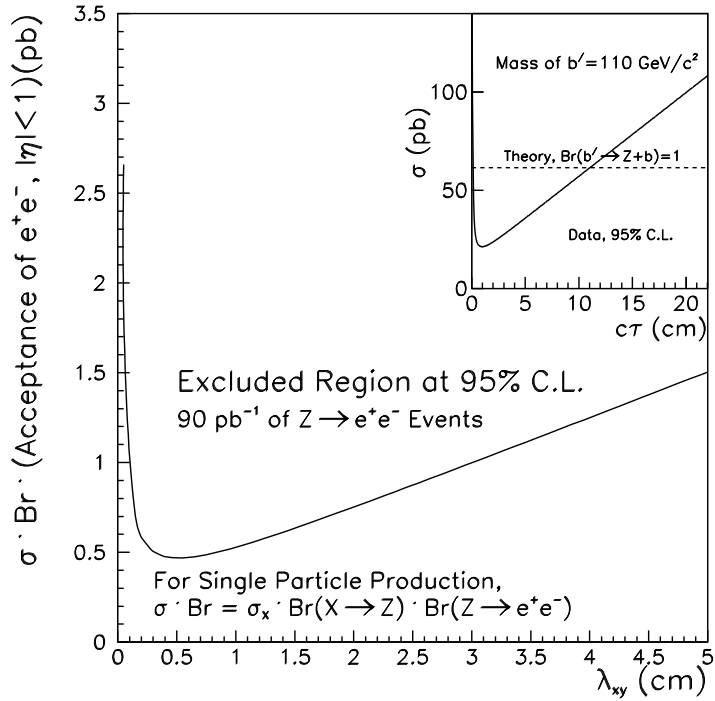}

\caption{The 95\% confidence level upper cross section limit for $\sigma.Br$
times the acceptance for an electron-positron pair to be within the detector as
a function of fixed $\protect\lambda_{xy}\protect\equiv
\protect\gamma\protect\beta_{xy}c\protect\tau$.  Cross sections above the curve
have been excluded at the 95\% confidence level.  The inset shows the exclusion
curve and the theoretical prediction for a
$b^{\protect\prime}$ quark of mass 110 GeV as a function of its lifetime,
assuming 100\% decay into
$bZ$.}

\end{center}
\end{figure}

The above discussions and figures deal with limits on the
production of a single parent with its subsequent decay into
a $Z$. We are, however, most interested in the detection of
a long-lived quark. As we have mentioned earlier, this
long-lived quark would be produced in pair. The CDF search
for a long-lived $D$ (or $b^{\prime}$) which
is pair-produced can be summarized as follows. The kind of
events that are searched for would include, besides the
$e^{+} e^{-}$ pair coming from the $Z$, two or more jets. For 
instance, this could come from the reaction: $q \overline{q}
\rightarrow D \overline{D} \rightarrow
b Z \overline{b} Z \rightarrow b e^{+} e^{-} \overline{b} q \overline{q}$.
The $L_{xy}$ distribution is shown in the inset of Fig. 25. There one would
expect less than one event for 
$L_{xy} \leq 0.01$ cm. The CDF collaboration found one event.
This was then translated into a cross section limit as
a function of $c \tau$, where $\tau$ is the lifetime of
$D$. Assuming that Br($D \rightarrow
b Z$) = $100\%$, the exclusion curve for the $D
\bar{D}$ production cross section as a function
of $c \tau$ is shown as an inset of Fig. 25 for
a $D$ quark of mass 110 GeV. The theoretical prediction
for the cross section for such a mass is shown as a horizontal
line. Clearly, this is ruled out for a wide range of lifetimes.
For other masses, the exclusion curves for the
cross section are not shown but are
instead translated into exclusion regions in the mass-lifetime
plane. (This is because the production cross section can be calculated
in QCD as a function of the $D$ mass as mentioned above.) The plot
shown in Fig. 26 assumes the above branching ratio.

\vspace{0.5in}

\begin{figure}
\begin{center}
\ \hskip -4cm \epsfysize 4in \epsfbox{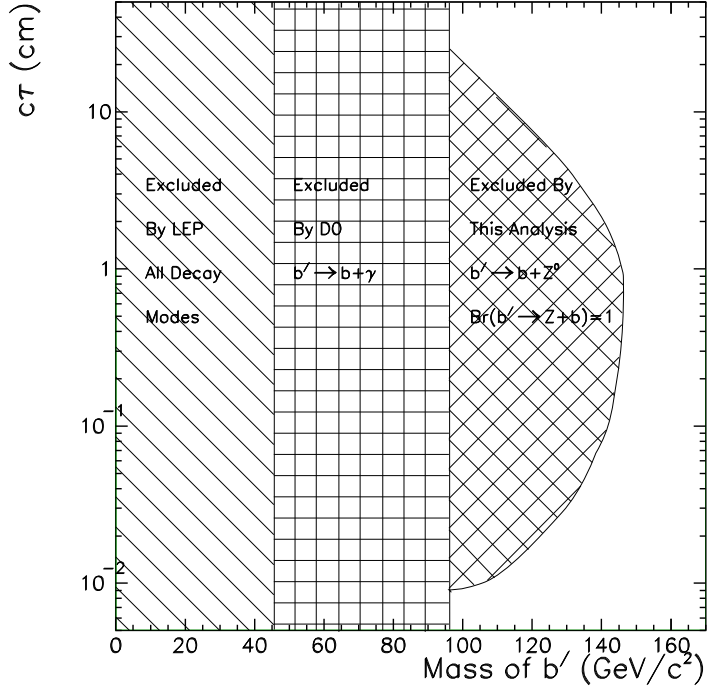}

\caption{The hatched areas in this plot represent the 95\% confidence-level
regions of $b^{\protect\prime}$ mass and lifetime that have been excluded. For
$c\protect\tau=1$ cm, CDF excluded up to a mass of $148$ GeV.}

\end{center}
\end{figure}

In Fig. 26, three forbidden regions are presented: the LEP,
the $D0$, and the CDF constraints. The most stringent constraint
comes, of course, from the CDF results. As can be seen from Fig. 26,
for every $c\tau$, there is a range of forbidden masses represented
by the shaded region. The largest forbidden range is for $c\tau$ =
1 cm corresponding to a lifetime $\tau \approx 3.3 \times
10^{-11}$ sec . This rules out the mass of the $D$ quark up to
148 GeV. For smaller or larger $c\tau$, we
can see that the lower bounds on the $D$ mass become somewhat
smaller than 148 GeV . If the $D$ quark,
happens to have a mass larger that 148 GeV, it
could escape detection for a large range of lifetimes as can be seen
in Fig. 26.
The mere fact that there 
exists unexplored regions of the detector, as shown in the unshaded
areas of Fig. 26, is reason to believe that there are
plenty of opportunities for future searches.

What does this result tell us about the long-lived quarks with the
mass range that we have discussed in Section IV?  First, as we have
mentioned earlier, we are especially concerned with the mass of the
heavy quarks larger than 150 GeV.  From Fig. 26, 
one can see that the CDF search based on the
decay mode $D \rightarrow b Z$ does not set any constraint
on long-lived quarks with a mass greater than 150 GeV. In other words,
if this long-lived quark exists and if it decays at a distance
$L_{xy} >$ 0.01 cm, it has yet to be discovered. This statement is,
of course, based on the assumption that the branching ratio for
$D \rightarrow b Z$ is $100\%$. As discussed in Section IV and in
Ref. \cite{framptonhung}, for $m_D \leq m_t$, there is
another possible decay mode for $D$, namely $D \rightarrow (c, u) W$
(the $c$ quark channel in any reasonable scenario dominates over
the $u$ quark channel). Whether or not 
$D \rightarrow b Z$ dominates over $D \rightarrow c W$ will depend
on a particular model for the mixing element $|V_{Dc}|$. 
As discussed in Ref. \cite{framptonhung}, even with a very naive assumption
$|V_{Dc}| \sim x^{3/2}$, with $x$ being the mixing parameter
between the third generation and the heavy quark, 
$D \rightarrow b Z$ can dominate over $D \rightarrow c W$ for
a certain range of mass and mixing $x$. For $|V_{Dc}| \sim x^2$
(or less), the $bZ$ mode will almost always be the dominant one.
Since this is a rather model-dependent statement, one should,
in principle, look for both modes when $m_D \leq m_t$. Unfortunately,
the mode $D \rightarrow c W$ would be rather difficult to detect.
We shall come back to this issue and others in the discussion of future
searches. 

The next question concerns the limits on a charged ``stable'' or
very long-lived massive
particle. Such a search is being carried out a CDF. Basically, this
search focuses on decay lengths larger than 1 m, i.e. larger than the
radius of the Central Tracking Chamber.
As of this writing,
preliminary results have only appeared in  conference talks,
\cite{hoffman,CDFbounds}. Therefore, what is described below will be considered
preliminary. A stable massive quark moving at a low velocity will
leave an ionization track in the tracking chamber because the energy
loss $dE/dx \propto 1/\beta^2$ (the Bethe-Bloch equation) and a 
low $\beta$ would imply a {\em large energy loss}. 
Furthermore, for such a stable massive quark to be detected, 
one would look for signals in the muon
detector after mesons formed from this particular quark have traversed
the calorimeters and reached the muon detector. To distinguish it from
a muon, one would have to correlate this signal with the large energy
deposited in the tracking chamber. The measurement of $dE/dx$ as a
function of $\beta\gamma = p/M$, combined with the momentum
measurement would allow for a determination of the mass of the
particle. The mass limits from CDF for such a very long-lived
quark are approximately 200 GeV.

In summary, the most stringent limits, so far. on long-lived quarks
come from the CDF collaboration \cite{abe1}. It excludes a long-lived, 
charge -1/3 quark of mass up to 148 GeV  for a 
lifetime $\tau \approx 3.3 \times 10^{-11}$ sec ($c\tau$ = 1 cm).
For other values of lifetimes, the excluded mass ranges are weaker
as can be seen in Fig. 26. This constraint was based on
the search for a displaced vertex for the decay $Z \rightarrow
e^{+}e^{-}$ which could come from the decay $D \rightarrow b Z$.
Furthermore, all of these constraints come from the search for
the decay of the $D$ quark inside the Central Tracking Chamber.
If the $D$ quark lives long enough to enter the calorimeters and
subsequently trigger a signal in the muon chamber, the constraint
(which is preliminary) is much stronger: a $D$ quark mass below
approximately 200 GeV is excluded. In short, under what circumstances will a
$D$ quark escape detection so far? First, if its mass is above
148 GeV (as referred to in Section IV) and if it decays inside
the Central Tracking Chamber. If the mass is above $\sim$ 200 GeV, $D$
is no longer required to decay in the tracking chamber: it simply
escapes detection regardless of where it decays. Most of the 
discussion in Section IV concerned these possibilities.

What kind of improvements should be made in order to be able to search
for these quarks heavier than 148 GeV which could either decay inside
the Central Tracking Chamber or, if heavier than 200 GeV, could 
also travel through the calorimeters? What about the $U$ quark? How
could one detect it? What if the dominant decay mode of the $D$ is $D
\rightarrow c W$? These are the kinds of questions that one would like to
address, at least qualitatively, in the next section.

\subsubsection{Future Searches}

The first kind of future searches would be based on present facilities
such as the Tevatron. In particular, one might ask what kind of 
improvement one can make by exploiting the present CDF RunI data with up to
120 $pb^{-1}$. One can then ask what light RunII with a large improvement in
luminosity and detector might shed on the search for long-lived quarks.

The above search has focused on the discovery of displaced vertices
for decay lengths greater than 100 $\mu m$, with the resulting constraints
as described above. What if the decay length is less than 100 $\mu m$?
The appropriate kind of experiment would be a counting experiment
which is not based on the search for displaced vertices \cite{stuart}.
How feasible this kind
of experiment might be will probably depend on the improvements 
planned for RunII.

These improvements include: a) a detector upgrade with, among several
things, added layers of silicon; b) an increase by a factor of 20 in the
luminosity. The radius of the silicon vertex detector is roughly
of the order of 22.3 cm. Added silicon layers would increase that radius
to about 28 cm \cite{stuart} and consequently the tracking ability of the detector.

For decay lengths between 100 $\mu m$ and 1 m, one of the most important 
tasks would be to improve the tracking efficiency of the Central
Tracking Chamber by adding,for instance, more silicon layers to the
vertex detector. Decay
lengths of a few tens of cm might be hard, although possible, to detect 
because of poor tracking efficiency in such a region. This would require
new reconstruction algorithm.

The search for very long-lived or ``stable'' quarks will also be 
improved by the detector upgrade and the increase in luminosity. 

One might ask what else could be done at CDF and D0 in the next run beside
those issues discussed above. In particular, one would like to know
how feasible might the detection of a signal such as $D \rightarrow
c W$ be if it happens to be the dominant decay mode of the $D$.
Needless to say, such a task would be much more daunting than the
detection of $D \rightarrow b Z$. Nevertheless, a feasibility study
would probably be extremely useful. As we have discussed at length
in Section IV and in Ref. \cite{framptonhung}, there is also the partner of the
$D$, namely the charge 2/3 quark denoted by $U$, which should not be 
forgotten. If $U$ is heavier than $D$- but not by much because of the
$\rho$-parameter constraint- it will decay into $D$ via
$U \rightarrow D + (l^{+} \nu, q_{2/3} \bar{q}{-1/3})$, where particles
inside the brackets denote light quarks or leptons or it can decay
via $U \rightarrow b W$, depending on how degenerate $U$ and $D$ are
and how large $|V_{Ub}|$ is. For the first mode, it was shown in Ref.
\cite{framptonhung} that $U$ practically decays near the interaction point, with
a decay length typically of the order of $10^{-5} \mu m$. The
$D$ will subsequently decay between $100 \mu m$ and 1 $m$. It will be
a challenge to be able to identify such a signal. For the second mode
$U \rightarrow b W$, one has to be able to distinguish it from a
signal coming from top decay. It would be extremely hard, if not
impossible, to be able to resolve the decay vertex to distinguish
$U$ from $t$. However, by comparing the predicted number of $t$'s
with the observed ones, one might rule out the mode $U \rightarrow
b W$ with a $U$ mass close to the top mass.

Turning our attention to the upcoming experiments at the LHC, we would
like to briefly describe the two main detectors which will be crucial
to the search for long-lived quarks (if such a search would be carried
out). They are the Compact Muon Solenoid (CMS) and a Toroidal LHC
Apparatus (ATLAS) detectors \cite{cmsatlas}. The layout for both detectors is 
generically very similar to CDF and D0. 

In the search for long-lived
quarks, the components which are crucial would be the central
tracking system of CMS and the inner detector of ATLAS. The
central tracking system of CMS consists of silicon pixels,
silicon and gas microstrip detectors with high resolutions. (The
resolution of the silicon pixels is about 11-17 $\mu m$ while the
outermost part of the detector, namely the gas microstrip detector,
has a resolution of approximately 2$mm$!) The silicon pixels and
silicon microstrips cover a radial region up to about 40 $cm$, a
marked improvement over the CDF vertex detector. The microstrip gas 
chamber covers a radial region to approximately 1.18 $m$ which is
roughly similar to that covered by the Central Tracking Chamber
of CDF.

The inner detector of ATLAS consists of a Semi-Conductor Tracker (SCT):
pixel detectors, silicon microstrips and GaAs detector, and 
Microstrip Gas Counters (MSGC). The pixel detectors have a spatial
resolution of about 14 $\mu m$ while the MSGC have a resolution of
about 1.8 $mm$, very comparable to CMS. The SCT part covers a radial
region of up to 60 $cm$ while the MSGC covers a radial region of up
to 1.15 $m$. Again one sees a marked improvement over CDF in the
region of interest.

In addition, as can be seen from Fig. 23,
the production cross section for a given mass is now increased by
at least two orders of magnitude at the LHC because the center of mass
energy is now 14 TeV. Such an increase in the cross section combined
with the increase in the radial distance covered by silicon detectors 
would, in principle, help in the search for long-lived quarks. One might
wonder if a similar analysis as the one preformed by CDF could be
carried over to the LHC experiments. Considering the fact that,
with a much higher energy and consequently larger cross section,
the number of events and background will be significantly higher as well.
This would probably require a different search algorithm.

Finally, concerning proposed but not yet approved colliders such
as the Next Linear Collider (NLC) with $\sqrt{s} = 500$ GeV, the
long-lived quarks with the mass range discussed in Section IV, would
be produced copiously and with little background. What kind of
signal would one search for will depend on the kind of detectors
involved. Whether or not the existence (or nonexistence thereof)
of these long lived quarks will be established by CDF, ATLAS, or
CMS by the time the NLC operates (if approved) remains an open
question.

\subsection{Lepton Searches}

Earlier in this Report, indirect bounds on the masses of heavy leptons arising
from violations of $e-\mu-\tau$ universality were discussed.  The bounds were
very sensitive to the mixing angle between the third and fourth generations. 
In this section, we discuss direct detection of heavy leptons.  

All current experimental bounds on heavy leptons come from experiments at
electron-positron colliders.   This is not surprising; the cross section for
heavy lepton production at hadron colliders is small and backgrounds are
large.   Of course, once LEP200 shuts down in a couple of years, the only
available colliders for searching for heavy leptons will be the Tevatron and
the LHC.  We will first examine the current  bounds on heavy lepton masses,
and then turn towards the future.

It is generally believed that charged heavy leptons can be excluded up to
the approximate kinematic limit of LEP.   This is not necessarily the case
however; the charged heavy leptons of many of the most interesting models have
not been excluded for masses above approximately 45 GeV.   Below 45 GeV, heavy
leptons (charged or neutral) would contribute to the decay width of the $Z$,
and such leptons can be excluded.

The strongest bounds on heavy lepton masses reported by LEP have been reported
by OPAL\cite{opalbounds} and by L3\cite{l3bounds}.   Both experiments assume
that the heavy leptons decay via the charged current decay---the decay
$E\rightarrow \tau Z$, which can be large in vectorlike models, is not
considered (since it seldom dominates the charged current decay, their bounds
are not affected).   In the OPAL analysis, they exclude charged leptons which
decay via $E\rightarrow \nu_l W$ with masses below 80.2 GeV, and those which
decay via $E\rightarrow N W$ with masses below 81.5 GeV.  Unfortunately,
this latter decay assumes that the mass difference between the $E$ and the $N$
is greater than 8.4 GeV.  As we have seen in earlier sections, vectorlike
models have mass splittings on the order of a few hundred MeV, and even in
chiral models, the splitting could also be small.  Note, however, that if the
mixing angle between the third and fourth generation is bigger than $10^{-6}$,
then the $E\rightarrow \nu_l W$ would occur near the vertex, and the OPAL
bound would apply.   The bound was obtained for LEP at
$\sqrt{s}=170-172$ GeV, and can be improved somewhat for the later runs.

In the L3 analysis, the mass splitting was assumed to be larger than in the
OPAL case, greater than 10 GeV, and similar bounds were obtained.
The L3 analysis also looked for long-lived charged leptons, which would exist
if the mixing angles with lighter generations were small (typically less than
$10^{-7}$) and the charged lepton is lighter than its neutral partner.  Of
course, such leptons must eventually decay, for cosmological reasons, but we
discussed a variety of such scenarios earlier in this Report.   L3 excludes
such leptons up to a mass of $84.2$ GeV.

Both experiments also looked for heavy neutrinos, which decay at the vertex
(mixing angle greater than $10^{-6}$ or so) into a charged lepton and a $W$. 
For both experiments, the bounds for Dirac (Majorana) neutrinos are
approximately $78$ $(66)$ GeV for decays into electrons or muons and $70$
$(58)$ for decays into taus.

So, summarizing the current situation, the bounds on the charged heavy lepton
are approximately at the kinematic limit of the collider if and only if this
lepton is either stable (i.e. with a lifetime greater than tens of nanoseconds),
has a large ($8$ GeV or greater) splitting with its neutrino partner, or has a
relatively large mixing angle ($10^{-6}$ or greater) with lighter generations.  
Note that one of the most interesting models is the $E_6$ motivated model with
a vectorlike doublet with very small mixing, and this lepton satisfies {\bf
none} of the above conditions.  The mass bound for such a lepton is still only
given by
$Z$ decays.  (A search for a nearly degenerate lepton doublet was
reported\cite{mark2} many years ago by the Mark II detector, but only applied
for leptons lighter than $10$ GeV.)

Could more analysis at LEP find such a charged heavy lepton? In vectorlike models, 
the principal decay of the $E$ is into the $N$ plus a very soft pion.  It
appears to be impossible to pick this pion out of the background from soft tracks
from beam-beam interactions.  Recently, Thomas and Wells\cite{thomas} proposed
a new signature---triggering on an associated hard radiated photon.  This is
similar to proposals for counting neutrino species through $e^+e^-\rightarrow
\nu\overline{\nu}\gamma$.  At LEP, one would look for  $e^+e^-\rightarrow
L^+L^-\gamma$.  There are backgrounds from the above neutrino process, but
they can be reduced by looking for a displaced vertex (the decay length is
of the oorder of centimeters) and for the soft pions.  Thomas and Wells plot the
cross section as a function of the
$L$ mass and the minimum photon energy.  With an integrated luminosity of $240$
pb${}^{-1}$ at $\sqrt{s}=183$  GeV, and a minimum photon energy of $8$ GeV,
they estimate that a doublet mass of up to $70$ GeV could be detected. 
Presumably, this reach will be considerably higher for the more recent higher
energy runs.  They note that this signature is unusual for $e^+e^-$ machines
since it is not only limited by the machine energy, but also by the luminosity,
and that higher luminosity can significantly extend the reach.

In the chiral case, there remains a ``hole" to be filled.  If the $E$ is either
very close in mass to or lighter than the $N$, it will primarily decay via
mixing.  If the mixing angle is greater than about $10^{-6}$, then it decays
near the vertex and can be detected at LEP up to the kinematic limit.  If the
mixing angle is smaller than about $10^{-7}$, it is effectively stable and can
be detected at LEP up to the kinematic limit.  For intermediate angles, the
decay length is of the order of tens of centimeters to a meter.  Of course,
some will still decay near the vertex, and some will decay well within or
outside the detector, and so it is possible that a complete analysis could
close this hole.  Doing so would be useful, since one of the plausible values
for the mixing angle discussed in Section III is given by
$\sqrt{m_{\nu_\tau}/M_N}$, which, using
$0.01$ eV for the $\nu_\tau$ mass, gives $3\times 10^{-7}$ for the mixing angle.

In the future, similar analyses to the above will give similar bounds at lepton
colliders near the kinematic limit of the colliders.  We have noted, however,
that  the detection of the vectorlike doublet leptons remains problematic and
can best be attacked looking for an associated hard photon.   There also seems
to be a window for decay lengths of the order of tens of centimeters which has
yet to be closed. 

  The decay mode
$E\rightarrow \tau Z$ is generally smaller than the charged current
decay--although it is a much cleaner mode, backgrounds are not the problem for
$E\rightarrow \nu W$, thus the latter would be detected first.   This neutral
current decay mode, however, provides a much cleaner signature for hadron
colliders, which we now discuss.

A study of searches for heavy charged leptons at hadron colliders was
performed by Frampton et al.\cite{fram}   They considered charged lepton
production at the SSC and at the LHC (at 17 TeV).   There are two main
productin mechanisms for heavy leptons at hadron colliders.  The first is gluon
fusion, through a triangle graph, into a Higgs boson or a Z-boson.  The second
is quark fusion directly into a Z-boson (the effects of photon exchange are
much smaller than those of the Z).   The cross section for lepton production
through quark fusion falls off very rapidly as the lepton mass increases, but
the cross section through gluon fusion does not fall off as rapidly, since the
matrix elements increases as the square of the lepton mass.

First consider the chiral case.  Here, gluon fusion dominates for lepton masses
above about 150 GeV, and the total cross section for masses between 100 and 800
GeV drops from 0.5 pb to 0.05 pb.   This will lead to many thousands of events
per year at the LHC.  The signature would be a conventional heavy lepton
signature. For those masses, and for chiral leptons, one can expect a
reasonably large splitting between the $N$ and the $E$, leading to standard
single lepton and missing momenta signatures; even if there was an unexpected
degeneracy (or if the $N$ were heavier) mixing would lead to clear signatures
(note, as discussed above, the importance of closing the window for mixing
angles near $10^{-6}$).

What about the vectorlike case?  Here, gluon fusion doesn't contribute, since
the leptons don't couple to the Higgs and the vectorlike coupling to the Z
gives no contribution due to Furry's theorem.  Thus, the contributions are only
through quark fusion, which fall off much faster.  As the lepton mass increases
from 100 GeV to 800 GeV, the cross section falls from 1 pb to 0.001 pb.   For a
400 GeV heavy lepton, this will give only 1000 events annually at the LHC.
This makes detection more difficult, however one should recall that these
leptons can decay via the neutral current: $E\rightarrow \tau Z$ which has a
branching fraction of at least a few percent (and in some models much larger).
This would give a very clear signature with very low background.  Even if the
decay is suppressed by very small mixing angles, and thus the $E$ passes
through the detector, stable lepton searches should see it (it can be readily
distinguished from a muon by time-of-flight using the velocity distributions as
given in Ref. \cite{fram}.).

\section{Conclusions.}

There are still several reasons to
believe that further quarks and leptons remain to be
discovered. Although the fourth
or further quark-lepton generations cannot be exactly
similar and sequential to the 
first three generations, there are plenty
of alternative possibilities which avoid
the experimental embarassment to the
invisible Z partial width of a 
fourth light neutrino. The additional quarks 
and leptons may be chiral as in the first three generastions
or non-chiral and vector-like.

The allowed masses are constrained by the
precise electroweak data particularly at the 
Z pole where the data now agree with the minimal standard
model at an astonishing 0.1\% level. The S, T, U
parameters then restrict what states may 
be added as discussed above in
Section (III). Also
the stability of the observed vacuum places  constraints
on additional fermions as does the (optional)
requirement of grand unification of the three gauge
couplings. Mixing angles for the new quarks and leptons
are relatively unconstrained, except by unitarity,
without new experimental data.

The lifetime and decay modes (see Section (IV)) of 
a heavy lepton depend critically
on whether the N or E state is the 
more massive. A similar dependence occurs
for heavy quarks which may have such small mixing
with the known quarks that at least
one new quark may have an exceptionally long lifetime.

In Section (V) we have considered the fascinating 
possibility that the Higgs boson is not elementary 
but rather some bound state of additional fermions which
transform under the standard gauge group.
The heaviness of the top quark has 
suggested to some that it plays a special role
in electroweak symmetry breaking, but even heavier 
fermions are more attractive candidates to 
participate in dynamical symmetry breaking.

CP symmetry violation has two disparate but likely related aspects
in the Standard Model: the strong CP problem and the weak CP
violation in kaon decay.
Strong CP can be addressed by addition of
extra quarks as explained in our Section (VI). Weak CP
violation by the KM mechanism requires at least three generations,
and acquires even more CP violating phases in the presence
of additional quarks.
We have illustrated this with the Aspon model which invokes spontaneous
CP violation to relate solution of the strong 
CP problem by extra vector-like quarks
to the violation of CP symmetry in kaon decay. 
The new vector-like quarks may have long
lifetime as mentioned in Section (IV).

Experiment is the final arbitor of everything 
we have reviewed. Long-lived quarks are being sought
at collider facilities. To some extent, detectors have not been
designed for such a possibility and 
this review may encourage further thought in
detector design. Similarly heavy leptons 
are being, and will be,
investigated at existing and future colliders. 

Discovery of a further quark or lepton would be 
revolutionary and propel 
high-energy physics in a new and exciting direction.

\bigskip
\bigskip
\bigskip

This work was supported in part by the US Department of
Energy under Grants No. DE-FG02-97ER41036 and DE-A505-89ER40518, and 
by NSF Grant No. PHY-9600415.   We would like to thank David Stuart
for many useful discussions about the CDF Collaboration.

\def\prd#1#2#3{{\rm Phys. ~Rev. ~}{D#1}, #3 (19#2)  }
\def\plb#1#2#3{{\rm Phys. ~Lett. ~}{B#1}, #3 (19#2)  }
\def\npb#1#2#3{{\rm Nucl. ~Phys. ~} {B#1}, #3 (19#2) }
\def\prl#1#2#3{{\rm Phys. ~Rev. ~Lett. ~}{#1}, #3 (19#2) }

\end{document}